\documentclass[twocolumn,trackchanges]{aastex63}
\usepackage{amsmath}
\usepackage{graphicx}
\usepackage[normalem]{ulem}
\usepackage{txfonts}
\usepackage{hyperref}

\submitjournal{AJ}
\shorttitle{MIRC-X: a highly-sensitive six telescope interferometric imager at the CHARA Array}
\shortauthors{N. Anugu et al.}
\graphicspath{{./}{Figures/}}

\begin{document}

\title{MIRC-X: a highly-sensitive six telescope interferometric imager at the CHARA Array}
\correspondingauthor{Narsireddy Anugu}
\email{nanugu@arizona.edu}

\author[0000-0002-2208-6541]{Narsireddy Anugu}
\altaffiliation{Steward Observatory Fellow for Instrumentation Development}
\affiliation{School of Physics and Astronomy, University of Exeter,  Exeter, Stocker Road, EX4 4QL, UK}
\affiliation{Steward Observatory, Department of Astronomy, University of Arizona, Tucson, USA}
\affiliation{University of Michigan, Ann Arbor, MI 48109, USA}

\author[0000-0002-0493-4674]{Jean-Baptiste Le~Bouquin}
\affiliation{University of Michigan, Ann Arbor, MI 48109, USA}
\affiliation{Institut de Planetologie et d’Astrophysique de Grenoble, Grenoble 38058, France}

\author[0000-0002-3380-3307]{John D. Monnier}
\affiliation{University of Michigan, Ann Arbor, MI 48109, USA}

\author[0000-0001-6017-8773]{Stefan Kraus}
\affiliation{School of Physics and Astronomy, University of Exeter, Exeter, Stocker Road, EX4 4QL, UK}

\author[0000-0001-5980-0246]{Benjamin R. Setterholm}
\affiliation{University of Michigan, Ann Arbor, MI 48109, USA}

\author[0000-0001-8837-7045]{Aaron Labdon}
\affiliation{School of Physics and Astronomy, University of Exeter, Exeter, Stocker Road, EX4 4QL, UK}

\author[0000-0001-9764-2357]{Claire L Davies}
\affiliation{School of Physics and Astronomy, University of Exeter, Exeter, Stocker Road, EX4 4QL, UK}

\author[0000-0001-9745-5834]{Cyprien Lanthermann}
\affiliation{Instituut voor Sterrenkunde, KU Leuven, Celestijnenlaan 200D, 3001 Leuven, Belgium}
\affiliation{Institut de Planetologie et d’Astrophysique de Grenoble, Grenoble 38058, France}

\author[0000-0002-3003-3183]{Tyler Gardner}
\affiliation{University of Michigan, Ann Arbor, MI 48109, USA}

\author[0000-0002-1575-4310]{Jacob Ennis}
\affiliation{University of Michigan, Ann Arbor, MI 48109, USA}

\author[0000-0002-3766-5204]{Keith J. C. Johnson} 
\affiliation{University of Michigan, Ann Arbor, MI 48109, USA}

\author[0000-0002-0114-7915]{Theo ten Brummelaar}
\affiliation{CHARA Array, Georgia State University, Atlanta, GA 30302, USA}

\author[0000-0001-5415-9189]{Gail Schaefer}
\affiliation{CHARA Array, Georgia State University, Atlanta, GA 30302, USA}

\author{Judit Sturmann}
\affiliation{CHARA Array, Georgia State University, Atlanta, GA 30302, USA}

\begin{abstract}

MIRC-X (Michigan InfraRed Combiner-eXeter) is a new highly-sensitive six-telescope interferometric imager installed at the CHARA Array that provides an angular resolution  equivalent of up to a 330\,m diameter baseline telescope in J and H band wavelengths ($\tfrac{\lambda}{2B}\sim0.6$ milli-arcseconds).  We upgraded the original MIRC (Michigan InfraRed Combiner) instrument to improve sensitivity and wavelength coverage in two phases. First, a revolutionary sub-electron noise and fast-frame rate C-RED ONE camera based on a SAPHIRA detector was installed. Second, a new-generation beam combiner was designed and commissioned to (i) maximize sensitivity, (ii) extend the wavelength coverage to J-band, and (iii) enable polarization observations. A low-latency and fast-frame rate control software enables high-efficiency observations and fringe tracking for the forthcoming instruments at CHARA Array.  Since mid-2017, MIRC-X has been offered to the community and has demonstrated best-case H-band sensitivity down to 8.2 correlated magnitude. MIRC-X uses single-mode fibers to coherently combine light of six telescopes simultaneously with an image-plane combination scheme and delivers a visibility precision better than 1\%, and closure phase precision better than $1^\circ$. MIRC-X aims at (i) imaging protoplanetary disks, (ii) detecting exoplanets with precise astrometry, and (iii) imaging stellar surfaces and star-spots at an unprecedented angular resolution in the near-infrared. In this paper, we present the instrument design, installation, operation, and on-sky results, and demonstrate the imaging capability of MIRC-X on the binary system $\iota$\,Peg. The purpose of this paper is to provide a solid reference for studies based on MIRC-X data and  to inspire future instruments in optical interferometry.
\end{abstract}


\section{Introduction} \label{sec:intro}

Long baseline optical and infrared interferometers can achieve unprecedented angular resolution down to a fraction of milli-arcseconds, but they have historically limited by the number of telescopes to make model-independent images and instrument sensitivity to observe fainter targets. Currently, CHARA Array\,\citep[California, USA;][]{tenBrummelaar2005} and VLTI\,\citep[Antofagasta, Chile;][]{Haguenauer2012} are the only two large facilities capable of combining four or more telescopes with telescope sizes of 1\,m or more. While, NPOI\,\citep[Arizona, USA;][]{Armstrong1998} is evolving towards extended imaging capabilities and sensitivity, and MROI\,\citep[New Mexico, USA;][]{Creech-Eakman2018} is under the construction phase. MIRC\,\citep{Monnier2006, Monnier2010, Che2010} was, the only six-telescope beam combiner in the near-infrared,  built to leverage the six available telescopes and largest optical baselines (up to $B=330$\,m) of the CHARA Array.  MIRC has achieved landmark results in stellar astrophysics, for instance, by imaging the fireball expansion phase of a nova explosion\,\citep{Schaefer2014}, the transit of an eclipsing binary system\,\citep{Kloppenborg2010} and the surfaces and spots of other stars\,\citep{Monnier2007, Monnier2012, Roettenbacher2016}.

However, to achieve the challenging goals of imaging disks around faint young stellar objects (YSOs), astrometric planet detection and other key science cases in stellar astrophysics, it was required to redesign and rebuild MIRC substantially for higher sensitivity and precision visibility ($V$) and closure phase ($C$) observations. The upgraded instrument, MIRC-X, features a revolutionary sub-electron noise and high-frame-rate near-infrared detector system. We also added new optical functions and operational modes to MIRC-X such as (i) simultaneous J and H band observations, (ii) control of polarization and dispersion effects, (iii) polarization-maintaining fibers, and (iv) a new beam combiner with crosstalk resistant scheme and higher-throughput photometric channels. Furthermore, the control software was upgraded to improve maintainability, increase observation efficiency, and enable simultaneous co-phased observations with future instruments, e.g., MYSTIC\,\citep{Monnier2018} and SPICA\,\citep{Mourard2017}.

The MIRC-X instrument was commissioned in two phases (June 2017 and September 2018) in order to minimize risks during the upgrade. In this manuscript, we provide a detailed overview of MIRC-X in its current state at CHARA Array, following more than two years of operations. Section\,\ref{sec:2} presents the main science drivers of MIRC-X and the corresponding high-level instrument specifications. Section\,\ref{sec:3} details the instrument concepts, its subsystems, and software control. Section\,\ref{sec:4} describes the MIRC-X operations, including daily activities, acquisition of flux fringe tracking, and raw data recording. It also quickly describes the data reduction pipeline and calibration procedures. Section\,\ref{sec:5} details the MIRC-X performance focusing on sensitivity and stability. It illustrates the MIRC-X image reconstruction capability on a well known spectroscopic binary, $\iota$\,Peg. The paper ends with brief conclusions and perspective for future upgrades.

\section{\label{sec:2}Science drivers and specifications}
MIRC-X combines the six-telescopes of the CHARA Array allowing the simultaneous measurements of 15 baselines and 20 closure phases. It is designed for model-independent aperture synthesis imaging by observing in J and H-band wavelengths with precision closure phases and visibilities.

\subsection{Science cases}
The key science cases for MIRC-X are the following:
\begin{itemize}
\item Imaging young stellar objects: the goal is to characterize the physical conditions of planet formation and to unveil time-variable structures in proto-planetary disks.  The wavelength dependent emission of disks requires broad wavelength coverage, including eventually simultaneous J+H+K observations. MIRC-X probes inner-disk features (down to a fraction of au) and complements the observations of SPHERE~\citep{Beuzit2019}, GPI~\citep{Macintosh2014}, and ALMA~\citep{ALMA2015}, which probes outer disk features of YSOs (i.e., approximately 20 to 500 astronomical unit, au) at near-infrared and sub-millimeter wavelengths.

\item Detecting exoplanet with high-precision astrometry: The first detection of  exoplanets by optical interferometry was achieved by VLTI/GRAVITY\,\citep{Lacour2019B,Nowak2020} with foreknowledge of the location of the exoplanets. A high-precision differential astrometry of binary systems could also allow detection of a planet by observing the `wobble' in the binary orbit caused by the gravity of the planet\,\citep{Gardner2018}.  The precision of a few micro-arcseconds ($\mu$as) is necessary to detect Jupiter-mass exoplanets at a separation of $\approx2\,$au or a 4 Jupiter-mass planet at $\approx0.5$~au separation\,\citep{Zhao2011}. This differential astrometry mode allow us to search exoplanets in regimes that are difficult to probe with transit and radial velocity surveys, such as around hot binary stars.
    
\item Imaging stellar surfaces and star spots: imaging stellar surfaces and spotted stars reveal fundamental information about stellar dynamos in comparison to the solar dynamo\,\citep{Monnier2007,Roettenbacher2016,Ohnaka2017,Paladini2018}. 

\item Spectro-interferometry in J and H-bands:  spectro-interferometry provides direct constraints on the spatial distribution of the line-emitting gas. For instance, H-band Br 6 to 12 line emission\,\citep{Kreplin2020}, He-I 1.08\,$\mu$m, and Pa-$\gamma$ 1.094\,$\mu$m are available (Pa-$\beta~1.282~\mu$m line is not accessible because the 1.319 $\mu$m metrology laser is filtered out). There are also interesting line tracers in the H-band from the higher-transition Br$\gamma$ line emission\,\citep{Weigelt2007,Malbet2007} and some forbidden metallic lines. However, these lines appear prominently only in very few YSOs and are typically also weaker than the J-band transitions.  


\item Polar-interferometry for scattered light imaging: polar-interferometry can measure astrophysical polarization signals that might be caused by dust scattering\,\citep{Ireland2005,LeBouquin2008}. 

\item Imaging binary-disk interactions in post-asymptotic giant branch binaries\,\citep[cf. ][references therein]{Hillen2016,Kluska2019}.

\item Carry out large-scale binary surveys: The snapshot modeling and imaging capability with many observables and high precision calibration allow for the full interferometric field-of-view to be searched down to high contrast ratios (500:1 or better) with high efficiency.  MIRC-X bridges the gap between radial velocity and speckle/adaptive optics techniques to  fill long-standing gaps in our statistics of binary demographics fully\,\citep{Sana2014}.

\end{itemize}

\subsection{Technical requirements}
To achieve the aforementioned science cases, the following are the technical requirements:

\begin{itemize}
\item Sensitivity boost of 2 magnitudes:  the former MIRC instrument's sensitivity was limited by an aged high-readout detector (PICNIC, $N_{\rm RON}\approx14~\rm{e^{-}px^{-1}}$ root-mean-square, RMS) coupled with the lower throughput of its optical fibers. The faintest stars that secured fringe detection on MIRC were in the range of 4-6 magnitudes in the H-band depending on seeing conditions. Consequently, observations of faint YSO disks, as well as observations requiring higher spectral resolution (i.e., for detecting exoplanet companions with precision astrometry) or polarimetric measurement could not be practically conducted with the former MIRC. These modes indeed require to spread the light over many more pixels, which reduces the signal-to-noise (SNR) ratio. As an example, a sensitivity improvement of two magnitudes increases the number of accessible YSOs from $4$ to about $40$.

\item J + H  simultaneous observations: Previous CHARA observations have shown that YSO objects are very resolved, in many cases, down to a raw visibility contrast $\leq10$\%\,\citep[from MIRC experience]{Tannirkulam2008}. We require additional J and K-band observing modes to supplement the existing H-band to enable simultaneous J+H+K observations covering objects with higher correlated flux in either J/H/K.   Furthermore, the simultaneous J+H+K observations cover the astrophysical object in different colors and extend the (u, v)-coverage. Therefore, we design H-band optimized for sensitivity, but are able to detect and track fringes in J-band as well. The MIRC-X twin instrument MYSTIC\,\citep{Monnier2018} will cover the K-band observations.

\item Simultaneous observations with MYSTIC: Coordinated simultaneous observations between MIRC-X and MYSTIC are required with a 95\% overlap of integration time between them, to allow efficient a-posteriori fringe-tracking in the pipeline. Simultaneous observations will include the flexibility of using MIRC-X or MYSTIC as a fringe tracker based on the correlated magnitude of objects depending on their baseline length and wavelength.

\item Precision and wavelength calibration: \citet{Gardner2018} demonstrated a few $\mu$as precision with MIRC on the 10~milli-arcseconds (mas) separation binary $\delta$ Del. However, for wider binary separations, $\geq100$\,mas, we need wavelength calibration stability at the level of $\Delta\lambda / \lambda\approx10^{-5}$ level for $\leq 10~\mu$as astrometric precision. The original MIRC was limited to a wavelength calibration stability of $\approx10^{-3}$\,\citep{Monnier2012}, 100-times larger than required for planet detection.

\item Spectro-interferometry in J and H-bands: Install a medium spectral resolution grism of R~$\geq1000$ to study velocity-integrated imaging of the line-emitting regions.  Smaller prisms are also required for sensitivity (R~=~50) and for high-precision astrometry (R~=~190) covering the interferometric field-of-view ($\approx R\frac{\lambda}{B}$) approximately of the diffraction limit of the CHARA 1\,m telescopes.

\end{itemize}

\subsection{Instrument high-level specifications and  trade offs}

The following upgrades were required in comparison to the MIRC instrument considering the aforementioned science requirements:

\begin{itemize}
\item Sensitivity: (i) Achieve higher sensitivity with the state-of-the-art sub-electron noise and fast-frame electron avalanche photodiode technology (eAPD) infrared detector camera, C-RED ONE\,\citep{Gach2016} (ii) Redesign of MIRC photometric channels to have higher throughput  and to be less sensitive to alignment drifts.  MIRC-X uses bulk optics  to re-image fibers onto the slit instead of MIRC multimode fibers as transport \citep{Che2010}.

\item J+H simultaneous observations: (i) Install new polarization-maintaining fibers with excellent performance over full J and H bands. (ii) Implement polarization birefringence control for minimizing birefringence due to fiber length mismatch and other polarization issues induced by the beam-train. (iii) Implement chromatic dispersion control for J-band as the CHARA Array delay lines are not entirely in a vacuum.

\item Precision: (i) Enable crosstalk resistant beam combination optimized for precision visibility and closure phases. (ii) MIRC-X uses a high-sensitive and high-dynamic range Leonardo SAPHIRA detector. By increasing the avalanche gain in eAPD, we can achieve a very low readout-noise by amplifying electrons produced by photons, but at the expense of a lower dynamic range \,\citep[e.g. PIONIER,][]{LeBouquin2011}. This lower dynamic range can limit the high SNR observations, but this is only $\leq30\%$ loss for our camera. For bright object observations such as Betelgeuse ($\rm{mH} = -3.7$) we use smaller avalanche gain and higher spectral resolutions to avoid saturating the detector. (iii) Accurate wavelength calibration using etalon.

\item Simultaneous observing with MYSTIC: Design of MIRC-X hardware and software to allow simultaneous observations with the forthcoming twin instrument MYSTIC working in K-band, to provide more extensive wavelength coverage and higher spectral resolution through fringe tracking.

\item Spectro-interferometry in J and H-bands: Install R~=~50, R~=~190 and  R~=~1035 dispersion elements for higher sensitivity mode, astrometry mode  and to study velocity-integrated imaging of the line-emitting regions. The higher spectral resolution  choices for MIRC-X are limited  ($R \leq 1350$ in H-band) due to the smaller size of the detector ($320\times256$ pixels) and obtaining high SNR necessitates longer integration times, which requires a co-phasing performance of the instrument. Considering these limitations, we acquired a trade-off medium resolution R=1035 in H-band commercial-off-the-shelf grism from Newport (catalog 53-*-880R). 

\item Polar-interferometry for scattered light imaging: Install a half-wave plate (modulator/retarder) for each beam and a Wollaston prism (beam analyzer) to measure the full Stokes parameters. This is accomplished by taking data at $0^\circ$, $22.5^\circ$, $45^\circ$, and $67.5^\circ$ half-wave plate rotations to measure Stokes parameters.  The Wollastron prism is a trade-off between how many pixels it occupies on the detector in combination with the spectral dispersion unit and the sensitivity required for YSO observations.  The selected 73 pixels occupying United Crystal $\pm0.5^\circ$ deviation Wollaston prism is suitable to work with spectral dispersions of R~=~50 and R~=~190 for J+H bands. 

\item Observing efficiency: Transition the real-time control and observing user interfaces to the CHARA standard, to allow efficient observations and remote observing capabilities for long-term follow-up such as precision astrometry or target of opportunity events.
\end{itemize}

\begin{figure*}
\centering
\includegraphics[width=0.95\textwidth]{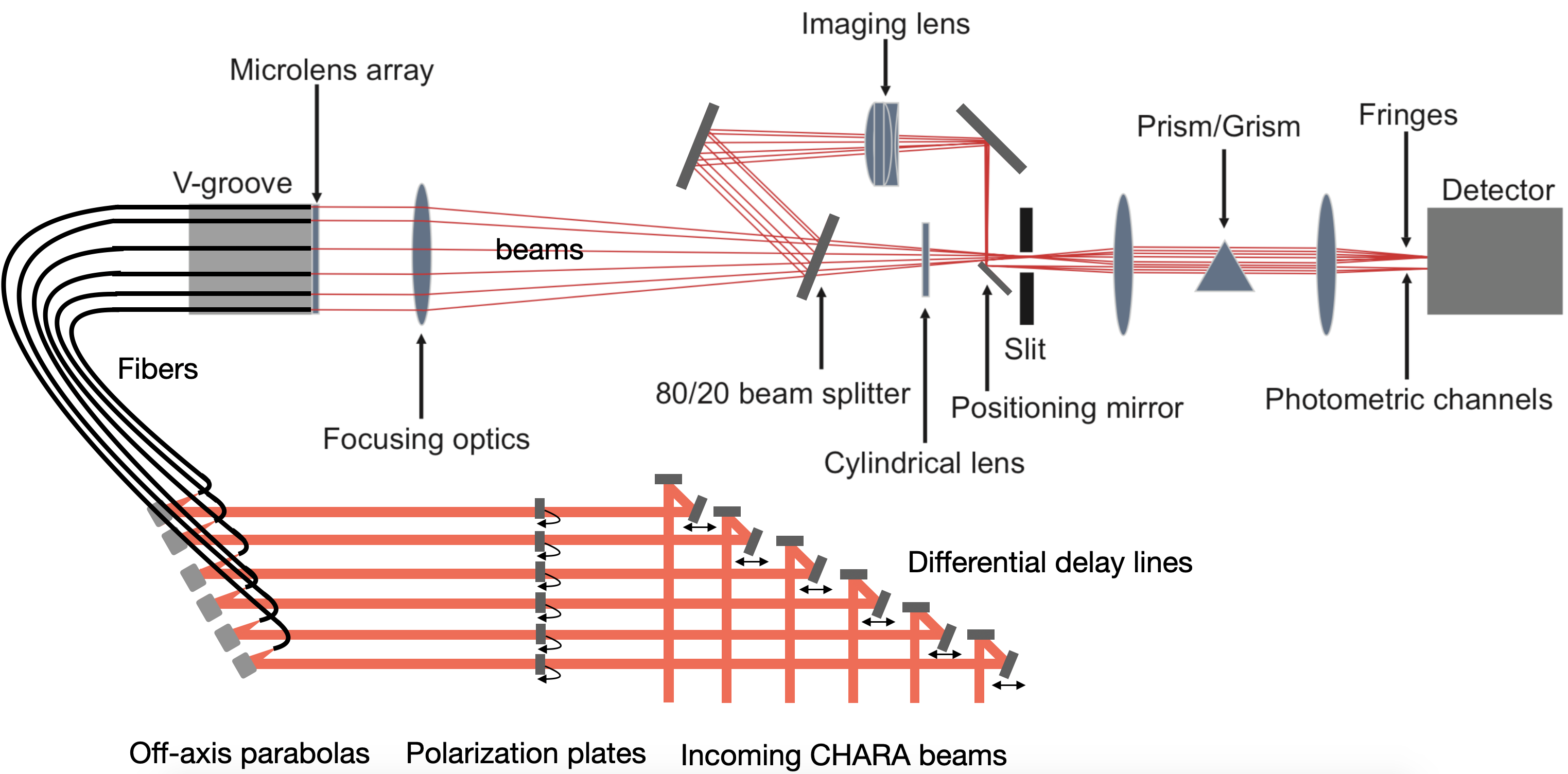}
\caption{Conceptual sketch of MIRC-X. The beams are first redirected by MIRC-X pick-off mirrors, which also act as internal delay lines. The beams cross the rotating birefringent LiNbO3 plates. Off-axis parabolas focus the light into single-mode fibers. The fibers are arranged non-redundantly in a V-groove and collimated with a micro-lens array. These beams are recombined together at the focus of a long-focal spherical mirror (here shown as a lens). A cylindrical lens compresses the fringe pattern in the spectral direction to define a slit. An 80/20 beam splitter extracts the photometric channels, which are re-imaged into the slit. The slit is 1:1 re-imaged on the camera through the spectrograph. \label{Fig1_mircx_layout}}
~\\


\centering
\includegraphics[width=0.95\textwidth]{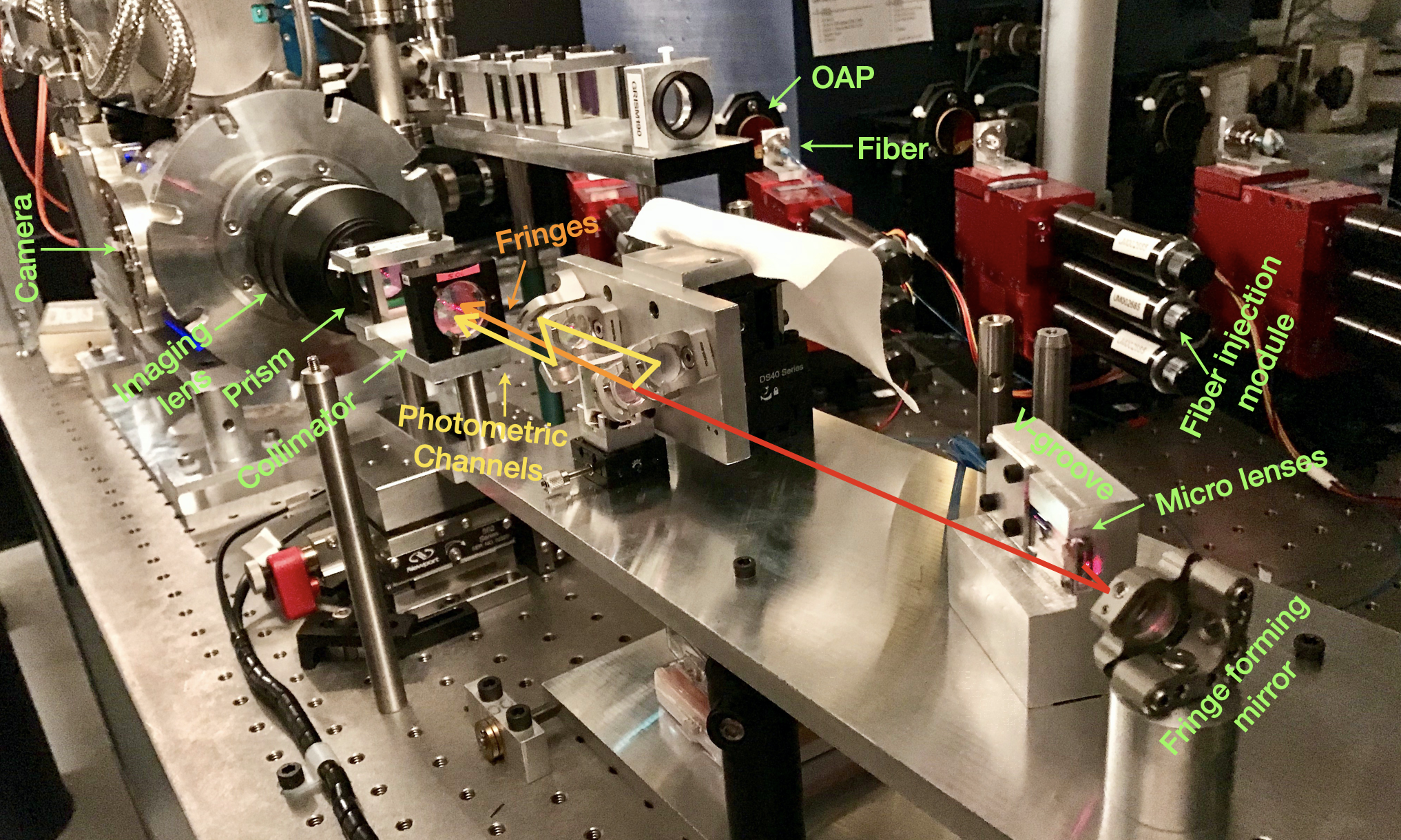}
\caption{Picture of MIRC-X showing the injection module, fibers, beam combiner, spectrograph and C-RED ONE camera. \label{Fig4_MIRC-X_photo}}
\end{figure*}

\begin{figure}
\centering
\includegraphics[width=0.45\textwidth]{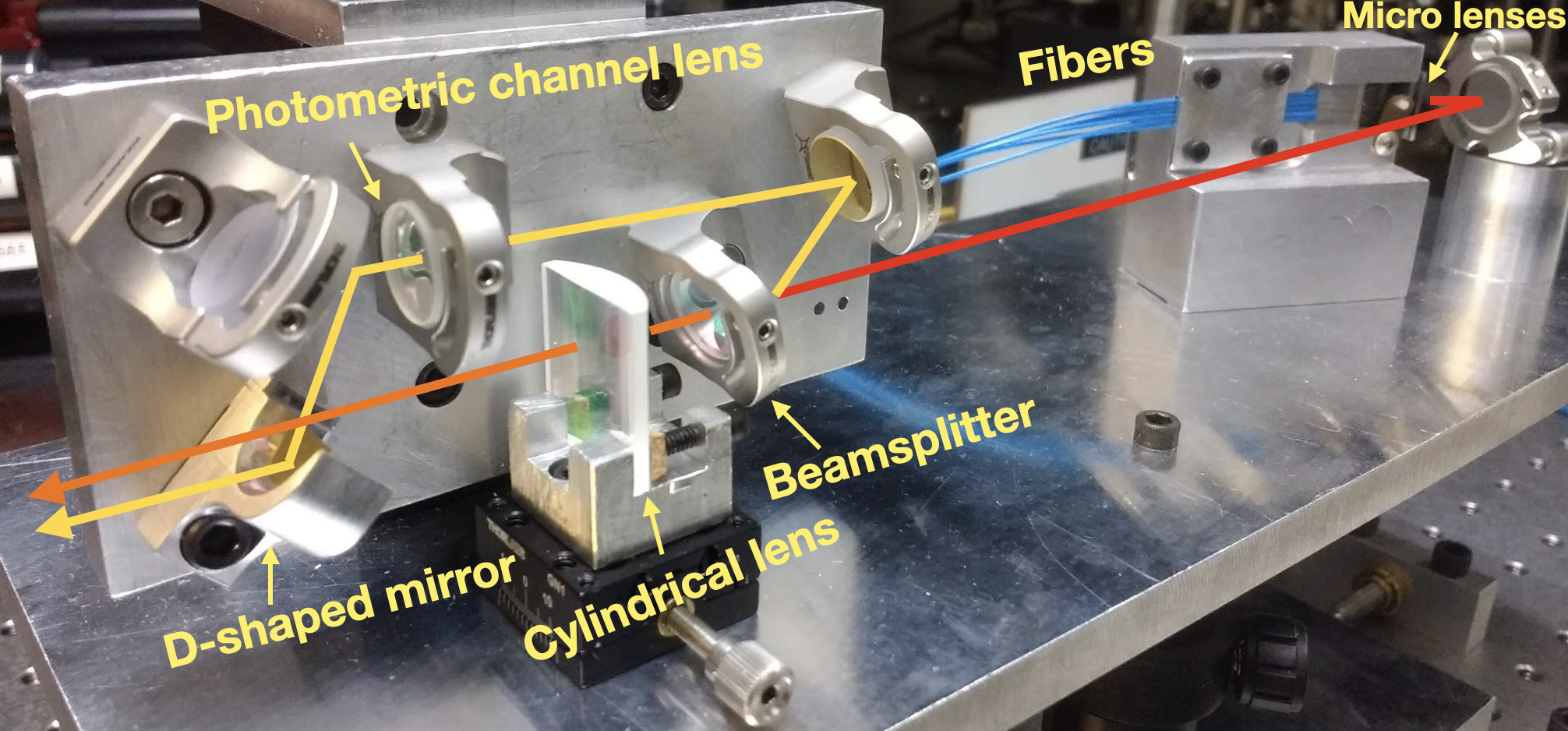}
\caption{Front view of the MIRC-X beam combiner.}
\label{Fig3_mircx_beamcombiner_frontview}
\end{figure}

\section{\label{sec:3}Instrument concept}
The MIRC-X instrument overview is illustrated in Figure\,\ref{Fig1_mircx_layout} and pictured in  Figure~\ref{Fig4_MIRC-X_photo} and \ref{Fig3_mircx_beamcombiner_frontview}. In the following, we describe the various sub-systems in more detail.

\subsection{\label{sec:Dispersion}Chromatic dispersion control}

The CHARA Array delay lines are not in a vacuum, so the beams travel through the open air inside the lab. Differential chromatic dispersion of air between the different beams requires correction, especially at shorter wavelengths where dispersion is dominant. When using the J-band, MIRC-X makes use of the CHARA built-in longitudinal dispersion compensator (LDC) optic made of a pair a glass prisms~\citep{Berger2003}.  The amount of dispersion correction necessary is a function of the wavelength (the lower the wavelength, the larger the correction), target star location on-sky, and environmental conditions such as temperature and humidity. This error is modeled and corrected at a 1\,Hz frequency.

\subsection{Pick-off mirrors, internal delay lines}

The 19-mm diameter collimated CHARA beams are extracted with a set of 1-inch mirrors that flip up into the beam path using Newport New Focus 8892-K Motorized Flipper Mounts. These mirrors also act as 
internal differential delay lines to co-phase with other instruments at CHARA Array. The flipper mounts are mounted on linear stages with Zaber T-LA28A motors and controllers. These delay lines have a 14\,mm-mechanical range with repeatability better than 4\,$\mu$m, and with a step size of 0.1\,$\mu$m. The dynamical equalization of the path-lengths traveled by the starlight to the instrument is accomplished with the main CHARA delay lines \citep{tenBrummelaar2005}.

The incidence angle on the pick-off mirror is a few degrees only, allowing to move the internal optical path delay by several millimeters without shifting the pupil significantly. The beams are then folded toward the instrument.

\subsection{\label{sec:Polarization}Polarization birefringence control}
One of the main technical hurdles for optical long-baseline interferometry is polarization control. If orthogonal polarizations (vertical or horizontal axes) have unmatched phases, the detected fringe contrast is reduced. In MIRC-X, this is corrected using rotating birefringent plates, following the approach implemented on PIONIER\,\citep{Lazareff2012}. MIRC-X uses 4 mm thickness, AR-coated z-cut Lithium Niobate (LiNbO3) plates from Crylight. The plates are motorized using rotation stage AY110-60 from OES with R256 controllers from Lin Engineering. The device induces a phase-shift between the vertical or horizontal axes of up to $5\,\lambda$ and with a resolution better than $\lambda/20$.

\begin{figure}
\centering
\includegraphics[width=0.45\textwidth]{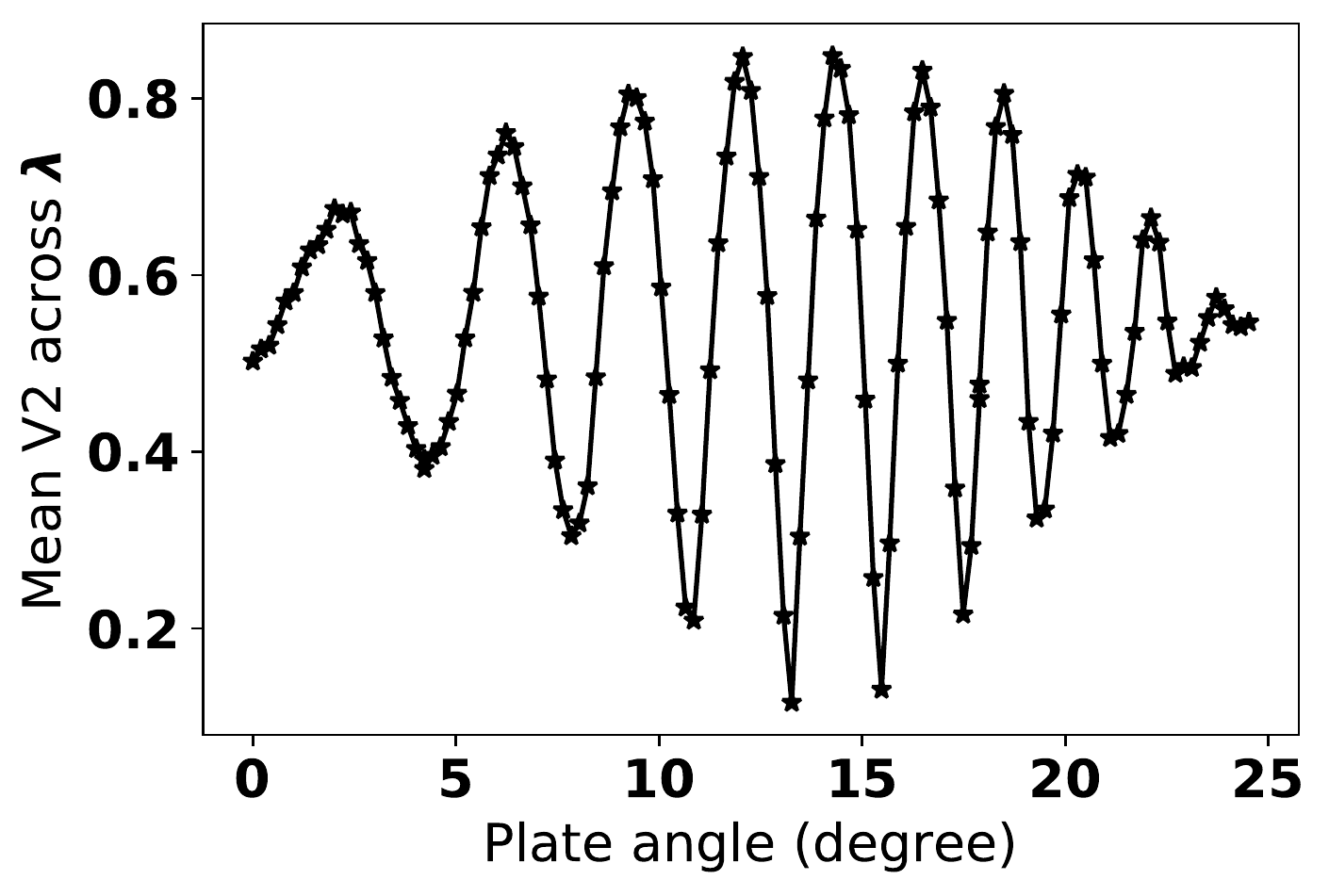}
\caption{MIRC-X internal source measured visibility square (V2) as a function of the LiNbO3 plate angle. The visibility is averaged over the wavelength band (prism R~=~50).}
\label{Fig4_Polarization_plate_vis2}
\end{figure}

\begin{figure}
\centering
\includegraphics[width=0.45\textwidth]{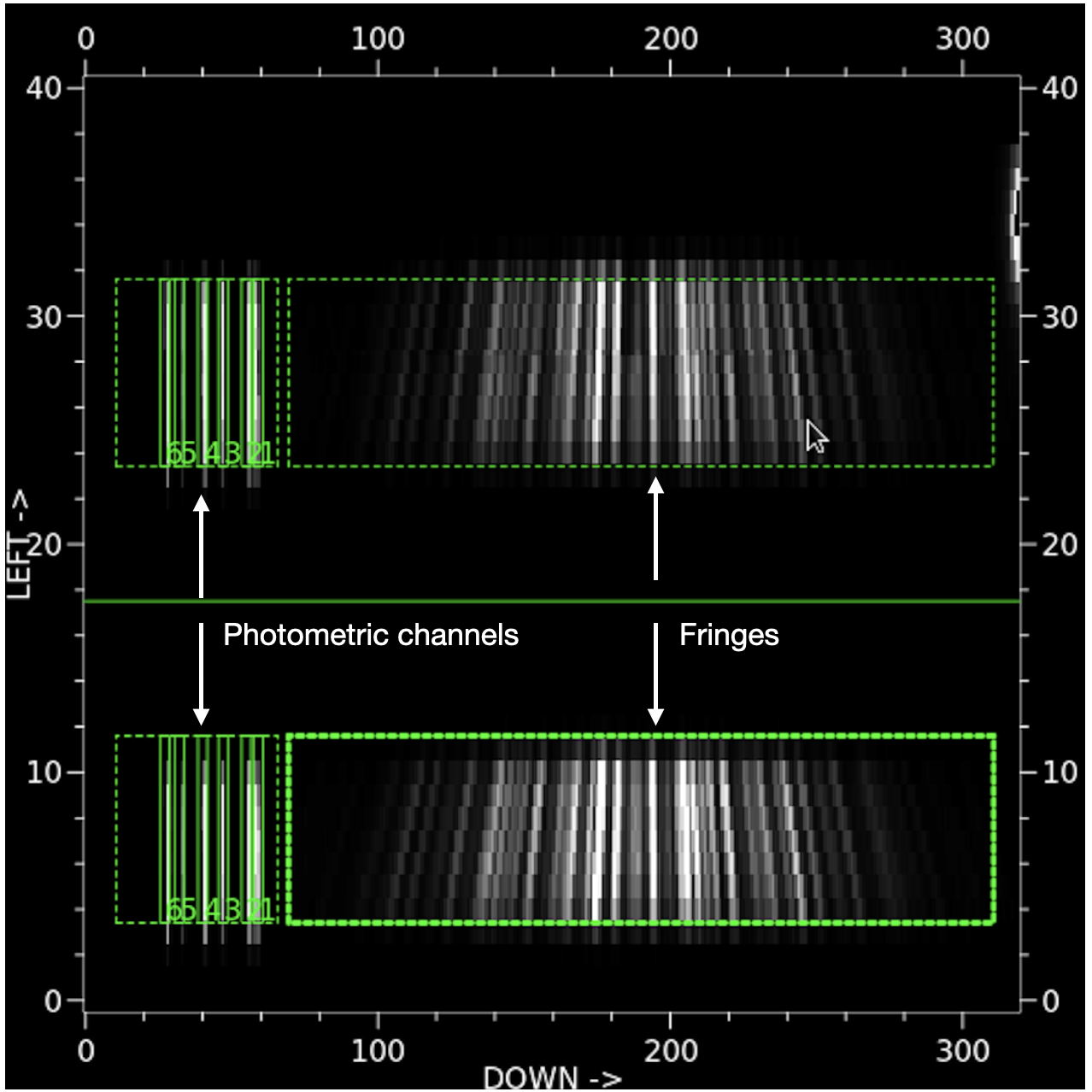}
\caption{H-band photometric channels and fringes (see Section \ref{sec:Beamcombiner}) obtained with a Wollaston prism.  The upper and lower panel present fringes and photometric channels in the orthogonal polarizations (horizontal/vertical).   The green numbers are the numbers of photometric channels. }
\label{Fig5_Wollastron_fringes}
\end{figure}

The optimized positions of LiNbO3 plates for a maximum contrast is found with a polarization explorer scanning (see Figure\,\ref{Fig4_Polarization_plate_vis2}). The fringe contrast is measured as a function of the LiNbO3 plate angle for all six internal beams. The plates are then rotated to the angles that achieve maximum contrast. We verified that the angle of the best contrast in natural light (method described above) matches with the differential phase angle measured between the two polarizations split by the Wollaston (see Figure\,\ref{Fig5_Wollastron_fringes}). This second method opens up the possibility to actually tune the LiNbO3 plates on-sky as  the first exploration method is extremely challenging because of the low SNR on faint stars, and the constantly changing raw contrast on bright and resolved stars.

Chromatic effects between J and H-bands add complications when trying to co-phase the J and H bands together simultaneously, so currently, we optimize the polarization controller to only one of the two bands. However, we acquired a better fiber length-matched (error  $<$0.7mm) V-groove bundle and plan to install it by the end of 2020 to lower the birefringence effects in J and H-bands.

For polar-interferometric science observations, a rotating half-wave plate from Wuhan Union Optic. inc. is installed upstream of the LiNbO3 plates on each beam. The combination of a half-wave plate (modulator/retarder) and Wollaston prism (beam analyzer) allows for the measurement of all Stokes parameters\,\citep{Ramaprakash1998, LeBouquin2008}. The polarization differential visibility measurements will allow for observations of asymmetric emission distributions such as dust scattering after proper calibration of instrumental effects.

\subsection{Injection modules}
\label{sec:fiber_explorer}

Next, the beams are redirected towards the fiber injection units, where the incident flux in each beam is focused by a $60$\,mm focal length off-axis parabola (OAP) onto the end of a mounted optical fiber. Each fiber is installed on 3-axis nano-precision positioning mounts (from Luminos), and the fiber tips are aligned for optimal flux coupling. These mounts have a repeatability error of 0.2\,$\mu$m in the horizontal (X) and vertical (Y) directions, and a few microns in the focus (Z) direction.

The initial alignment of the fiber in the injection module has been done by optimizing the image quality, so that the fiber core is centered at the focus of the OAP. This is done in the lab, with a dedicated setup, and using the manual screws of the Luminos mounts. The motorized axes of the mount are used for fine-tuning the injected flux, namely by scanning the fiber mounts over a small field of view. The resulting map is fitted with a Gaussian model to find the position that results in the best flux injection. This fiber explorer can use the flux measured in the photometric channels (defined hereafter), so that all beams are explored simultaneously; or, alternatively, use the flux in the fringe window, which is useful when the positions of the photometric channels on the detector have been lost entirely.

The fiber can be moved in the focus direction, to correct the non-common-path error between MIRC-X and the CHARA guiding system. But this procedure is not automated.

\subsection{Single-mode fibers}

MIRC-X uses single-mode fibers to spatially filter the effects of wavefront errors caused by atmospheric turbulence \citep{Shaklan1987} to enable stable visibility and closure phase measurements~\citep[e.g., FLUOR demonstrated calibration better than 1\%]{Coude1997}. The preservation of coherence in single-mode fibers should take of two effects:
\begin{itemize}
    \item Wavelength dispersion causes a reduction of the broad-band fringe contrast. Tackling this problem requires matching the length of the fibers in all six beams to few millimeters for observations at a spectral resolution of R=50.
    \item The output polarization state of the light varies due to the birefringence and/or polarization rotation,  reducing the fringe contrast when polarizations are not split (natural light). This is tackled by selecting highly birefringent polarization-maintaining fibers and installing a birefringence compensation device in the instrument.
\end{itemize}  

To summarise, the considerations which drove the selection of optical fibers for MIRC-X are (i) single-mode in J and H-band wavelengths, (ii) high throughput, (iii) polarization-maintaining, and (iv) numerical aperture matched to the current MIRC fibers so the existing injection module could be used as such. We studied a large sample of fibers from 14 vendors to find the one best matching our requirements\,\citep{Kraus2018}. Ultimately, we selected Fibercore company HB1000C (6/125) for MIRC-X.

The fibers were assembled by OZ Optics into a silicon V-groove array, including $<$1$\mu$m absolute positioning of the fiber cores with a $0^{\circ}$ flat polish. Slow polarization axes are guaranteed by OZ Optics to be perpendicular to the X-axes of the front face of the V-groove to within $\pm3^{\circ}$. The fibers were then length-matched and connectorized (FC/PC) by Coastal Connections. The 1\,m fibers are jacketed in a Hytrel protection of outer diameter 0.9\,mm, and the largest difference of length is $2.38\,$mm between fibers 5 and 6.  We have recently acquired a second V-groove bundle with fibers lengths-matched even better to $<$0.7mm, as measured by a Luna optical backscatter reflectometer.

\subsection{\label{sec:Beamcombiner}All-in-one beam combiner}

The six fibers are arranged on a Silicon V-groove. A micro-lens array is glued to the V-groove with Norland 61 glue, similar to the procedure used for the VISION combiner \citep{Garcia2016}. It collimates the outputs into six beams of $\approx 150\,\mu$m in full-width-half-maximum (FWHM). These diffracting beams reflect into a 200\,mm focal spherical mirror, and interfere at its focus, where their FWHM is now 2.1\,mm. A cylindrical lens, with a focal length of 30\,mm, compresses the interference pattern in the direction perpendicular to the fringes, hence forming a pseudo-slit\footnote{By pseudo-slit we mean an image with an extreme aspect ratio between the long spatial direction that encodes the interference pattern and the narrow spectral direction that is eventually dispersed. Note that MIRC-X has no mechanical slit defining the aperture of the spectrograph entrance because the image is single-mode in the spectral direction.}. The FWHM of the pseudo-slit in the spectral direction is smaller than the pixel size ($<25\,\mu$m).

\begin{figure*}
\centering
\includegraphics[width=0.9\textwidth]{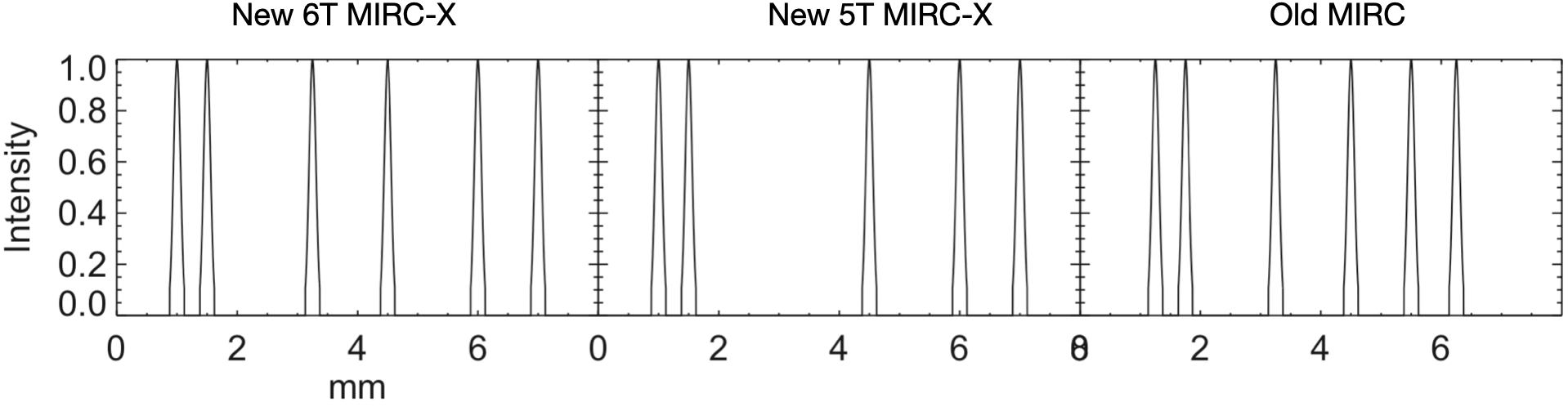}
\includegraphics[width=0.9\textwidth]{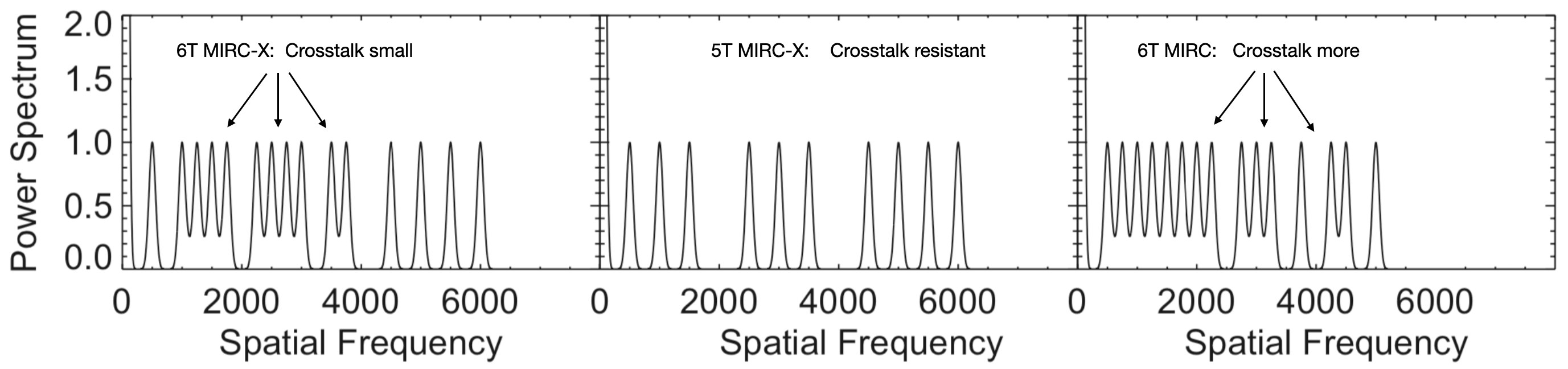}
\caption{Arrangement of fibers on the V-groove (top) and corresponding power spectrum (bottom), for the new MIRC-X V-groove (left), the new MIRC-X V-groove when dropping fiber 3 (crosstalk resistant mode; middle), and the original MIRC V-groove (right). The new MIRC-X uses a wide frequency range, and thus more pixels, than the original MIRC, but it provides (i) a better crosstalk resilience for the highest frequencies and (ii) a fully crosstalk resilient mode when dropping fiber 3.}
\label{Fig6_Fibers_arrangement}
\end{figure*}

The fibers are placed in slots 4, 6, 13, 18, 24, and 28 of the V-groove with pitch $250\,\mu$m (see Figure\,\ref{Fig6_Fibers_arrangement}). This non-redundant spacing of the fibers creates a unique spatial frequency signature for each pair. The positions were selected based on the following criteria: (i) to minimize the crosstalk between the shortest baseline frequency and the fringe envelope frequency (so-called DC spike), (ii) to allow a crosstalk-resistant design with 5-telescope combinations -- dropping one of the beams ensure a double-spacing between all coding frequencies in the power spectrum (see Figure\,\ref{Fig6_Fibers_arrangement}). The default mapping between beams and telescopes allows switching, for instance, configurations between S1+E1E2W1W2 and S2+E1E2W1W2  by dropping the beams 2 and 3. Since S1 and S2 are the closest telescopes, gathering these two 5-telescope configurations, sequentially, still offers an acceptable (u, v)-coverage. 

The sampling of the fringe on the detector is optimized for the sensitivity in H-band ($SNR \propto {N_{\rm RON}}^{-1} {N_{\rm px}}^{-1/2}$) with 2.7\,pix/fringes for the fringes created by fibers 1 and 6 at $\lambda=1.55\,\mu$m. Where, $N_{\rm px}$ is the total number of pixels used for recording fringes. Because of this choice, the J-band observations are only possible for 4-beams with the fibers 1-2-3-4; the fringes at highest spatial frequencies are indeed under-sampled (sub-Nyquist). This design choice gives a limited (u, v) coverage of 4-telescopes in the J-band. However, we are free to choose those 4-telescopes.

\begin{figure*}
\centering
\includegraphics[width=\textwidth]{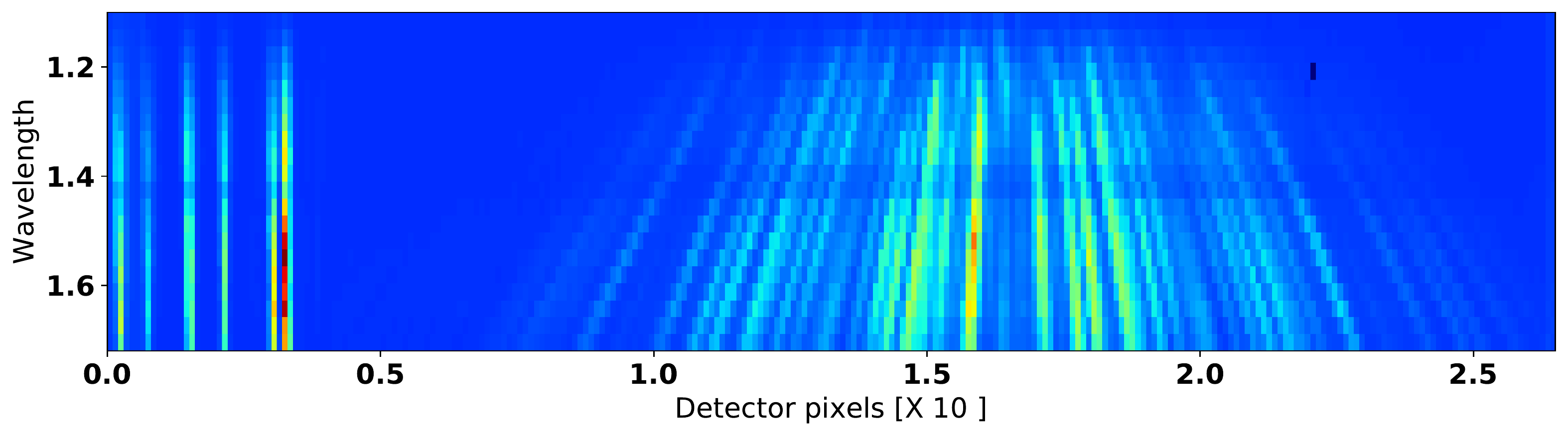}
\caption{Six beam J+H band images recorded with the internal light source (STS). The spatial direction is horizontal and the spectral direction is vertical. Wavelength is in $\mu$m. The photometric channels on the left are the images of the cores of the non-redundant fibers in the V-groove. The  photometric channels of beam 2 and 1 (right-most) are the closest one to each other, but are separated by more than one pixel. The  fringe pattern (right) encodes the coherence of the 15 pairs at different spatial frequencies. This image is with the R~=~50 prism and no Wollaston. The Wollaston creates two of such images separated in the vertical direction, one for each polarization.}
\label{Fig7_JH_band_fringes}
\end{figure*}

\subsection{Photometric channels} 

Rapidly varying flux losses due to the modal filtering of atmospheric turbulence by the fibers are calibrated by recording real-time, simultaneous photometric signals for each of the six-beams. Compared to the former MIRC, the photometric channels in MIRC-X implement a higher sensitivity and alignment-resistant state-of-the-art design. MIRC consisted of extracting a fraction of the light just after the micro-lenses, and injecting it into multimode fibers whose outputs are rearranged in the slit \citep{Che2010}. This design has a $45^{\circ}$ incidence on the beamsplitter, which is not ideal for polarization. Also, that was prone to vignetting problems, which often limited the amount of flux in the photometric channels.

MIRC-X extracts 20\% of the flux further out from the micro-lenses using a weakly-polarizing ($<$10\%) custom beamsplitter from Omega\,Filters (see Figure \ref{Fig3_mircx_beamcombiner_frontview}). This new beamsplitter corrects a flaw in the original MIRC, which used a beamsplitter coating with a 3:1 difference between the s- and p-wave reflectivity, leading to calibration difficulties as the linearly-polarized CHARA beams rotate in the lab during a night of observations. The light of the six individual beams is then re-imaged with a 30\,mm focal-length lens and positioned in the pseudo-slit alongside the compressed fringe pattern by a D-shaped mirror (Figure\,\ref{Fig7_JH_band_fringes}). The photometric channel spots in the pseudo-slit have an FWHM of less than one pixel in both the spectral and spatial direction. The closest spots, from fibers 1 and 2, are separated by 3 pixels in peak-to-peak. Having the measured beam size $\sim0.6$\,pixel FWHM and separation of nearest beams of 3-pixels produces no cross-talk in the photometric channels.

The flux allocation between fringes (80\%) and photometric channels (20\%) is chosen to reach a similar peak pixel value on both of them, considering the fringes are spread over a larger area on the detector. Compared to the MIRC design, the improvement in SNR in the photometric channels allows aligning the stars into fibers much quicker, thus reducing the overheads.

\subsection{Spectral dispersion and polarization-splitting optics}

The pseudo-slit containing the fringe pattern and the photometric spots is collimated, dispersed, and re-imaged on to the detector through a non-magnifying pair of 100\,mm focal length optics. Spectral dispersion is required in MIRC-X for many reasons: 

\begin{enumerate}
\item The spectrally-dispersed fringes allows for the reconstruction of the fringe visibility envelope as a function of delay offset, permitting fringe-tracking.
\item The MIRC-style image-plane combination requires spectral dispersion so that peaks in the Fourier spectrum do not overlap\footnote{Said differently: the temporal coherence should be longer than the number of fringes imaged on the detector. See for instance fringe 45 in Figure~\ref{Fig15_Contrast_beams} where the edges of the spatial envelope have reduced contrast because of the temporal coherence. Such reduced interferogram width in direct space translates into a larger Fourier peak in the frequency space.} (roughly R $>$ maximum fiber spacing / minimum fiber spacing $\equiv$ 24).
\item Spectral dispersion obviously allows studying emission of astrophysical objects across the continuum (flux ratio between components of different colors) or spectral lines (e.g., hydrogen recombination lines).
\item  The interferometric field-of-view increases with spectral dispersion ($\tfrac{R\lambda}{B}$), which is important for observing wide binaries and extended structure.
\end{enumerate}

For calibration precision, interferometric field-of-view, and spectroscopy use,  we always want higher spectral resolution; however, this comes at a signal-to-noise cost.  Therefore, it is important for the observer to choose the appropriate spectral model to achieve their science goal.

MIRC-X has spectral resolution elements R~=~22, R~=~50, R~=~102, R~=~190 (Newport 53-*-135R),  R~=~1035 for H-band (Newport 53-*-880R), and R~=~1170 for J-band (Newport 53-*-870R). The R~=~22 prism is intended to be used with a future pairwise, integrated optics beam combiner. The R~=~50 prism is the workhorse mode for sensitive observations. The R~=~50 and 102 prisms can be used in dual-band interferometry (J and H). The R~=~190 grism is the basic mode for high-precision astrometry because it delivers an interferometric field-of-view that entirely covers the diffraction limit of the 1\,m telescopes.  

The R~=~1035 grism will enable studies of velocity-integrated imaging of the line-emitting regions (R=1170 for J-band; R=1035 for H-band;). Note that the whole J+H will not fit on the detector for $R\sim1035$ grism, but we can observe in either the J or H-bands. The observation of fainter stars requires longer integrations, and that will be possible with the MYSTIC fringe tracker.  This grism is not yet tested.  



The collimator optics with focal lenth of 100 mm of the spectrograph is mounted in a 3-axes stage, motorized with pico-motors.   This allows for fine-tuning the alignment of the slit image in the detector,  mostly to correct for small non-repeatability when inserting the dispersing optics.

\subsection{Fast and low noise detector }\label{sec:Detector}

\begin{figure}
\centering
\includegraphics[width=0.45\textwidth]{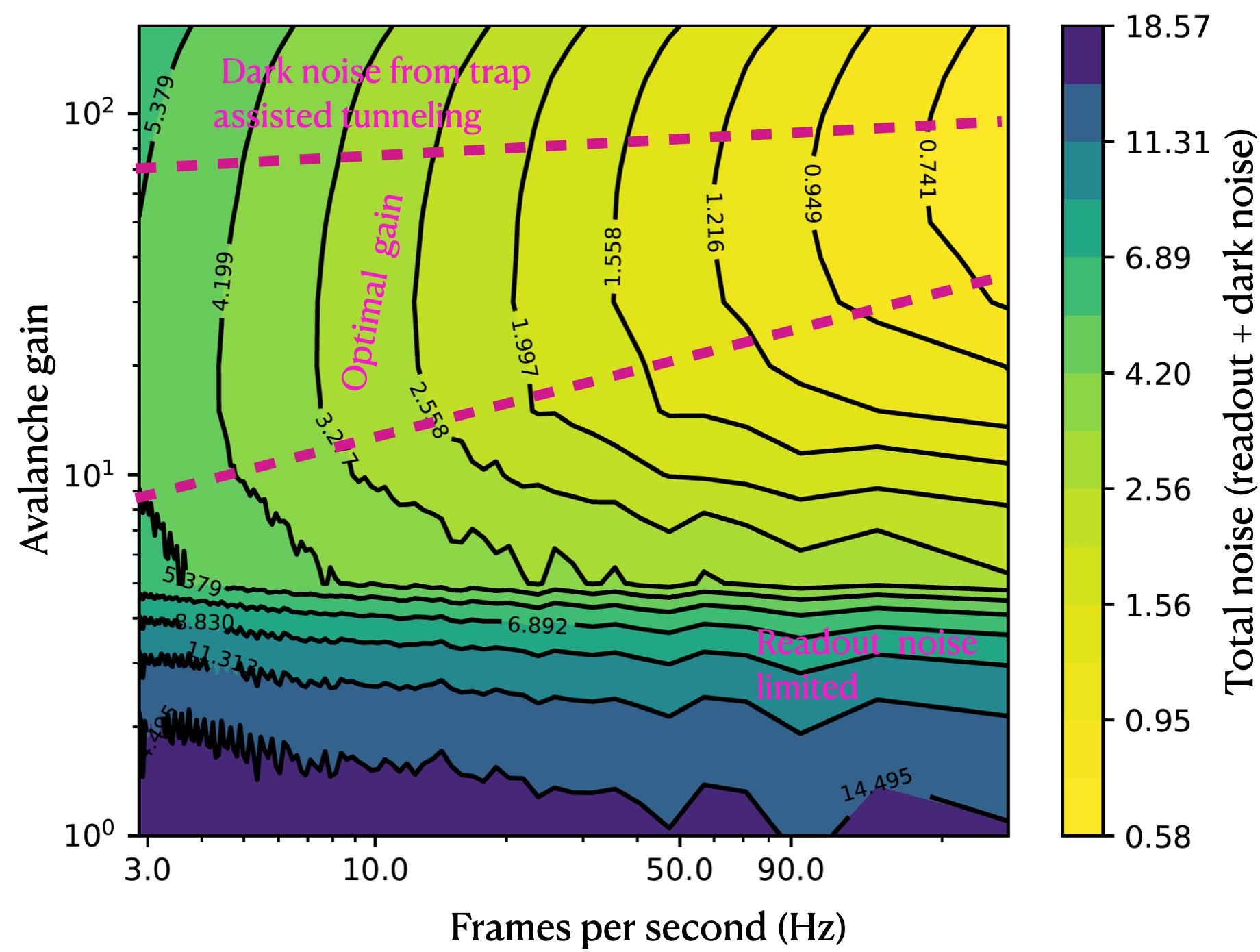}
\caption{Total noise ($\rm{e^{-}}$), the sum of readout noise + dark noise, as a function of the avalanche gain and frame rate. The axes are in a log scale. At gain values below 10, the performance of the detector is readout-noise limited. The optimal gains as a function of frame rate are shown in the gap between two dashed lines. At gain values above 60, trap assisted tunneling defects come into place and add additional dark noise.}
\label{Fig8_Detector_noise_as_function_gain_fps}
\end{figure}
   
MIRC-X is equipped with an ultra-low read-noise  C-RED ONE camera from the company First Light Imaging \citep{Gach2016}. The basic specifications are summarized in Table~\ref{table_det}. This camera incorporates a Leonardo SAPHIRA $320\times256$ pixel detector (MCT SWIR Mark13), with a pixel size of $24\,\mu$m and a $2.5\,\mu$m cutoff. The detector was developed in collaboration between Selex (now Leonardo) and the European Southern Observatory (ESO) to enable low readout-noise near-infrared sensors, operating at frame-rates exceeding 1000\,Hz \citep{Finger2014, GRAVITY2017,Gach2016}. The detector is based on the electron avalanche photodiode technology, where the electron in the pixel goes through an avalanche multiplication stage that amplifies the signal before it is stored in the capacity and readout.

\begin{table}
\caption{C-RED ONE camera specifications.}             
\label{table_det}      
\centering                          
\begin{tabular}{c c }        
\hline\hline                 
Parameter &  value  \\
\hline                        
Detector spectral response  & 0.8 to $2.5~\mu$m \\
Quantum efficiency & 48\% \\ 
Background blocking filters transmission & 81\% \\
Operating temperature & 80\,K\\
Excess Noise Factor & 1.47 \\
Read noise at gain 1 & $35\rm{e^{-} px^{-1}}$ RMS\\
Avalanche gain available to use & 1 to 100\\
Background noise & $90~\rm{e^{-}px^{-1}s^{-1}}$\\
Detector size & $320 \times 256$ pixels \\
Pixel pitch & $24~\mu$m\\
\hline                                   
\end{tabular}
\end{table}

The camera operates at $80$\,K with a total transmission of 48\%, including detector Quantum efficiency and absorption in cold filters. The MIRC-X camera has an $f/4$ cold stop and four low-pass cold filters to remove the thermal background from higher wavelengths --  two with a lower cutoff wavelength at $1.739\,\mu$m, and two at $2.471\,\mu$m.  The camera permits avalanche gains between 1 to 100. The detailed characterization of the MIRC-X camera in-terms of system gain, excess noise, transmission, readout noise as a function of frame rate can be found elsewhere \citep{Lanthermann2018,Lanthermann2019, Anugu2018}. At avalanche gains $\ge10$, the read noise is $\le 1 \rm{e^{-} px^{-1}}$  RMS, and the gain-independent statistical excess noise factor from the amplification process is $\approx1.47$. Excess noise factor indicates the increase in Poisson noise in the avalanche multiplication process as compared with the ideal multiplier, which is noiseless \citep{Lanthermann2019}.

Figure\,\ref{Fig8_Detector_noise_as_function_gain_fps} presents the total noise as a function of the avalanche gain and frame rate.  For the gains between 5 and 60 and frames rates below 60\,Hz, the detector is dark noise limited. For the gains above 10 and the frame rate above 60\,Hz, sub-electron readout noise is achieved. At gains above 60, trap assisted tunneling events add additional dark noise \citep{Finger2014}.

The C-RED\ ONE camera is cooled down using an in-built pulse-tube cryocooler with  heat-dissipating to two coolant hoses connected to a water chiller maintained at $20^\circ$\,C. It operates at $10^{-5}$\,mbar vacuum achieved using a HiPace 80 turbopump from Pfeiffer ($\leq\,10^{-6}$\,mbar). Once the camera is cooled down, the camera is sealed and the vacuum is maintained by an ion-pump from the Gamma vacuum (SPC PN 900026). The ion-pump is sufficient to maintain the required vacuum (achieves $\leq\,10^{-7}$\,mbar) for several weeks when the camera is kept cold. The turbopump is mainly used to pump the camera before and during cool down, and during warming-up sequences.

\subsection{Readout scheme and frame-grabbing}

The SAPHIRA detector has 32 parallel video channel outputs in 10 column blocks. Each channel output reads out 32 adjacent pixels in a row at a time. The pixel clock of the detector is set to 10\,MHz, which enables a readout speed of about $640$\,Mpixels per second. Full $320\times256$ pixel frame can be read at $3500$ frames per second (FPS). Furthermore, the C-RED ONE camera supports a sub-window readout mode, in the row and column directions, which increases the frame rate. 

MIRC-X uses a customized readout mode to reduce the readout noise further. This readout mode was originally implemented for a PICNIC detector at the IOTA interferometer\,\citep{Pedretti2004}. It consists of reading a pixel $N_{\rm reads}$ times before moving to the next pixel on the row. A row is also read $N_{\rm loops}$ times before moving to the next row. The effective pixel value is the mean of the $N_{\rm reads} \times N_{\rm loops}$ readout. The frame rate is reduced by $N_{\rm reads} \times N_{\rm loops}$, and the readout noise goes approximately ${ ( N_{\rm reads} \times N_{\rm loops} ) }^{-1/2}$. Several frames are read non-destructively a predetermined number of times before the detector is reset, i.e., ``up the ramp” sampling. Finally, a given exposure image is the difference between any two successive frames in a ramp.  

The most common MIRC-X on-sky observing modes are H-band prism R~=~50 and grism R~=~190. The prism R~=~50 mode uses a $320 \times 17$ pixel window on the detector  with $N_{\rm reads}=12$  and $N_{\rm loops}=8$. The grism R~=~190 mode uses a $320 \times 44$ pixel window with $N_{\rm reads}=8$ and $N_{\rm reads}=6$.   Both  modes provide a frame rate of $\approx350$\,Hz. This is sufficient to freeze the atmospheric turbulence and reduce the total size of the saved data. The detector is typically reset for every 100 frames, though this is reduced for brighter targets.

The data acquisition is implemented with a Matrox Radient eV-CL frame grabber\,\citep{Anugu2018}. The camera link cables are extended to 20\,m using a noiseless Thinklogical camera-fiber-link extender system (CFL-4000) using  LC/LC duplex multimode fiber optic patch cables (3\,mm PVC jacket; PC-KK5D30V20M) between the camera (inside the CHARA  beam-combiner laboratory) and the data acquisition computer (placed outside the lab) for the data acquisition.  Two communications happen simultaneously with the camera firmware: (i) the Ethernet house-keeping link allows monitoring and configuring the camera, and (ii) the fast Camera Link allows fast and low-latency frame grabbing.

\subsection{\label{sec:STS}The six telescope simulator}

\begin{figure}
\centering
\includegraphics[width=0.45\textwidth]{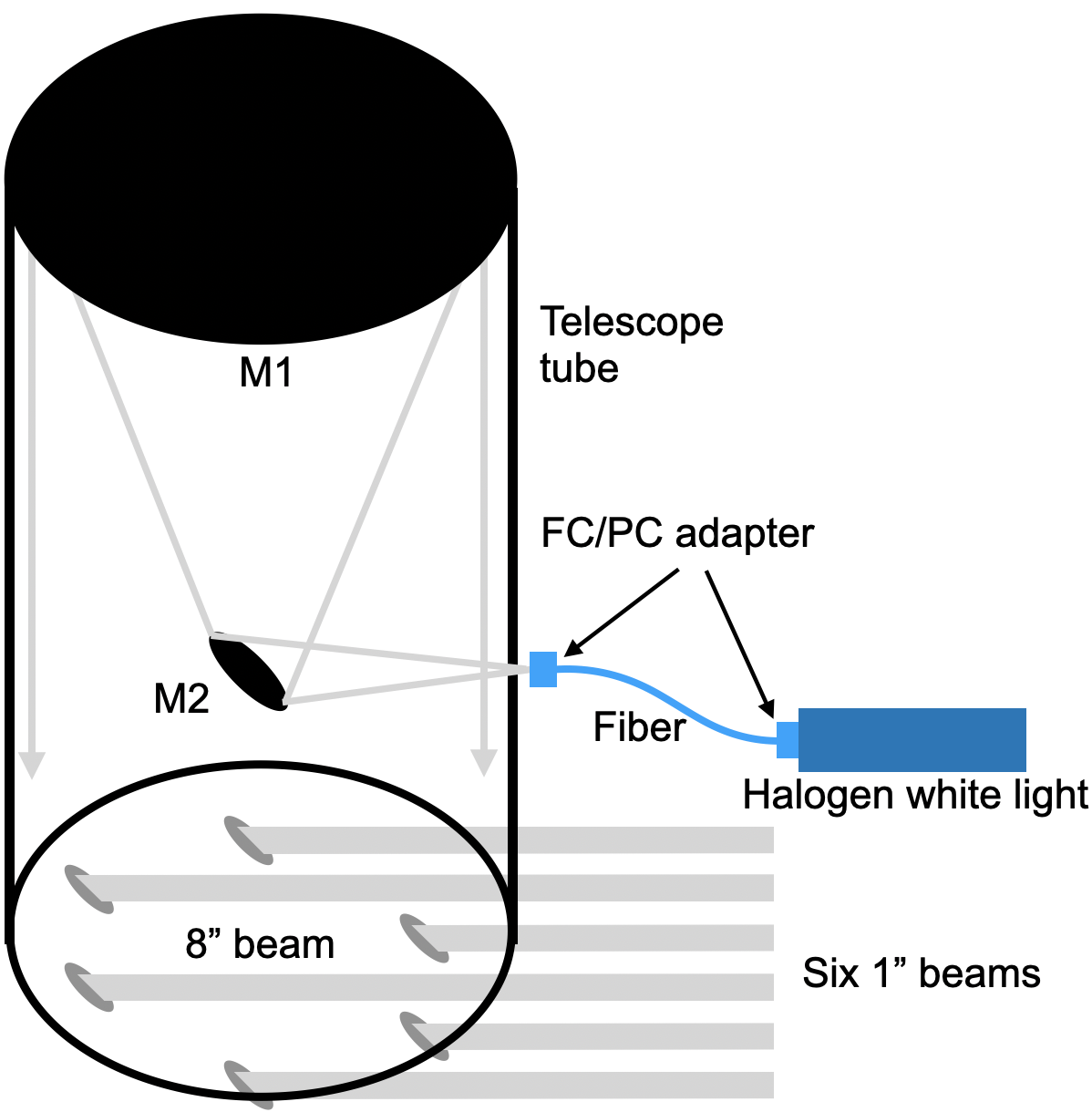}
\caption{Layout of the six telescope simulator (STS, not to scale). Six coherent beams are extracted from an 8-inch collimated beam created by a Classical Dobsonian telescope acting in reverse. The telescope eyepiece is fed by a single-mode fiber, which is injected light from a halogen lamp.}
\label{Fig9_STS_layout}
\end{figure}

The six telescope simulator (STS,  Figure\,\ref{Fig9_STS_layout}) is a new calibration source built by the MIRC-X team to allow laboratory verification and health check of MIRC-X and other instruments at CHARA Array. The STS delivers six coherent beams whose size, angles and delay match the CHARA telescope beams. A 8” coherent, collimated beam is created using an off-the-shelf Cassegrain telescope (Orion SkyQuest XT8) acting in reverse. The telescope eyepiece is replaced with an FC/PC free-space adaptor fed by a single-mode fiber illuminated by an Ocean Optics halogen lamp. Six 1-inch beams are extracted from the coherent, collimated beam. The STS is installed just before the CHARA shutters in the beam-combiner lab. The six STS beams are injected into the CHARA beams via a motorized set of pick-off mirrors such that no alignment is required. The STS is fully remotely controllable. It takes about 3\,minutes to switch from stellar light to STS light or reverse, which makes it a very useful last-resort alignment check facility.

The STS has been upgraded in March 2020 by the Observatoire de la C\^{o}te d'Azur with a fast-scanning piezo, allowing fast and accurate modulation of the optical delays over 20\,$\mu$m.  Furthermore, we have recently acquired photonic crystal fibers (manufactured by NKT Photonics, connectorized by ALPhANOV) that we hope will allow the STS to feed six-coherent beams to SPICA, MIRC-X and MYSTIC covering 0.5 to 2.5\,$\mu$m all simultaneously.

\subsection{Control software}

\begin{figure}
\centering
\includegraphics[width=0.45\textwidth]{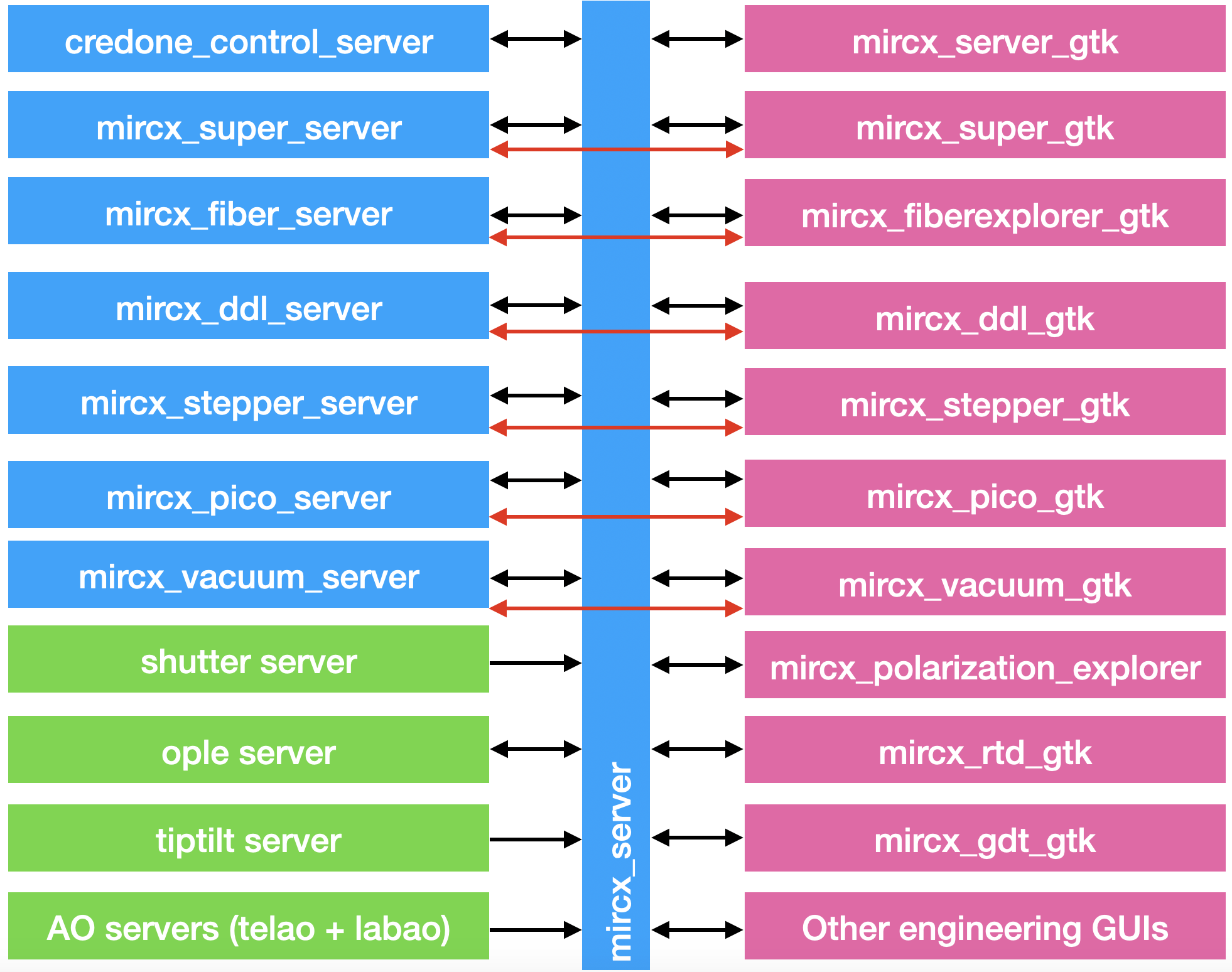}
\caption{Architecture of MIRC-X instrument software. It relies on the CHARA client/server architecture. A dedicated server interfaces each specific type of hardware. The blue color servers are specifically developed for MIRC-X. The green color represents the CHARA servers. The pink color represents the MIRC-X GUIs. The arrows indicate the different communications. All the servers and GUIs connect to the instrument \texttt{mircx\_server}.}
\label{Fig10_mircx_servers_GUIs}
\end{figure}

\begin{figure*}
\centering
\includegraphics[width=0.7\textwidth]{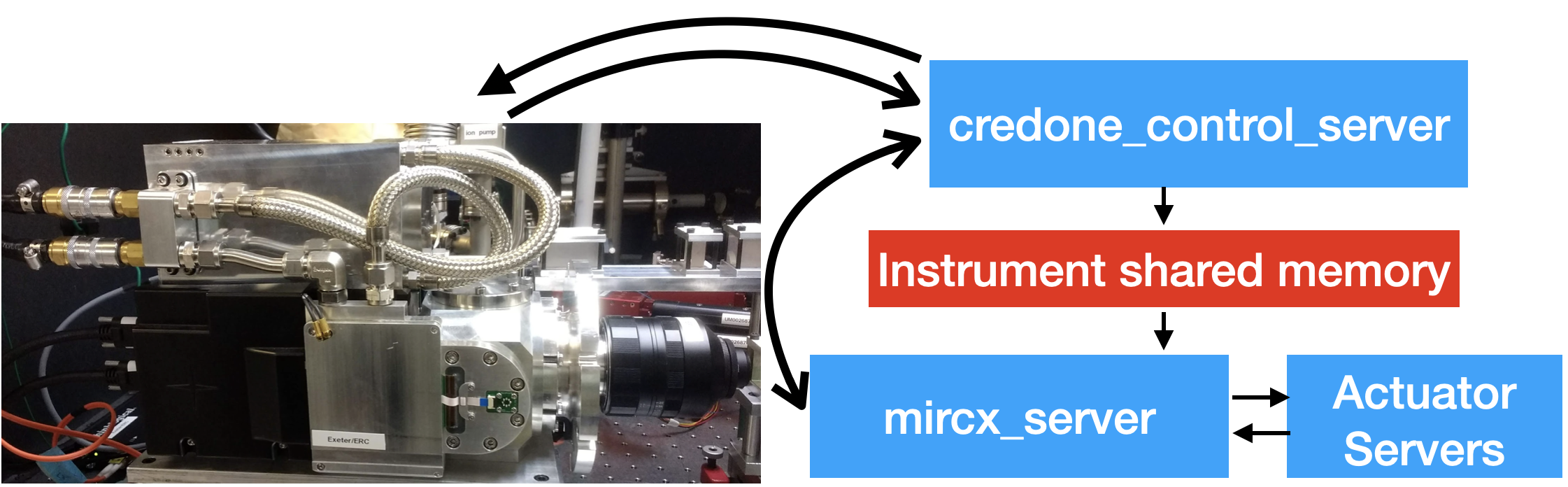}
\caption{Architecture of MIRC-X data acquisition.  The C-RED ONE camera images are grabbed in the MIRC-X computer and written in a shared memory by the \texttt{credone\_control\_server}. The \texttt{mircx\_server} reads the shared memory and implements image processing, fringe tracking, and data recording to an solid state drive. The \texttt{mircx\_server} uses a standard client/server model to monitor and update the C-RED ONE configuration and to talk to actuators.}
\label{Fig11_mircx_software}
\end{figure*}

The successful operation of an interferometric instrument depends on the ability to communicate between several systems (see Figure~\ref{Fig10_mircx_servers_GUIs}). The upgraded MIRC-X control software follows the CHARA compliant architecture, based on client/server architecture and GTK GUIs. The software runs on Linux Xubuntu operating system. The following requirements drive the design of the improved MIRC-X control software:

\begin{itemize}
\item Low-latency image acquisition that shall not miss frames even at a maximum frame rate of 3500\,FPS in the full window mode. The requirement for maximum frame rate is for future fringe tracking.

\item Real-time processing of the incoming frames delivers feedback for fiber injection, polarization optimization, fringe tracking, and LDC tracking.

\item Control all actuators and motors such as flipper motor mounts, shutters, stepper motors, pico-motors, pressure gauges, and the water chiller via standard CHARA servers.

\item MIRC-X and MYSTIC shall use the same copy of software and executables (startup flags different) to minimize the development and maintenance of the code. During the development of MIRC-X and MYSTIC at the University of Michigan, these two instruments and CHARA Array  software environments are created to minimize the coordinated and co-phasing observations testing.

\item Ensure 95\% of overlap of integration time between MIRC-X and MYSTIC when they are used simultaneously, to allow efficient a-posteriori fringe-tracking in the pipeline.

\item Remote observing capability.
\end{itemize}

Figure~\ref{Fig10_mircx_servers_GUIs} presents the list of servers and GTK GUIs used by the MIRC-X instrument for its operation. The actuators and motors used in MIRC-X are from a number of different companies, and consequently accept different protocols. The actuator and motor controllers are connected to a MOXA NPort network-ready serial port device using a DB9-to-RJ45 conversion cable connector and commanded using the TCP protocol. For each type of hardware, a server program is made to unify and simplify the command system, and log the positions of actuators. Each hardware server has an engineering GUI to perform moving, homing and calibration actions. 

The \texttt{credone\_control\_server} reads frames from the camera and writes them to an instrument shared memory based on circular-buffer without losing frames (see Figure\ref{Fig11_mircx_software}).  The instrument server, \texttt{mircx\_server}, reads the frames from this shared memory, performs the real-time image processing, provides feedback to actuators, and saves the raw frames to the disk.  For fast data writing, a 2\,TB SSD disk is mounted via an M.2 port (max write speed $\sim1.6$\,GB/s) on the computer. From the SSDs, the data is copied to HDDs for staging and later archived.
 
The \texttt{mircx\_server} also implements the logic of group delay tracking, fiber exploration, polarization optimization, and chromatic dispersion. It is connected to all other servers, at least to gather the house-keeping information that are logged in header. Many specialized GUIs are connected to it:  \texttt{mircx\_server\_gtk} configures the instrument, \texttt{mircx\_rtd\_gtk} displays the real-time images, \texttt{mircx\_polarization\_explorer} performs the polarization optimization, and \texttt{mircx\_fiber\_explorer} optimizes the light injection into the fibers.

\subsection{Top-level operational concept}

\begin{figure}
\centering
\includegraphics[width=0.45\textwidth]{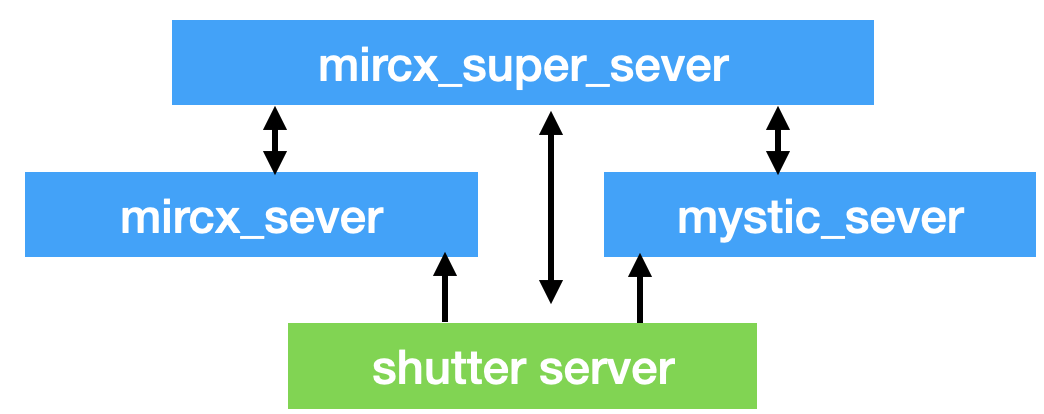}
\caption{Observing coordination of MIRC-X and MYSTIC instruments. \texttt{mircx\_super\_server} server coordinates the data taking process for the MIRC-X and MYSTIC instruments. For example, it opens the CHARA shutters and starts the integration on both instruments.}
\label{Fig12_mircx_mystic_super_servers}
\end{figure}

The top-level server \texttt{mircx\_super\_server} coordinates the actions to the CHARA Array and the data recording with MIRC-X and MYSTIC (see Figure \ref{Fig12_mircx_mystic_super_servers}). For example, the \texttt{mircx\_super\_server} opens the required CHARA shutters for the desired exposure type (background, beam sequence, or fringes), then configures MIRC-X and MYSTIC with adequate parameters, and starts the exposures. There is no synchronization or external trigger between the MIRC-X and MYSTIC instruments. However, the frames are individually time-tagged in a common reference frame with an accuracy better than 1\,ms thanks to the Network Time Protocol (NTP) in the Linux operating system. Consequently, as long as the integration period mostly overlaps, it will be possible to resynchronize the signals of both instruments in the off-line pipeline.

One of the objectives of the software upgrade was to provide remote observing capabilities. This mode is enabled by the new server-client scheme, where all real-time operations and commands are executed by the servers that run on the on-site instrument computer. On the remote machine, the observer runs the Linux GTK GUIs that communicate to the server via ssh-tunneling using the remote SSH port forwarding protocol\footnote{\href{http://www.chara.gsu.edu/observers/remote-observing}{More details http://www.chara.gsu.edu/observers/remote-observing}}. This communication protocol also permits the user(s) to open multiple instances of the operation GUIs, possibly at multiple remote locations (e.g., to enable real-time monitoring of ongoing observations, or to provide support and training). GUIs can be closed and restarted without affecting or interrupting the real-time operation.

The safety of the costly C-RED ONE system is critical. The background-blocking filters inside the camera can break near the detector if the internal pressure rises sharply. To mitigate this risk, the \texttt{mirx\_vacuum\_server} controls the electromagnetic valve of the camera by checking the camera pressure, the ion pump pressure, and the pressure maintained by the turbo-pump.  Instrument parameters, such as temperatures and pressures, are logged to the disk once every second. Plots are generated and emailed to interested parties, once every 12 hours, for easy monitoring. Additionally, automatic alert emails are generated and sent to the MIRC-X team when something suspicious occurs within the instrument. 

\subsection{Fringe tracking with MIRC-X}
\label{sec:fringe_tracking}

The fringe motion due to the atmospheric turbulence can be tracked in two ways, namely with group-delay tracking and phase-delay tracking. Currently, MIRC-X implements group-delay tracking, which consists of tracking the delay position of maximum contrast, by the mean of spectrally dispersed fringes. Phase-delay tracking is much more demanding as it consists of locking the actual phase of the fringe. It involves estimating the phase at a fraction of $\lambda$, at high processing rates, and with a complex state machine \citep{Lacour2019}. The phase tracking software is currently in development by the Observatoire de la C\^{o}te d'Azur (France) inspired by VLTI/GRAVITY \,\citep{Lacour2019}, with several commissioning runs planned for 2020 (Mourard et al. in prep.).

The MIRC-X real-time SNR and group delay estimators use spectrally dispersed fringes to reconstruct the fringe visibility envelope as a function of delay offset. A conceptual summary of the algorithm is presented in Appendix~\ref{sec:GDT}. The fringe images are first integrated coherently within a coherence time and then added incoherently within the typical fringe drift time to maximize the sensitivity. The 2D Power Spectrum of the fringes image show peaks at the spatial frequencies defined by the non-redundant fiber arrangement. As the slope of the fringe changes, the Fourier peak moves along a line with constant spatial frequency; and that displacement is the group delay. The amplitude of the peak above the noise is the SNR.

Two operational schemes are foreseen to implement fringe tracking in the future to enable co-phased, multi-instrument operation:

\begin{itemize}
    \item In the Primary-Secondary scheme, one instrument controls the CHARA delay lines to correct for the fast atmospheric optical path difference (OPD). The second instrument controls the differential delay lines to correct for the slow instrumental drifts and atmospheric dispersion. This simple scheme is sufficient when one combiner (primary) has a higher SNR than the other (secondary) systematically, for whatever instrumental or astrophysical reasons. This mode is already implemented, and can be used as soon as MYSTIC and/or SPICA are installed at CHARA Array.

    \item For very resolved objects, where baselines might reach zero visibility depending on their length and wavelength, the previous scheme is non-optimal. Ideally, we want to combine all the available information and control both the main and the differential delay lines from all this information. This mode is not yet developed.
\end{itemize}

\begin{figure}
\centering
\includegraphics[width=0.45\textwidth]{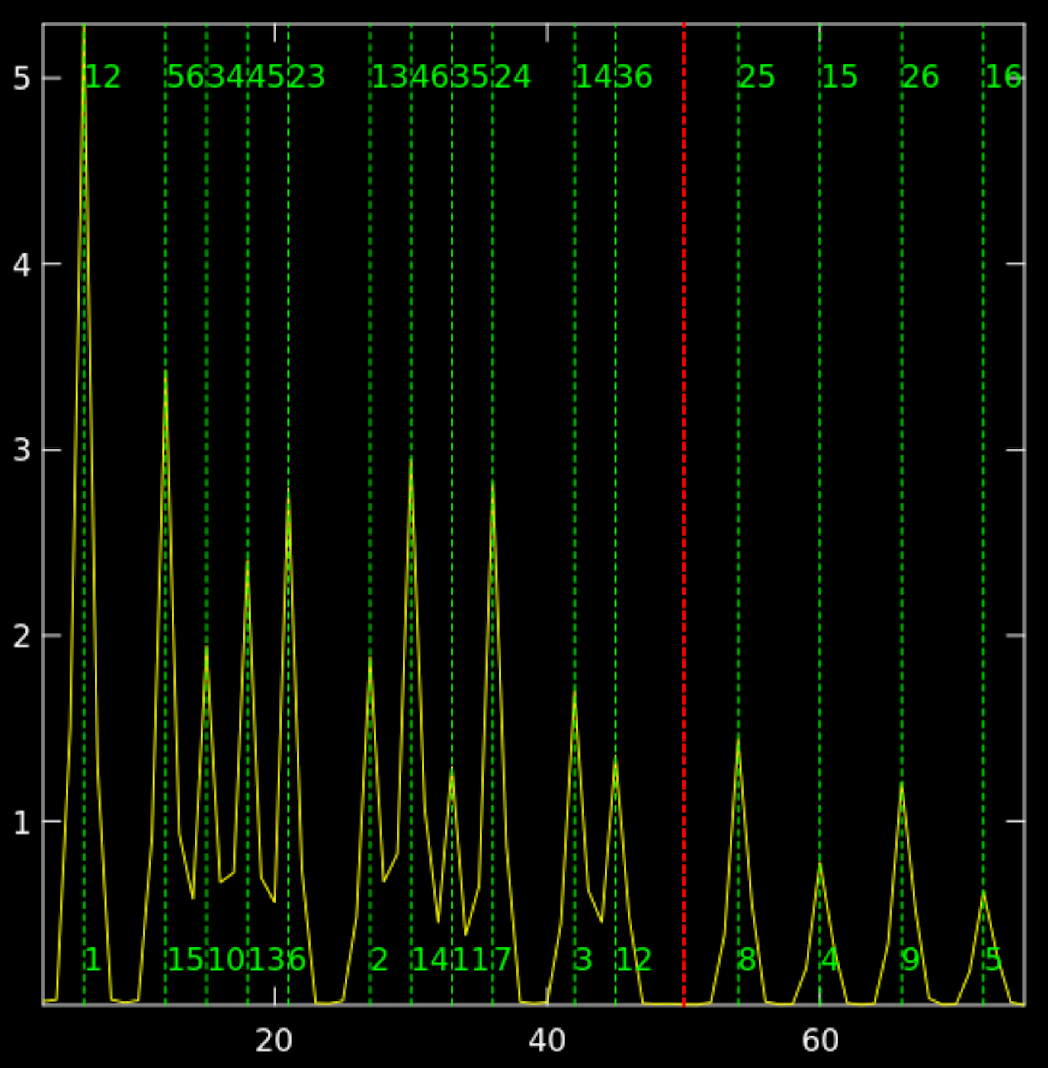}
\caption{Real-time display of power spectrum of MIRC-X fringes. The \texttt{mircx\_rtd\_gtk} GUI displays the power spectrum peaks at the expected frequencies (vertical dashed green lines) for the 15 pairs of beams. The red line represents the background noise pixels considered in the estimation of power spectrum peaks.   The green numbers represent the beam pair (top) and the baseline number (bottom). The arrangement of beam pair 1 and 2 being the lowest spatial frequency and 1 and 6 being the highest spatial frequency (see the design in Figure \ref{Fig6_Fibers_arrangement}). The unequal fluxes of beams degrade the contrast of raw fringes.  The fringe contrast is low for highest frequency fringes in a comparison because of pixel under-sampling and uncorrected camera vibrations, which affect the under-sampled fringes the most.}
\label{Fig13_Fringe_peaks}
\end{figure}

\section{\label{sec:4}Instrument operation}

MIRC-X is supported by the observation preparation software \texttt{Aspro2} \footnote{\href{https://www.jmmc.fr/aspro}{https://www.jmmc.fr/aspro}} from JMMC. It allows optimizing the CHARA POP configuration (pipe of pan, static delay lines) to maximize the observability of the desired targets, to compute the resulting (u, v)-plane, and even to simulate observations assuming a known object geometry. A step-by-step guide on how to operate MIRC-X is described in great detail in the user manual available  online\footnote{\href{http://chara.gsu.edu/wiki/doku.php?id=chara:instruments}{http://chara.gsu.edu/wiki/doku.php?id=chara:instruments}}.

\subsection{Preparations before the on-sky observations}

CHARA offers the possibility to change the mapping from instrument beam to telescope before the night. The standard MIRC-X beam order is chosen such that low-frequency fiber fringes are paired with longer-baselines (i.e., fibers 5 and 6 with S1 and E2 telescopes), and high-frequency fiber fringes are paired with shorter-baselines (i.e., 1 and 6 with E1 and E2 telescopes). The procedure of this alignment for every night is described on the CHARA wiki page\footnote{\href{http://chara.gsu.edu/wiki/doku.php?id=chara:operating\_procedures}{http://chara.gsu.edu/wiki/doku.php?id=chara:operating\_procedures}} and is dynamically improving with the commissioning of CHARA Adaptive Optics (AO) systems\,\citep{tenBrummelaar2018}.

Before the night, the alignment of MIRC-X is optimized by using the STS coherent beams (see Section\,\ref{sec:STS}). The goal is to ensure that the light is well centered within the predefined coordinates on the MIRC-X detector. While the instrument calibration is performed daily, it has proven to be stable at least over the timescale of a week unless the observing mode has been changed. First, the photometric channels are moved into the predefined pixel columns in the spatial direction, using the remotely controlled collimating optics of the spectrograph. Next, the power spectrum fringe peaks are checked to make sure they fall on their predetermined positions (see. Figure\,\ref{Fig13_Fringe_peaks}). This calibration is accomplished through remotely adjusting the spectrograph collimating optics in the spatial direction.

When the J+H dual-band setup is desired, it is necessary to manually insert a notch filter in front of the camera to filter out the CHARA metrology laser wavelength ($1.3\,\mu$m). It is also necessary to control the CHARA LDC in the CHARA beams in order to correct the atmospheric dispersion across J+H (see Section \ref{sec:Dispersion}).

A polarization exploration with the Lithium Niobate plates (see section~\ref{sec:Polarization}) can be executed to verify that the instrument delivers the optimal contrast in natural light. This step is not executed every day, but mostly only in occasional commissioning or when experimental setups are being tested.

\subsection{On-sky target acquisition and flux injection into fibers}

\begin{figure*}
\centering
\includegraphics[width=0.95\textwidth]{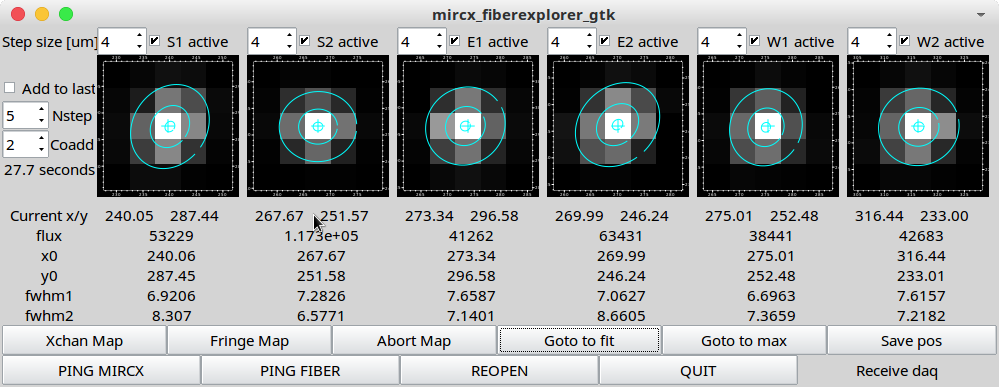}
\caption{\texttt{mircx\_fiberexplorer\_gtk} interface. The flux injection into the fiber is optimized by scanning the fiber in X and Y directions. The resulting map is fitted with a Gaussian model to find the position that results in the best flux injection. The GUI shows the Gaussian fit centers (location of fiber motors where the maximum flux can be found), flux, and FWHM. \texttt{Xchan Map} performs a fiber explorer map by searching for the flux using the six photometric channels simultaneously.  \texttt{Fringe Map} performs a search using the flux in the fringe window and searches for the flux in one beam at a time.}
\label{Fig14_FiberExplorer}
\end{figure*}

At the beginning of each night, observations are started with a bright star for three reasons:
\begin{itemize}
    \item To align the CHARA AO system.

    \item To perform an initial fiber explorer map with a large field of view on a bright star in order to align the MIRC-X with the CHARA beams.

    \item To find the fringe offsets on a bright calibrator. These initial fringe offsets serve later on as initial guess for fringe searches on fainter science targets.
\end{itemize}   

The star is first acquired by the CHARA guiding system (either the old Tip/Tilt system, or the new Adaptive Optics system). At that point, an image of the star is formed and guided near the MIRC-X fiber core.

The flux injected into the fibers is optimized by doing a fiber exploration (see Figure\,\ref{Fig14_FiberExplorer} and Section~\ref{sec:fiber_explorer}). The position of best flux is generally stable over the 30\,mins required to complete the sequence. However, it is required to redo the fiber exploration mapping after slewing to each new target as the mean tip/tilt in the visible (guiding) and in the infrared are different. This offset is due to non-common path offsets between the tip-tilt camera and MIRC-X and atmospheric differential refraction\,\citep{Filippenko1982}. The differential refraction photo-center shifts between J and H-bands are much smaller than the 1-m diffraction-limited point spread function (PSF) in H-band (60 mas atmospheric differential shift at a zenith angle of $40^\circ$ or 17\% in size). Given the practical larger adaptive optics corrected PSF ($\approx1.5\times$ than diffraction-limited PSF), the differential refraction does not pose problems for our flux injection strategy in J+H bands.

\subsection{Fringe acquisition and group delay tracking}

The MIRC-X real-time SNR and group delay estimators use spectrally dispersed fringes to reconstruct the fringe visibility envelope as a function of delay offset (see Section~\ref{sec:fringe_tracking}).

The fringe search is initiated by moving the CHARA delay lines in regular steps until the SNR of the fringes exceeds a threshold, typically set to a factor of $\approx2$ above the off-fringe threshold. The off-fringe thresholds are estimated when flux, but no fringes, are recorded (for this purpose, the delay line carts are moved far away from the expected fringe position). Once fringes are found, they are locked by sending the measured OPD offsets to the CHARA delay lines at typically few Hz. The visibility, and thus the SNR, can be arbitrarily small on some baselines (often the longest ones), depending on the geometry of the observed target. The fringe tracking algorithm implements baseline bootstrapping\,\citep[e.g.,][]{Armstrong1998a} based on all closing triangles, in order to efficiently track the most resolved baselines.

The initial fringe search is done with respect to a reference telescope, usually W2 as it is located closest to the center of the array. Frequently, the search encounters fringes between pair of telescopes not including the reference telescope (e.g., fringes S1S2, S1E1, etc.). By using this information, a cross-fringe algorithm predicts the zero-OPD position for the other delay lines.

The fringe tracking algorithm runs on the instrument server and the user interacts with it using a dedicated group delay tracking GUI (\texttt{mircx\_gdt\_gtk}, Appendix~\ref{sec:GDT}), arguably the most complicated of the entire MIRC-X operation system.
Fringe search is a tedious part of the acquisition process at CHARA Array. The fringes are often found several millimeters from the prediction of the baseline model, for still poorly understood reasons. The drifts are likely due to thermal expansion and contraction of the telescopes and beam paths on monthly timescales. This adds or removes an unknown path length from each baseline, meaning the solution fails to predict the correct fringe location resulting in large offsets. The only way to effectively account for seasonal changes in the baseline solution is with regular observation of calibration stars on all baselines across the whole sky and is currently planned with two observing nights for each observing semester.

\subsection{Observations sequence}\label{subsec:sci-cal}
At the start of the MIRC-X observing run, finding fringe offsets takes 10-30 mins depending on how far the offsets are from the previously known positions. Finding fringe offsets is implemented with a bright fringe finder star close to the target star sky coordinates.  Night-to-night mechanical drifts in the CHARA optical path introduce a maximum offset of $\sim$ 1 mm, and finding these offsets at the beginning of subsequent nights takes approximately 10 mins.

The standard observing sequence is as follows:
\begin{itemize}
 \item 5~minutes of acquiring star at the telescopes,
 \item 3~minutes of performing a fiber explorer map to align light into the fibers,
 \item 2$-$5~minutes of searching for fringes,
 \item 10~minutes of DATA, with all shutters open and fringes tracked,
 \item 1~minute of BACKGROUND with all shutters closed,
\item 1~minute of each BEAM where the shutter of only one of the six beams is open sequentially, and
\item 3~minutes of FOREGROUND frames where all shutters are open but the optical path is set to a large value to ensure that no fringes are present in the data set.
\end{itemize}

During 2017$-$2019, the total time for an observation of a science target or calibrator took $\sim$ 30 mins, including the star acquisition, flux acquisition into MIRC-X fibers, searching for fringes, locking fringes, and saving the data and shutter sequences. Observations of faint targets or resolved stars with low visibility contrast can take longer to find and track fringes.  Other beam combiners at CHARA have a faster observing cadence: 5-15 min per star for the two-telescope combiners CLASSIC\,\citep{Ten2013} and PAVO\,\citep{Ireland2008}, 10-20 min per star for the three-telescope combiners CLIMB\,\citep{Ten2013} and VEGA\,\citep{Mourard2009}. However a single observation with MIRC-X provides significantly more data on all 15 baselines and 20 closure triangles simultaneously.

Currently, the CHARA adaptive optics commissioning is underway.  During this transition, the star acquisition, including closing the adaptive optics and tip-tilt loops on all six-telescopes takes longer, up to 10-30 mins compared to 5 min prior to the adaptive optics upgrade. This overhead adds significantly to the time required for an observation, but is expected to decrease as the performance of the adaptive optics systems are optimized.

The BEAM sequence is necessary to calibrate the flux ratio between the fringe window and the photometric channels, hereafter called the $\kappa$-matrix. FOREGROUND data are used to calibrate the empirical de-biasing coefficients for the bispectrum.

All data are saved in the \texttt{FITS} format with an extensive primary header containing several hundred keywords that document the configuration and status of MIRC-X, the CHARA Array, Adaptive Optics systems, delay lines, actuators and motors. Time-stamped telemetry data from CHARA systems are also saved automatically. This includes (i) residual aberrations, AO loop status, loop frequency, and loop gains using in post-processing for better calibration of errors; (ii) Shack-Hartmann pupil images to measure the pupil shifts as it is critical for precise astrometric measurements\,\citep[cf. ][references therein]{Anugu2018a}.

The raw visibility should be calibrated using calibrator stars with known diameter. Typical observations execute a standard calibrator-science-calibrator (CAL-SCI-CAL) cycle. Calibrators can be chosen, using the \texttt{SearchCal}\footnote{\href{https://www.jmmc.fr/search-cal}{https://www.jmmc.fr/search-cal}} software to be (a) close to the science target, both in terms of sky position and magnitude and (b) have a smaller angular diameter so that their visibility on a given baseline is less dependent on the diameter. Observers should strive to use the same detector configuration throughout the night (frame rate, gain, readout mode, etc.), as far as it is feasible so that all calibrators can be shared for the science targets.

The closure phase is, in principle, a self-calibrated measure, though the science star closure-phase is also corrected with a calibrator to reduce instrumental residual artifact (see Section~\ref{sec:internal_carac}).

\subsection{\label{sec:Pipeline}Data reduction pipeline}

The MIRC-X data reduction pipeline produces science-ready visibilities and closure phases written in \texttt{OIFITS} format \citep{Duvert2017}. These \texttt{OIFITS} files are compatible with standard interferometric software such as \texttt{macim} \citep{Macim2006}, \texttt{squeeze}\footnote{\href{https://github.com/fabienbaron/squeeze}{https://github.com/fabienbaron/squeeze}} \citep{Baron2010}, \texttt{MiRA}\footnote{\href{https://github.com/emmt/MiRA}{https://github.com/emmt/MiRA}} \citep{Thiebaut2008},  \texttt{LITpro}\footnote{\href{https://www.jmmc.fr/litpro}{https://www.jmmc.fr/litpro}}, and \texttt{CANDID}\footnote{\href{https://github.com/amerand/CANDID}{https://github.com/amerand/CANDID}}\citep{Gallenne2015} for image reconstruction, modeling and binary detection.

The MIRC-X pipeline is  implemented in \texttt{Python}~3.7, leveraging some of the key concepts from the MIRC pipeline written in \texttt{IDL}, and is hosted by CHARA in a publicly available git repository\footnote{\href{https://gitlab.chara.gsu.edu/lebouquj/mircx\_pipeline}{https://gitlab.chara.gsu.edu/lebouquj/mircx\_pipeline}}. The main steps of the pipeline are the followings:

\begin{enumerate}
\item  The raw detector frames are pre-processed for background subtraction, bad pixel removal, and flat fielding. An electronic 90~Hz parasitic signal is removed using edge columns.

\item The spectral calibration is achieved by comparing the observed fringe-frequency in 1/pix to the expected spatial frequency based on the optical magnification. The magnification is checked regularly toward a stable reference (see Section~\ref{sec:etalon}).

\item The real-time photometry of each of the six beams $P_{i}(\lambda, t)$, where $i$ is  the number of the beam, $t$ is the time (frame), $\lambda$ the spectral channel, are estimated from the photometric channels and the $\kappa$-matrix.

\item The 15 coherent fluxes $I_{ij}(\lambda, t)$ are the Fourier components of the interferometric window, in the spatial direction, at the spatial frequencies of the corresponding pair (where $i$ and $j$ are the numbers of the considered beam pair).

\item Averaged squared visibilities for the baseline $ij$ are computed as follows:
\begin{equation}\label{Eq.2}
    V^2_{ij}(\lambda) = \frac{| \langle \; I_{ij}( \lambda, t) \times I_{ij}(\lambda, t-1)^*  \,-\,\beta(\lambda) \;\rangle |} { 2 \langle P_{i}(\lambda, t) \times P_{j}(\lambda, t) \rangle }
\end{equation}
Because of the time-shift in the cross-product, the read-noise does not contribute an additive bias term in the square visibility though we still need to correct for photon-noise bias.  This latter bias is corrected by assuming a flat spectrum and measuring the bias at high spatial frequencies where no astrophysical or instrumental signals can be present.  Note, this small time-shift does introduce a seeing-dependent decoherence in the transfer function.

\item Bispectrum $C$ of closing triangle $ijk$ is calculated by:
\begin{equation}
C_{ijk}(\lambda)=\langle I_{ij}(\lambda, t) \times I_{jk}(\lambda, t) \times I^{*}_{ik}(\lambda, t)\;-\;\beta_{ijk}(\lambda)\rangle
\end{equation}
The average bispectrum is corrected from a real additive bias term $\beta_{ijk}$, because of the all-in-one combination of the interferograms. The coefficients to compute this bispectrum bias are estimated by fitting all the BACKGROUND, FOREGROUND, and DATA files of the night, inspired by a suggestion in Appendix C2 of \citet{Basden2004} and similar in spirit to the method used by the VISION combiner described in~\citet{Garcia2016}. We describe the process of subtracting the bispectrum bias in Appendix\, \ref{sec:bbias}.

\item The instrumental transfer function is estimated by dividing the calibrator's observed visibilities by their expected visibilities assuming a central uniform disk (UD) model. UD diameters may be input manually or can be automatically retrieved from the JMMC Stellar Diameters Catalogue\,\citep[JSDC;][]{Bourges2017}.

\item The transfer function estimates are interpolated at the time of the science observation with a Gaussian-weighted average, typically with an FWHM of $1\,$hr. The science visibilities are calibrated using this interpolated transfer function from all calibrators sharing the same setup.
\end{enumerate}

\subsection{End of night activities}

At the end of each observing night, the data are duplicated to two different disks on the MIRC-X computer for redundancy. Later, they are archived to the repositories at Ann Arbor (Michigan, USA), Atlanta (Georgia, USA) and Exeter (UK). The pipeline runs automatically following the archiving procedure. Upon completion of the reduction and calibration steps, automatic checks on the instrument performance, data quality, and the transfer function stability are undertaken. 

As long as more than one calibrator star is observed on any given night, the pipeline iteratively inspects each standard star for signatures of binarity using \texttt{CANDID}. The reduced data for each calibrator star are calibrated in turn using the other calibrator stars, ignoring differences in detector modes (gain, number of coherent frames, frames-per-reset, etc.). The calibrated data are run through the \texttt{CANDID} \texttt{fitMap} routine with the star's UD diameter (retrieved from the JSDC) set as a fixed parameter in the fit. The closure amplitudes are not used in the fitting procedure. It is planned to report bad calibrators to the JMMC badcal database\footnote{\href{http://www.jmmc.fr/badcal/}{http://www.jmmc.fr/badcal/}}

A report with (i) plots of the reduced data, the calibrated data and the performance checks, and (ii) a tabulated summary of the calibrator inspections (reporting best fit solution parameters and the number of $\sigma$ corresponding to the detection) is distributed via email to the MIRC-X team members and the observer(s).

\begin{figure*}
\centering
\includegraphics[width=0.24\textwidth]{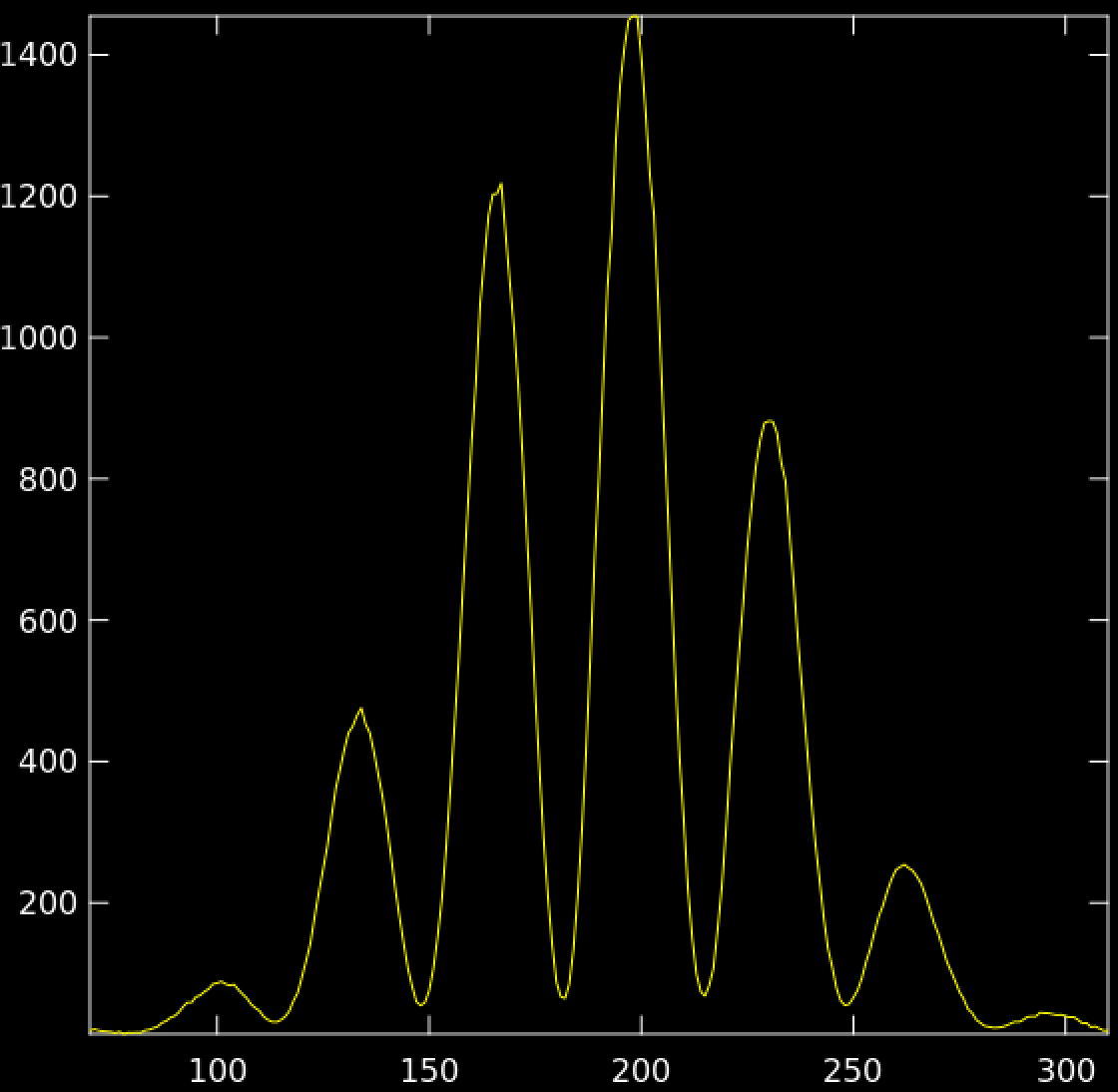}
\includegraphics[width=0.24\textwidth]{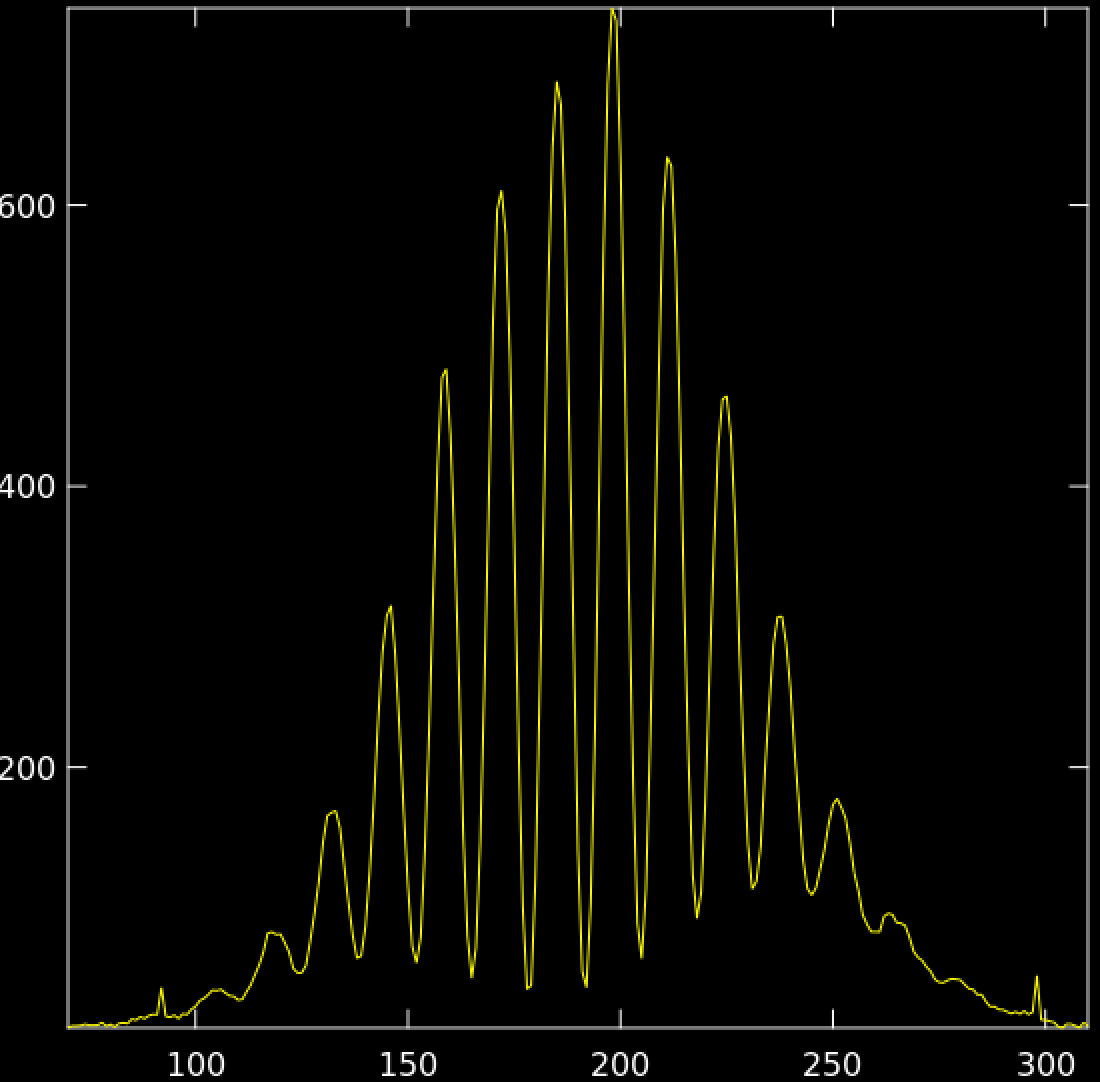}
\includegraphics[width=0.24\textwidth]{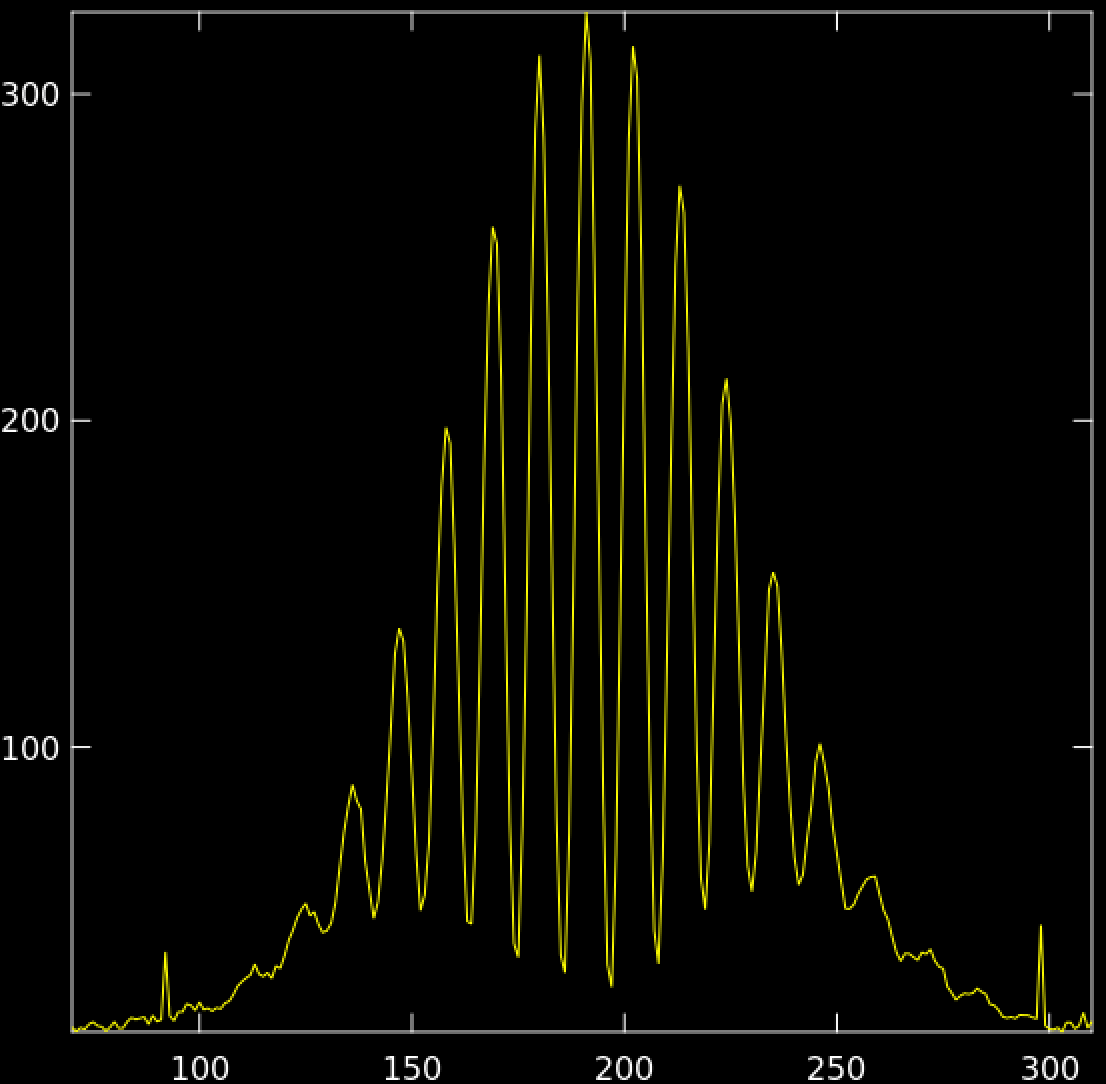}
\includegraphics[width=0.24\textwidth]{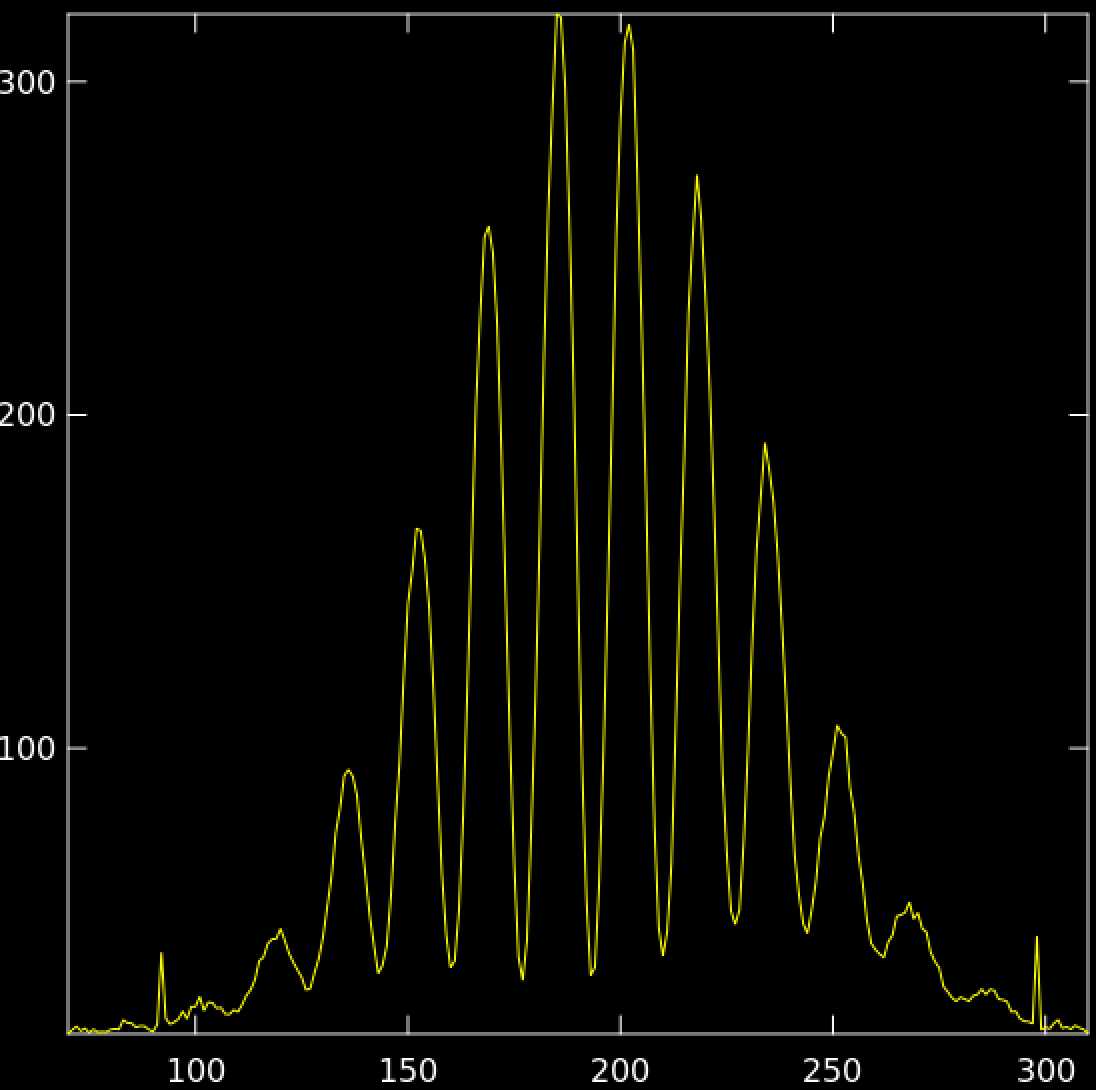}
\caption{Contrast of H-band fringes observed for nearest beam pairs 12, 34, 45, and 56 using the internal light source. A single spectral channel line is presented.  The non-calibrated visibility contrasts for the nearest pairs are 85 to 95\%. The X-axis is the spatial frequency direction in pixels. Y-axis is flux in ADU. The different beam pairs have different spatial frequencies because the separation between the two beams in each pair is non-redundant. }
\label{Fig15_Contrast_beams}
\end{figure*}

\begin{figure*}
\centering
\includegraphics[width=0.45\textwidth]{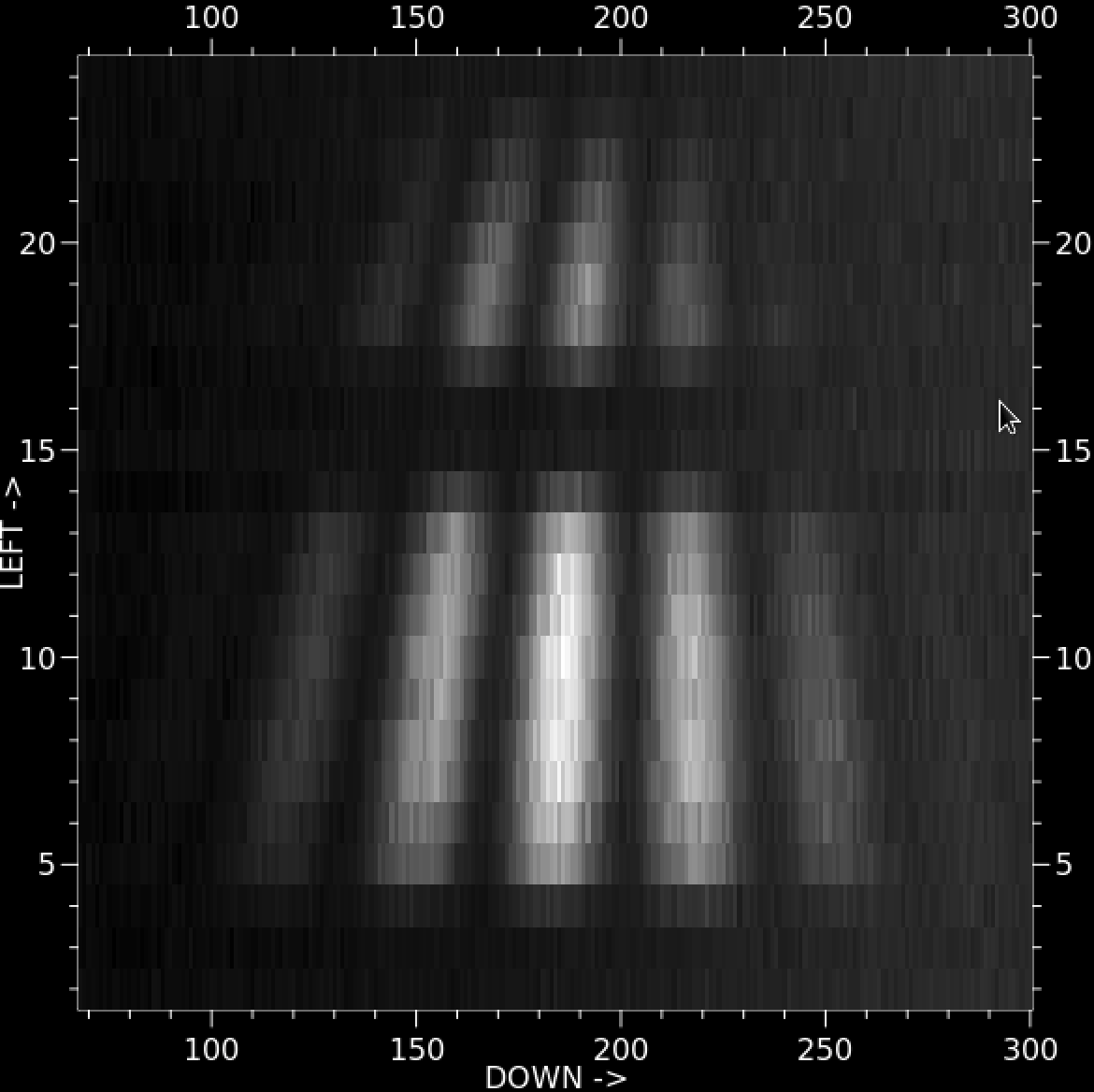}
\includegraphics[width=0.45\textwidth]{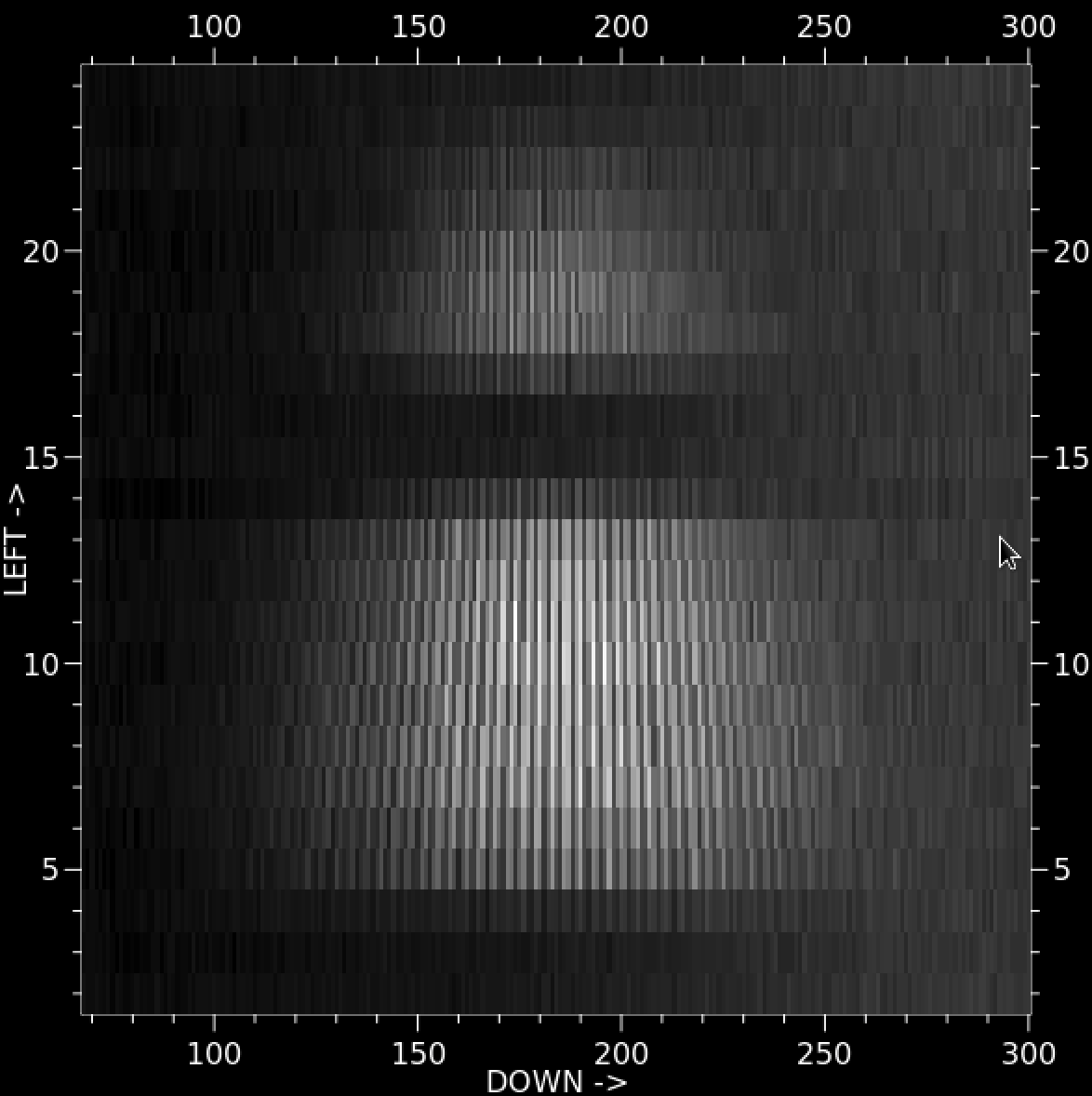}
\caption{Fringes in J+H band fringes observed for beam pairs 1 \& 2 (smallest spatial frequency) and 1 \& 6 (highest spatial frequency) using the internal light source. The horizontal axis and vertical axis are the spatial and spectral directions, respectively, in detector pixel coordinates. The dark patch in between the J and H-band fringes is due to the notch narrow-band filter used to filter out the CHARA metrology laser.}
\label{Fig16_JH_beams}
\end{figure*}

\subsection{Accurate wavelength calibration using etalon}
\label{sec:etalon}

One of the science goals of MIRC-X is to achieve $10\,\mu$as differential astrometric precision for wider binary systems. For this, we need wavelength calibration stability at the level of $\Delta\lambda / \lambda\approx10^{-5}$ level. The original MIRC was limited to a wavelength calibration stability of $\approx10^{-3}$\,\citep{Monnier2012}, 100-times larger than required for planet detection. This was tackled by installing state-of-the-art 6-beam optical etalon (Gardner et al. in prep). The etalon consists of six thin, $\approx2$\,mm-thick pieces of glass with parallel sides and 50\% reflective coatings on both sides, but with slightly different thickness ($\Delta x = 6\pm2\,\mu$m) for each beam. By inserting these pieces of glass in the optical path, the combiner detects multiple interference packets that mimic a binary star. Fitting the separation of this fake binary is a stable anchor of the spectral calibration of the instrument. This robust scheme measures the spectrometer wavelength using the same data pipeline and methodology as we use for binary science targets. The etalon is held in a unique, thermally-stable mount that can be inserted in the CHARA beams remotely. It can be used with the internal STS source, and can also be used when observing a (bright) scientific target.

MIRC-X can deliver astrometry of a maximum field of view of 1-m telescope diffraction ($\approx300$\,mas). We detect approximately $\leq10~\mu$as residuals on 100 mas binary star orbits after applying our etalon wavelength calibration (Gardner et al., in prep). It suggests we are reaching 0.01\% precision in wavelength calibration, though an upcoming program will compare binary orbits of MIRC-X directly with GRAVITY observations to confirm this precision.

MIRC-X delivers on-axis single field mode astrometry similar to PIONIER  observations. PIONIER can be wavelength cross-calibrated \,\cite[e.g., ][precision of $\sim0.04$\%]{Gallenne2018} with GRAVITY, which uses a dedicated internal reference laser source for the wavelength calibration. On other hand, with dual field phase referenced interferometry, GRAVITY demonstrates routine astrometry with precisions at the level of  $\sim50~\mu$as for a field of view of $2^{\prime \prime}$ or $4^{\prime \prime}$ based on either using UTs or ATs\,\citep{GRAVITY2017}. The first detection of exoplanets in optical interferometry was achieved with foreknowledge of the location exoplanets by VLTI/GRAVITY\,\citep{Lacour2019B,Nowak2020}. In their work, the astrometric positions of exoplanets $\beta$~Pic b  and HR~8799e with their respective stars are found with precisions of $40$ and $100~\mu$as, respectively.

\section{\label{sec:5}Performance and sensitivity limits}

The performance of MIRC-X was characterized using the MIRC-X internal source and on-sky observations. In these tests, the standard observation configuration was used unless stated otherwise. The standard observation mode uses prism R~=~50, a $320 \times 17$ pixels window, the IOTA readout mode with $N_{\rm reads}=12$, $N_{\rm loops}=8$, and a 350\,Hz frame rate. 

\subsection{C-RED ONE performances}

\begin{figure}
\centering
\includegraphics[width=0.45\textwidth]{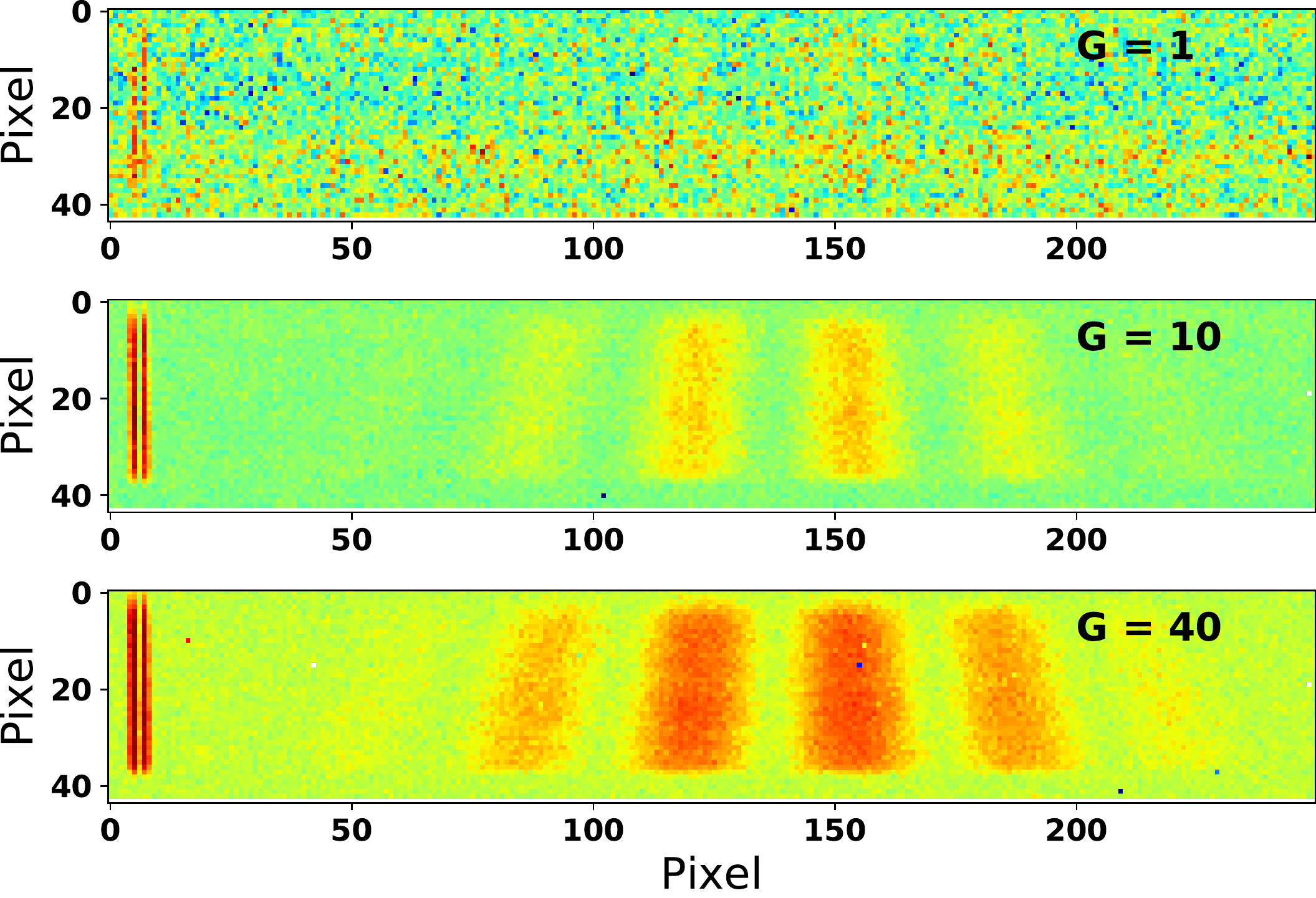}
\includegraphics[width=0.45\textwidth]{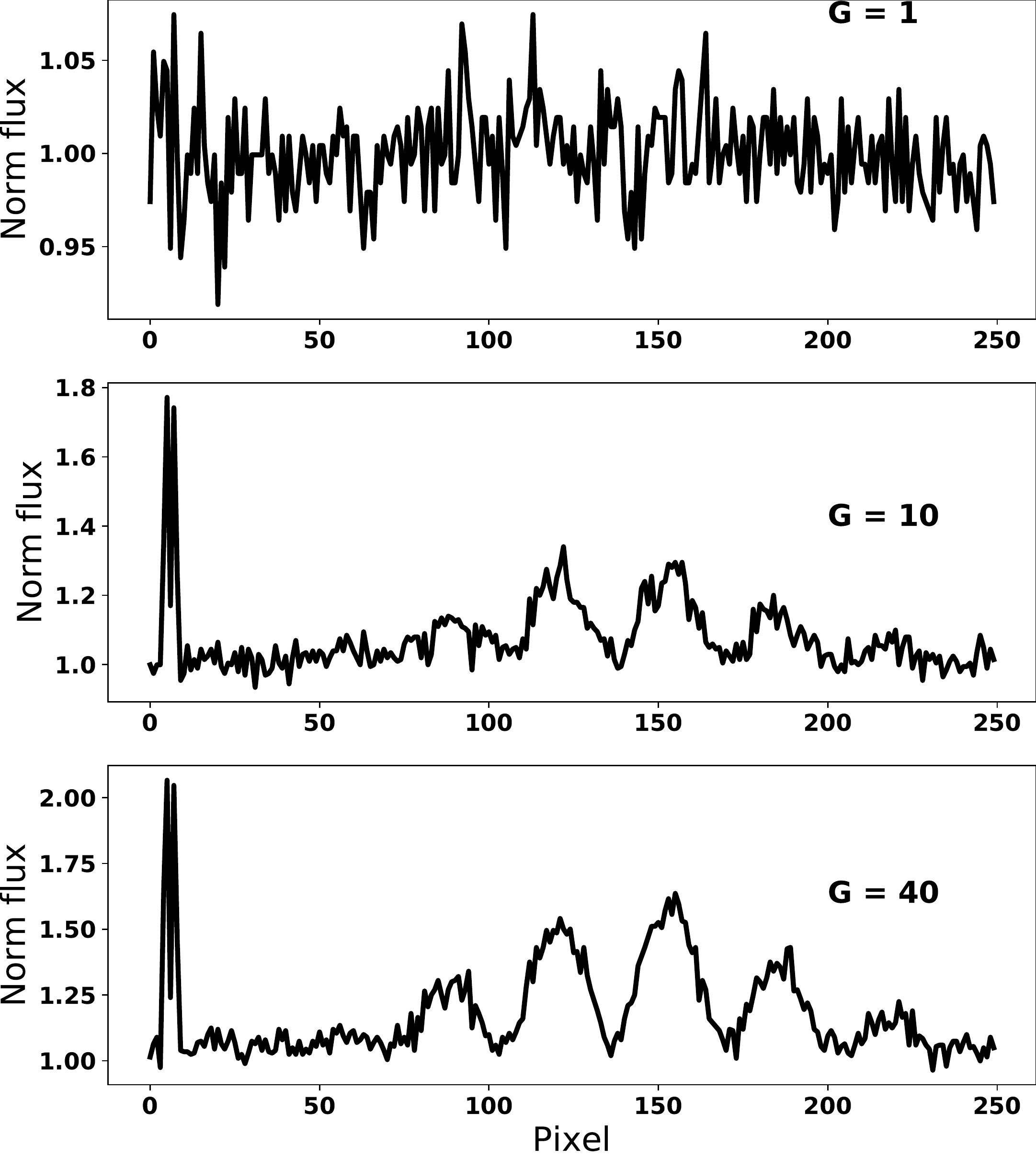}
\caption{MIRC-X two-beam fringes obtained with the internal source at avalanche gain 1, 10, and 40. The top three panels are the 2D images. The bottom three panels are for one spectral channel. Flux is normalized with respect to maximum flux in gain 1. In each panel, left are the photometric channels, and right are the fringes. For the avalanche gains above 10, the detector is in the sub-electron readout noise regime. For comparison, the MIRC-X performance at gain = 1 is approximately equal to the performance of MIRC's PICNIC detector. The photometric channels appear close but they are separated by 3 pixels peak-to-peak with beam size of $\sim0.6$\,pixel in FWHM.}
\label{Fig17_Fringe_SNR_at_Gain1_Gain40_IMAGE}
\end{figure}

\begin{figure}
\centering
\includegraphics[width=0.45\textwidth]{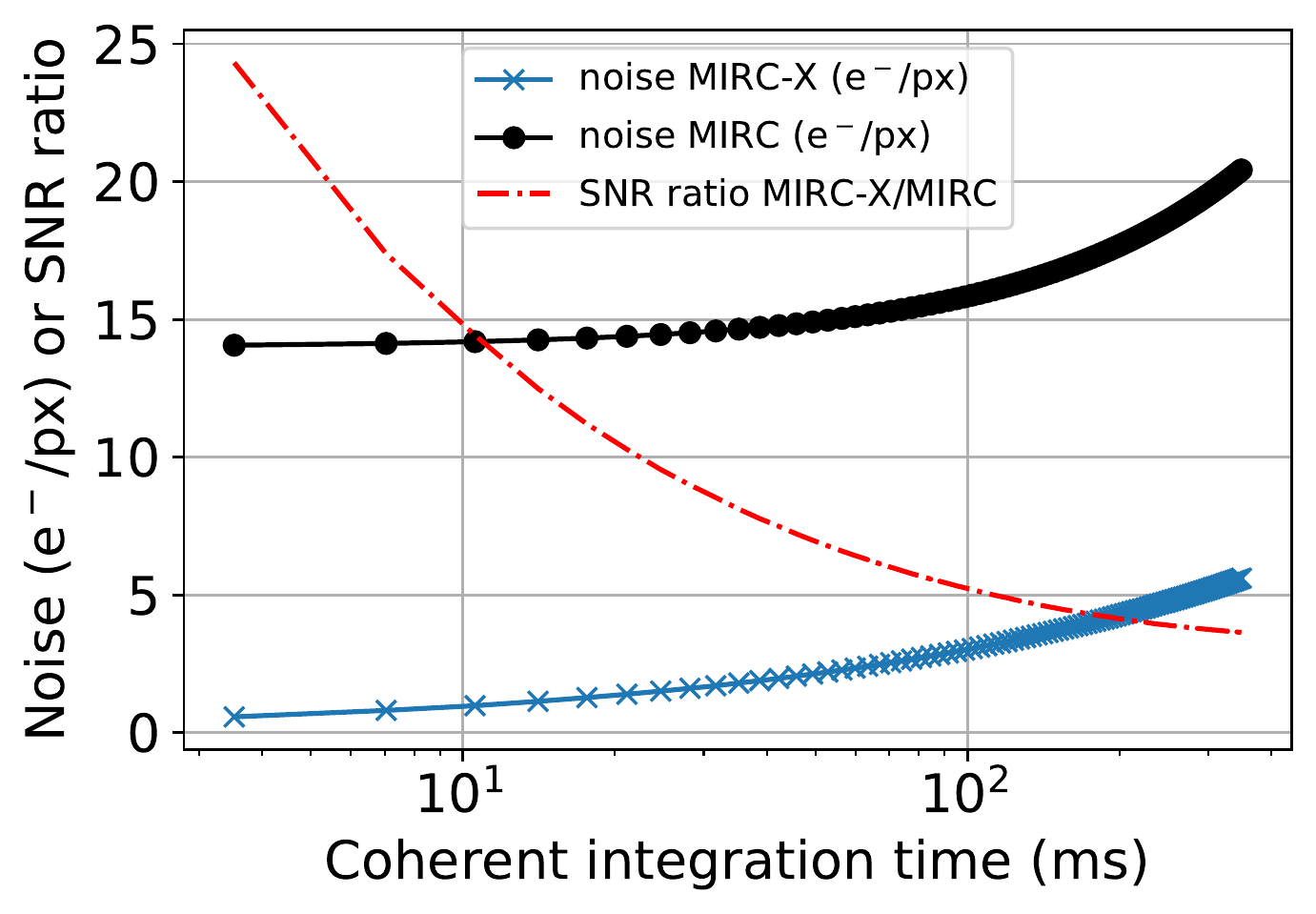}
\caption{Total noise model for MIRC-X camera at gain 40 (readout noise of $N_\mathrm{RON}\approx0.5\rm{e^{-}px^{-1}}$ RMS and background of $\approx90\rm{e^{-}px^{-1}s^{-1}}$) and for the MIRC camera (readout noise of $N_\mathrm{RON}\approx14\rm{e^{-}px^{-1}}$ RMS and background of $\approx340\rm{e^{-}px^{-1}s^{-1}}$). The MIRC-X camera delivers $14.5$ ($5.5$) better SNR than MIRC for 10\,ms (100\,ms) coherent exposures.}
\label{Fig18_MIRC-X_vs_MIRC}
\end{figure}
 
Figure\,\ref{Fig17_Fringe_SNR_at_Gain1_Gain40_IMAGE} showcases the revolutionary performance of the MIRC-X camera compared to the PICNIC. For comparison, the MIRC-X performance at gain$ =1$ is approximately equal to the PICNIC performance. For avalanche gains above 10, the detector readout noise is sub-electron per pixel. The camera has background noise as low as $90\,\rm{e}^-\rm{s}^{-1}$, depending on the avalanche gain. 

Figure\,\ref{Fig18_MIRC-X_vs_MIRC} presents a sensitivity estimation of MIRC and MIRC-X instruments based on their readout and background noise numbers.  For typical coherent integration of $\approx10$\,ms, MIRC-X has 15 times better SNR than MIRC. For longer coherent integration times, the MIRC-X camera is dark noise-limited and the improvement with respect to MIRC is less, but still significant.

\subsection{Characterization using internal light}
\label{sec:internal_carac}
The raw instrumental contrast is mostly affected by instrument vibrations, and polarization birefringence and wavelength dispersion in fibers. Instrument vibrations on the optical table have been proved challenging to correct. A dominant source of vibration came from the camera's pulse-tube compressor while cryocooling. These vibrations have since been reduced by upgrading the electrical wires used (Sep 2018). Differential polarization between beams is corrected with the polarization controller (see Section\,\ref{sec:Polarization}). The non-calibrated visibility contrasts in the H-band for the fringes with the smallest spatial frequency on the detector (beams 1 \& 2) is 95\% on the internal light source (see Figure\,\ref{Fig15_Contrast_beams}). For the highest spatial frequency beam pair (beams 1 \& 6), the fringe contrast is only $65$\% while about $75$\% is expected from loss due to the poor sampling. This reduced contrast is due to the residual vibrations on the optical bench. Figure\,\ref{Fig16_JH_beams} shows the fringes across the J+H bands for the smallest and highest spatial frequency. The instrument chromatic dispersion across the spectral bands is acceptable.

\begin{figure}
\centering
\includegraphics[width=0.45\textwidth]{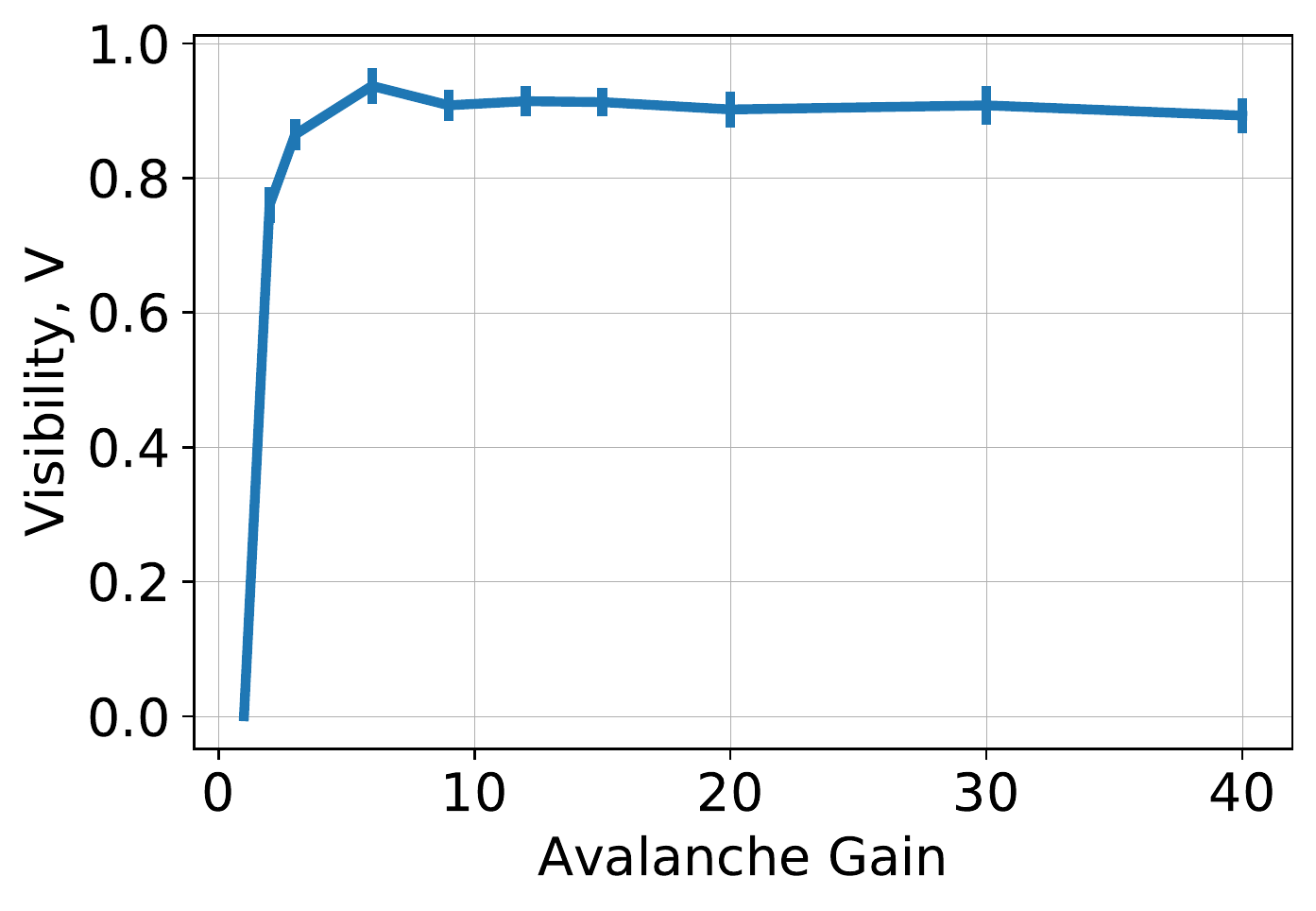}
\includegraphics[width=0.45\textwidth]{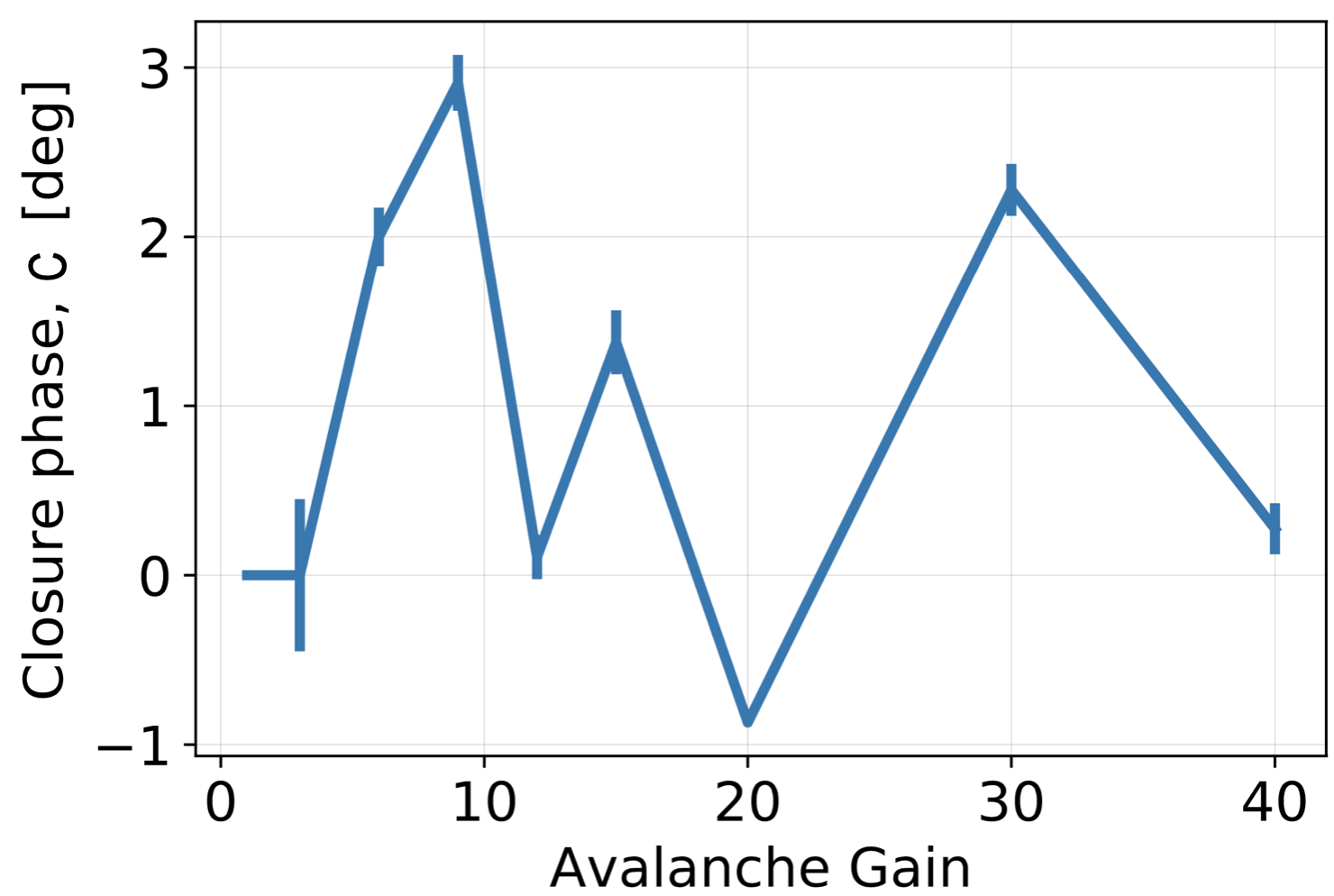}
\caption{Visibility (top) and closure phase (bottom) averaged across all wavelength channels (prism R~=~50) measured using the internal light source while varying the detector avalanche gain. The visibility variation is less than 2\%, and closure phase variation is less than $3^{\circ}$. The bottom plot data is very noisy because a lack of good flat fielding reference for each avalanche gain.}
\label{Fig19_Visibility_vs_STS_GAIN}
\end{figure}

The typical statistical precision measured on the STS is about 0.5\% for the squared visibilities and about $0.2^{\circ}$ for the closure phase. We investigated the dependency of the measurement to the amplification gain, because this parameter is the most susceptible to be modified between a bright, resolved science target and a fainter, less resolved calibrator. Figure\,\ref{Fig19_Visibility_vs_STS_GAIN} presents the visibility and closure phase stability as a function of detector avalanche gain. Measurements with gain$\,<5$ are affected by the detector electronic artifact -- interference noise\,\citep[see Figure 1 of ][]{Lanthermann2018}. Consequently, the camera is always operated with gain$\,>5$. For these gains, $\,<5$, the visibility variation is less than 2\%, and closure phase error is less than $3^{\circ}$. This variation is due to (i) varying hot pixels as a function of avalanche gain and (ii) a lack of good flat fielding reference for each avalanche gain. We are planning to install a diffuser in the filter-wheel, which spreads the light on the detector uniformly to make good flat fields for each gain. Since the change of avalanche gain introduces a few percentage variations, the standard calibration strategy is to observe calibrator and target objects using the same gain.

\subsection{Commissioning and high-priority science observations}

The MIRC-X instrument was commissioned in two phases in order to minimize risks during the upgrade. In Phase 1, in June 2017, we commissioned: (i) C-RED ONE camera in place of the MIRC PICNIC camera, (ii) CHARA compliant control software architecture, and (iii) polarization controllers. In Phase 2, in September 2018, we commissioned (i) the new crosstalk-resistant beam combiner optics, which are optimized for the pixel-scale of the new camera and (ii) the redesigned high throughput photometric channel relay optics with new J+H band fibers and a new custom non-polarizing beamsplitter.

Since June 2017, MIRC-X has made numerous nights of observations for a variety of science cases --  young stellar objects,  star spot imaging, astrometric planet detection, post-AGB circumbinary imaging, binary orbit studies, etc. The MIRC-X image reconstruction capability is demonstrated on an Asymptotic giant branch star CL Lac,  by detecting convection-related variability, in the first accepted paper\,\citep{Chiavassa2020}. The astrometric capability of MIRC-X is demonstrated on GW\,Ori, a YSO triple system \citep{Kraus2020}.  The other observational data is currently under various stages -- data analyzing, modeling, manuscript preparation and journal review (Thomas et al. submitted). It is now the most demanded instrument at CHARA. The first results of J+H bands and polar-interferometric observations of YSOs are in preparation (Labdon et al.; Setterholm et al.).

\subsection{Sensitivity limits on-sky}

\begin{figure}
\centering
\includegraphics[width=0.45\textwidth]{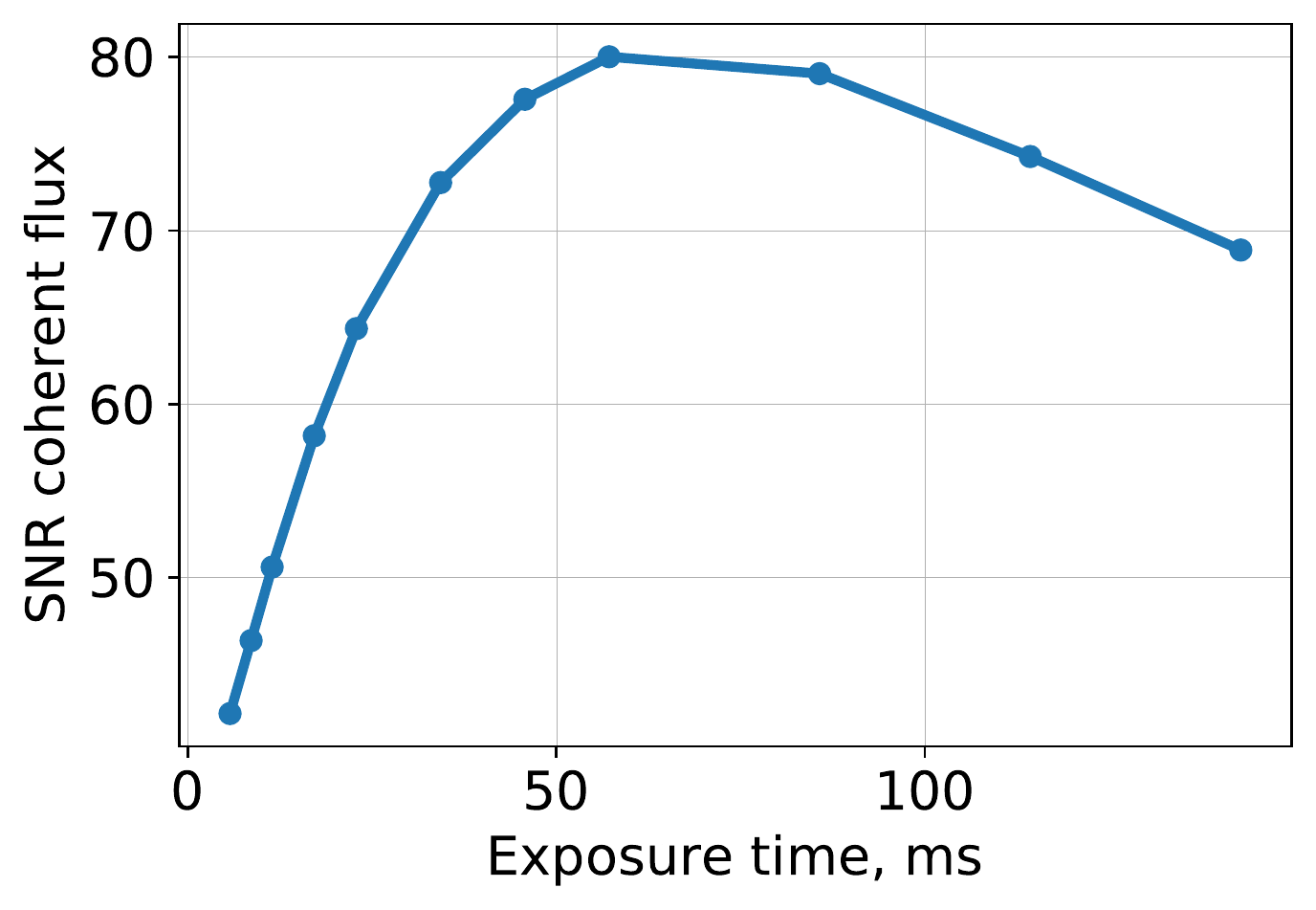}
\caption{Wavelength-averaged SNR of the coherent flux as a function of coherent integration time for UT 2019-08-27. The SNR of the coherent flux increases with coherent integration time until it equals the atmospheric coherence time of $\tau_0 \approx 60\,$ms. }
\label{Fig20_SNR_with_ncoherent}
\end{figure}

The limiting magnitude of MIRC-X is defined by the capability of tracking the fringes during the observations (SNR\,$> 2.5$). The SNR increases with the number of photons received from the object over coherence patch $r_{0}^2$ and during the time $\tau_0$ (the so-called coherence volume). Figure\,\ref{Fig20_SNR_with_ncoherent} presents the SNR of the coherent flux when varying the number of frames co-added coherently by the pipeline (coherent exposure time). The SNR of the coherent flux increases because the signal in integrated linearly while most of the noises are integrated in a random walk, and then decreases because of the limited atmospheric coherence time. The optimum value depends on the object brightness and the atmospheric coherence time. Nevertheless, the range of acceptable values for a loss of 20\% of SNR is rather large (from $0.3\times$ to $3\times$ the optimal coherent integration time).

For a typical coherent integration time of $25\,$ms, the usage of avalanche gain 40 enables an increase of 15 SNR (see Figure\,\ref{Fig18_MIRC-X_vs_MIRC}).  For the best possible coherent integration time of $60\,$ms, the total noise is dominated by the camera dark and background ($90\,\mathrm{e^{-}/px/s}\times 0.06\,\mathrm{s} \approx 5.4\mathrm{e^{-}/px}$). Consequently, using a higher spectral resolution reduces the sensitivity limit as it spreads the amount of light used for fringe-tracking over more pixels.

\begin{figure*}
\centering
\includegraphics[width=0.9\textwidth]{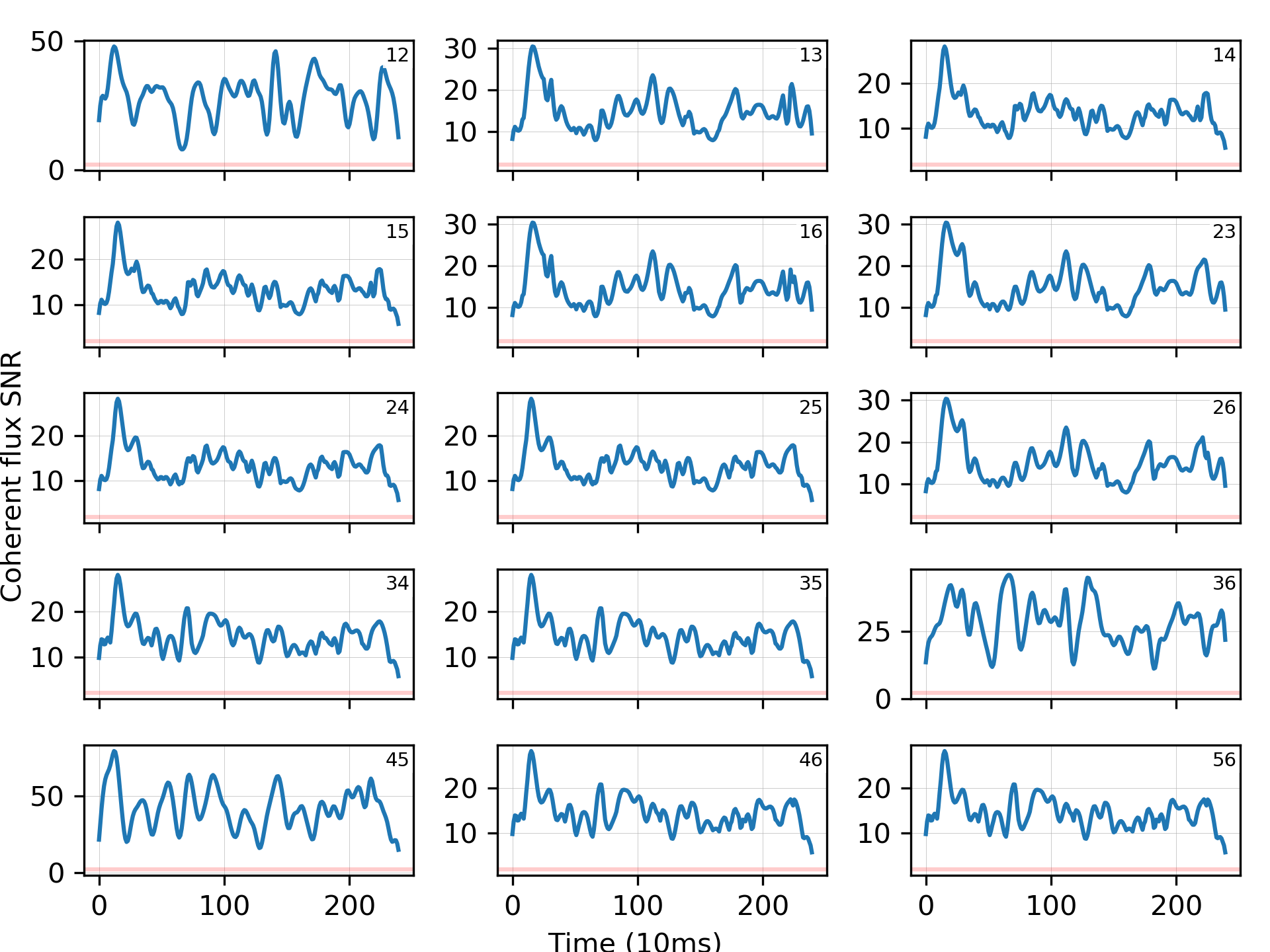}
\caption{SNR of fringes recorded on a faint YSO, GW\,Ori, mH~=~7.1 on UT 2019-08-27 under exceptional seeing conditions ($r_0\approx11$\,cm, wind speed $\approx2.7$\,m/s). The red color is the SNR threshold above which fringes are considered. The 15 panels are for the 15 baselines. The baseline pair numbers are indicated in each panel. The data are reduced with a coherent integration time of 30\,ms.}
\label{Fig21_SNR_faintest_YSO_GW_Ori}
\end{figure*}

Since the commissioning of MIRC-X, the faintest target on which fringes have been tracked is the H-band magnitude $\rm{mH}\approx8.2$  (correlated magnitude $\approx8.2$) star HD\,201345 using prism R~=~50 with averaged SNR on all baselines of approximately 25. The faintest YSO observed so far is GW\,Ori with $\rm{mH}\approx7.1$ (correlated magnitude $\approx7.34$), also using prism R~=~50 \citep{Kraus2020}. Figure~\ref{Fig21_SNR_faintest_YSO_GW_Ori} shows the SNR of fringes for GW\,Ori observed on UT date 2019-08-27 ($r_0\approx11$\,cm, wind speed $\approx2.7$\,m/s).

\subsection{On-sky precision and accuracy}

\begin{figure}
\centering
\includegraphics[width=0.45\textwidth]{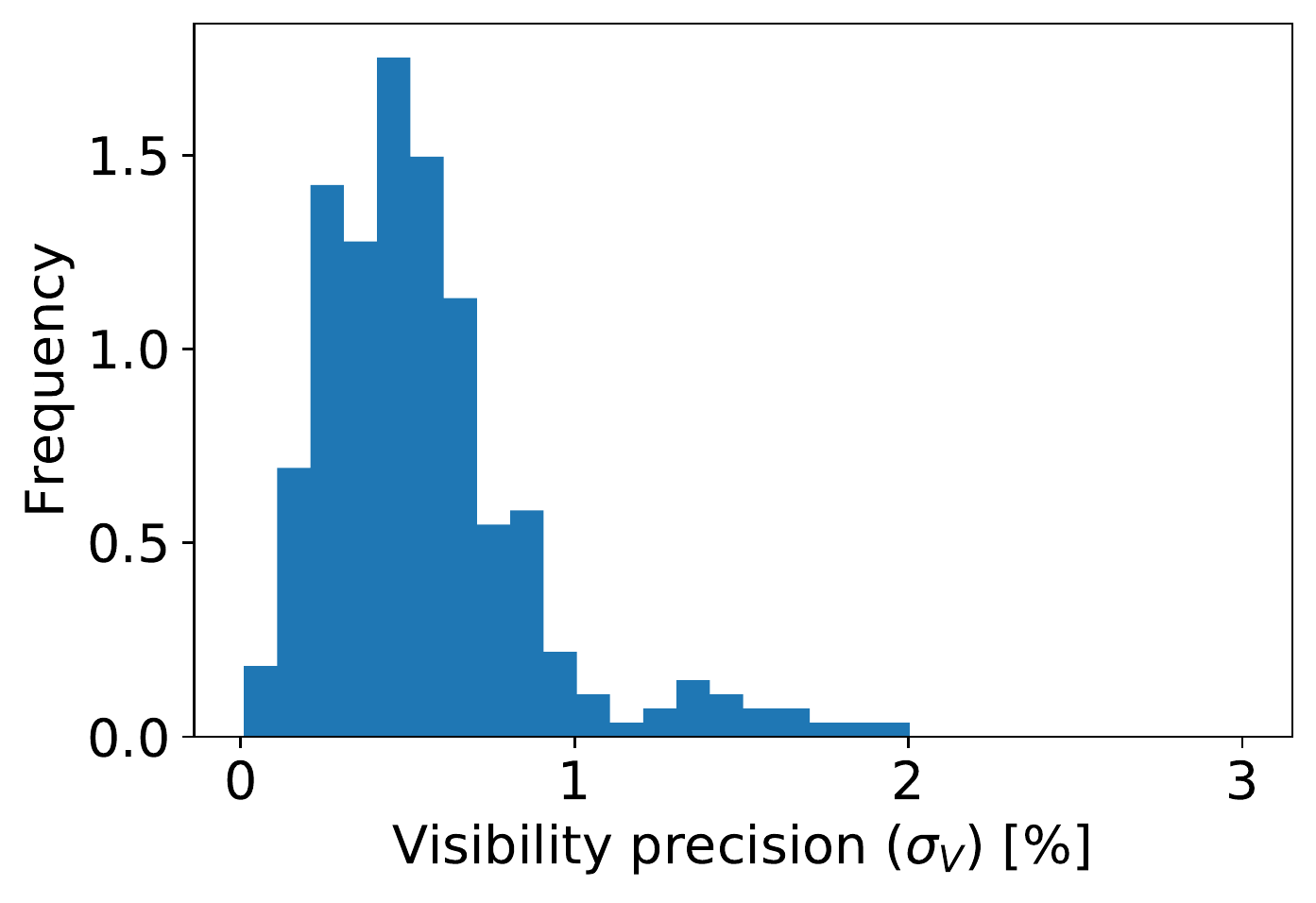}
\includegraphics[width=0.45\textwidth]{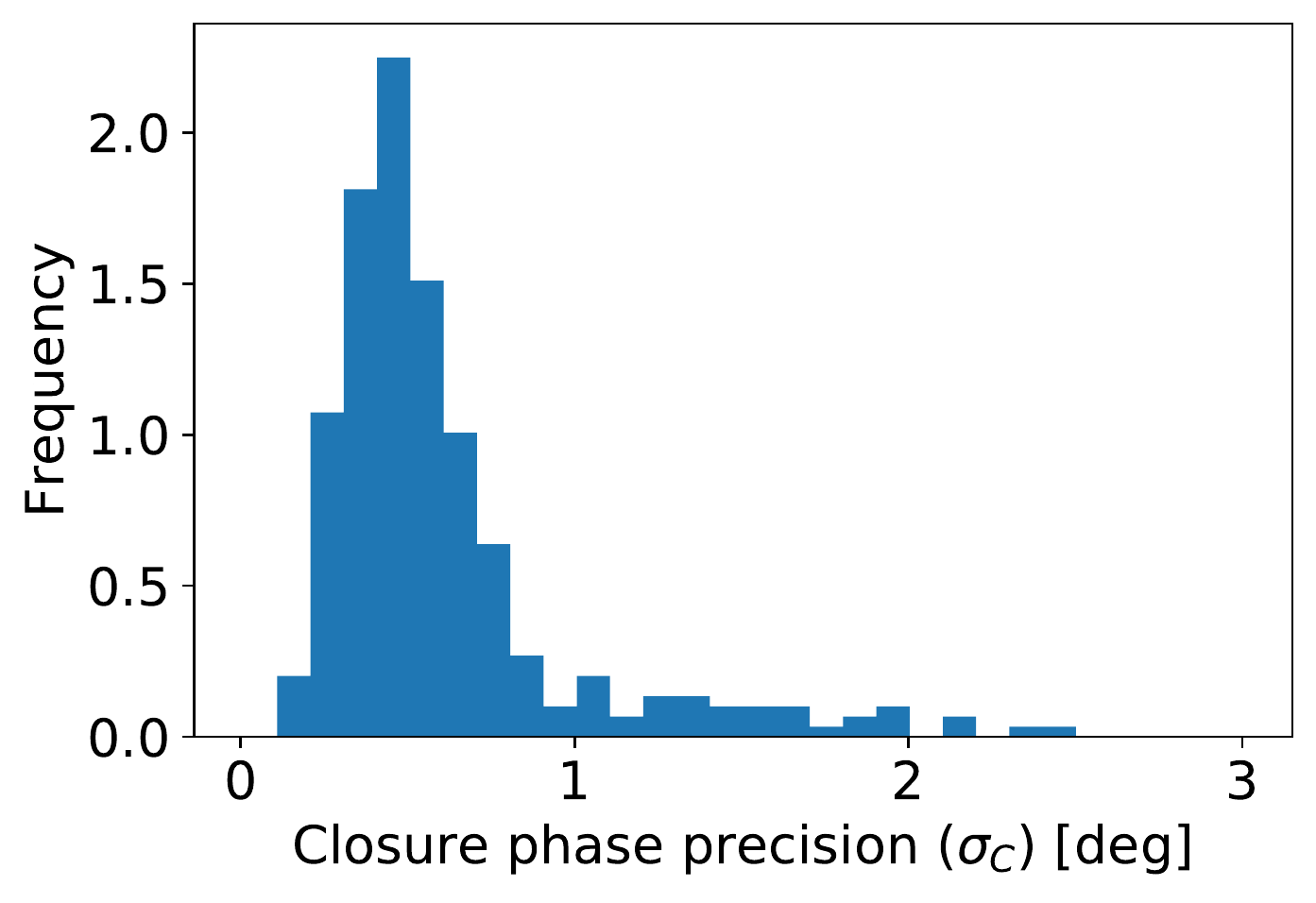}
\caption{Histogram of the wavelength-averaged visibility (top) and closure phase (bottom) precision estimated from on-sky observation of bright calibrators on UT\,2019-09-04. The precision is the statistical 1-$\sigma$ uncertainty estimated by the pipeline.}
\label{Fig22_On_sky_v_error_date_2019Aug29}
\end{figure}

Figure\,\ref{Fig22_On_sky_v_error_date_2019Aug29} presents the histograms of the standard deviations of the raw visibility and closure phase observed on bright calibrators on UT\,2019-08-29. The data are reduced with 30\,ms coherent integration. The statistical precision of the visibility is about 0.5\%, and about $0.5^{\circ}$ for the closure phase. Observations of Betelgeuse ($\rm{mH} = -3.7$) with MIRC-X show visibility square in the range of $2\times10^{-3} - 1\times10^{-4}$ demonstrates high dynamic range enabled precision (manuscript in prep.)  

\begin{figure*}
\centering
\includegraphics[width=0.95\textwidth]{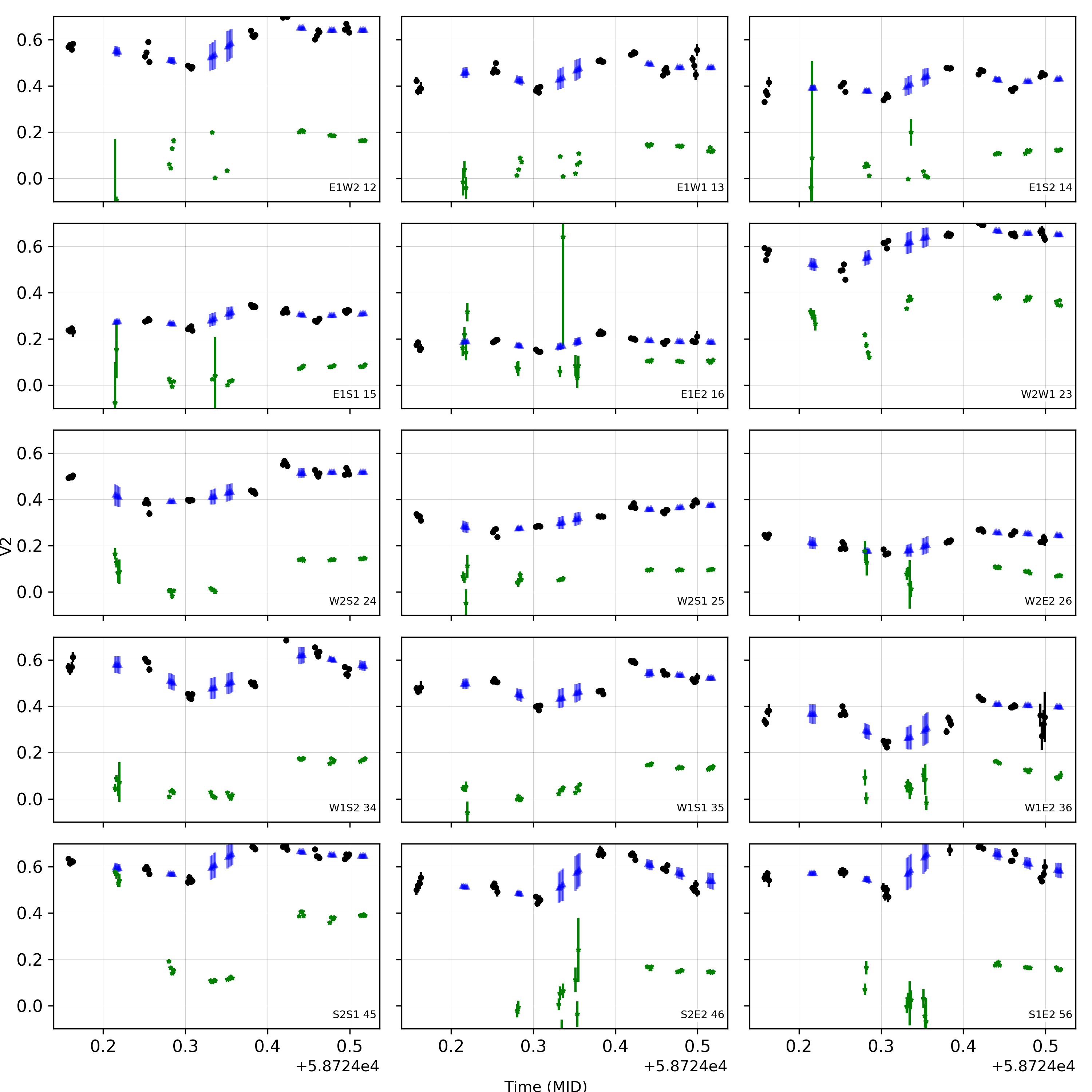}
\caption{Transfer function squared visibility measurements for MIRC-X observed on UT\,2019-08-29.  The black color represents the transfer function estimates for calibration stars. The blue color represents the interpolation of the transfer function at the time of science-star observations. The green color represents the observations of scientific objects. The observed calibrators are HD\,199580 (0.50\,mas), HD\,228852 (0.53\,mas), HD\,204050 (0.42\,mas), HD\,21820 (0.47\,mas), HD\,28855 (0.30\,mas) from left to right and with H-band angular diameters in parentheses \citep[JSDC;][]{Bourges2017}. The majority of the visibility variation seen in the calibrators is due to variations in the system visibility due to seeing and different pointings on the sky (polarization changes).}
\label{Fig23_TransferFunction}
\end{figure*}

The transfer function of MIRC-X observed on UT\,2019-08-29 ($r_0\approx14$\,cm, wind speed $\approx3.3$\,m/s) is shown in Figure\,\ref{Fig23_TransferFunction}. This data set consists of observations of several calibrators and science targets. The variations in visibility between different observations are not fully understood. They are much larger than the internal MIRC-X stability, and somewhat larger than the expected effect of atmospheric instability from MIRC experience. They can possibly be related to the varying polarization of the telescope and beam transporting optics of CHARA for different pointing locations on the sky. The visibility of science targets are calibrated by the standard method of interpolating the transfer function at the observing time of the science targets (see data reduction pipeline Section\,\ref{sec:Pipeline}).

\subsection{On-sky verification: $\iota$\,Peg binary star}

\begin{figure*}
\centering
\includegraphics[width=\textwidth]{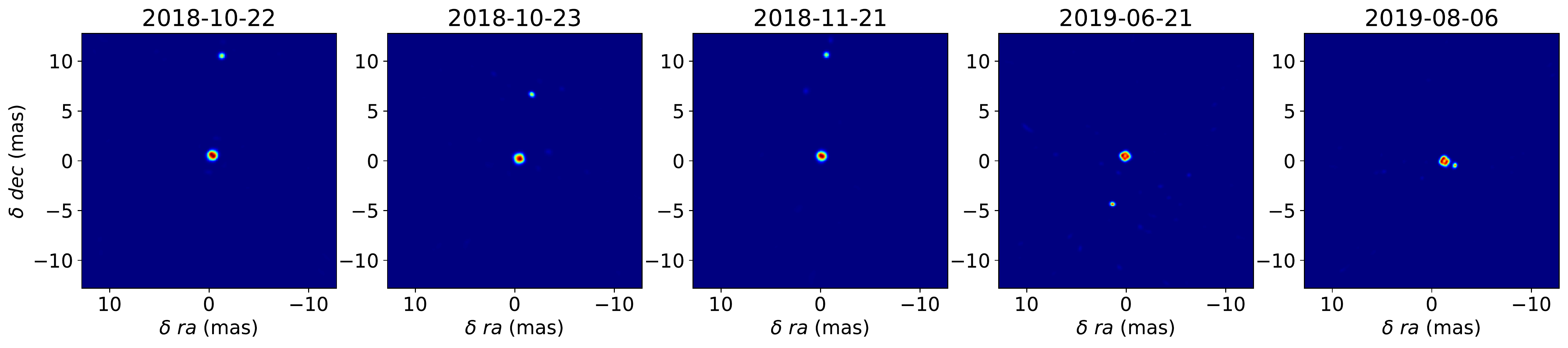}
\caption{Model-independent image reconstruction of $\iota$\,Peg at several epochs using the \texttt{squeeze} software.  Observations of each night are reduced independently, and images are reconstructed.  North is up and East to the left. The image maximum is normalized to unity, and the color scale is linear. The primary star is centered in the image window.}
\label{Fig24_Iota_peg_images}
\end{figure*}

We illustrate the imaging capability of MIRC-X using the well known binary star $\iota$\,Peg \citep[HD\,210027, period 10.2 days, ][]{Konacki2010}. The observations and resulting (u, v)-coverage are summarized in the Appendix\,\ref{sec:IotaPegObs}. The calibrated OIFITS files are available on the Optical interferometry DataBase\footnote{\href{http://oidb.jmmc.fr}{ http://oidb.jmmc.fr}}. For all epochs, $V^2$ significantly below unity  meaning that at least a fraction of the near-infrared emission is spatially resolved in all of them. The closure phases show a nonzero signal indicating asymmetry of the binary system. The visibility and closure phase data are analyzed with three methods: (i) model-independent image reconstruction with \texttt{squeeze}; (ii) binary detection with the \texttt{CANDID} algorithm; (iii) model fitting developed by our group (software available upon request). 

\texttt{squeeze} is an interferometry imaging software that allows for the simultaneous fitting of the squared visibilities and closure phases. Figure\,\ref{Fig24_Iota_peg_images} shows the image reconstruction of $\iota$\,Peg for different epochs. We used a ``field-of-view" regularizer with a weight of 1000. \texttt{squeeze} uses a simulated annealing Monte-Carlo algorithm as the engine for the reconstruction. We created five chains with 500 iterations each to find the most probable image. We used a $256\times256$ pixel grid with a scale of $\tfrac{\lambda}{8B}\sim 0.1$\,mas per pixel as it gave among the best $\chi^2\approx2$. The pixel scales less than $\tfrac{\lambda}{4B}$ gave almost the same binary separations ($\rho$) and position angles ($PA$), which are approximately reproduced with other methods described below. The final images are the average of the frames from the different Monte-Carlo chains that converged to fitting residual of $\chi^2\approx2$ for the visibilities and closure phases (see fitting residuals in Appendix \ref{sec:IotaPegObs}).  These observations and images illustrate the snapshot imaging capability of MIRC-X. The orbital motion of the binary is observed after just 40~minutes, detecting a motion of about 0.13\,mas. 

\begin{figure*}
\centering
\includegraphics[width=0.9\textwidth]{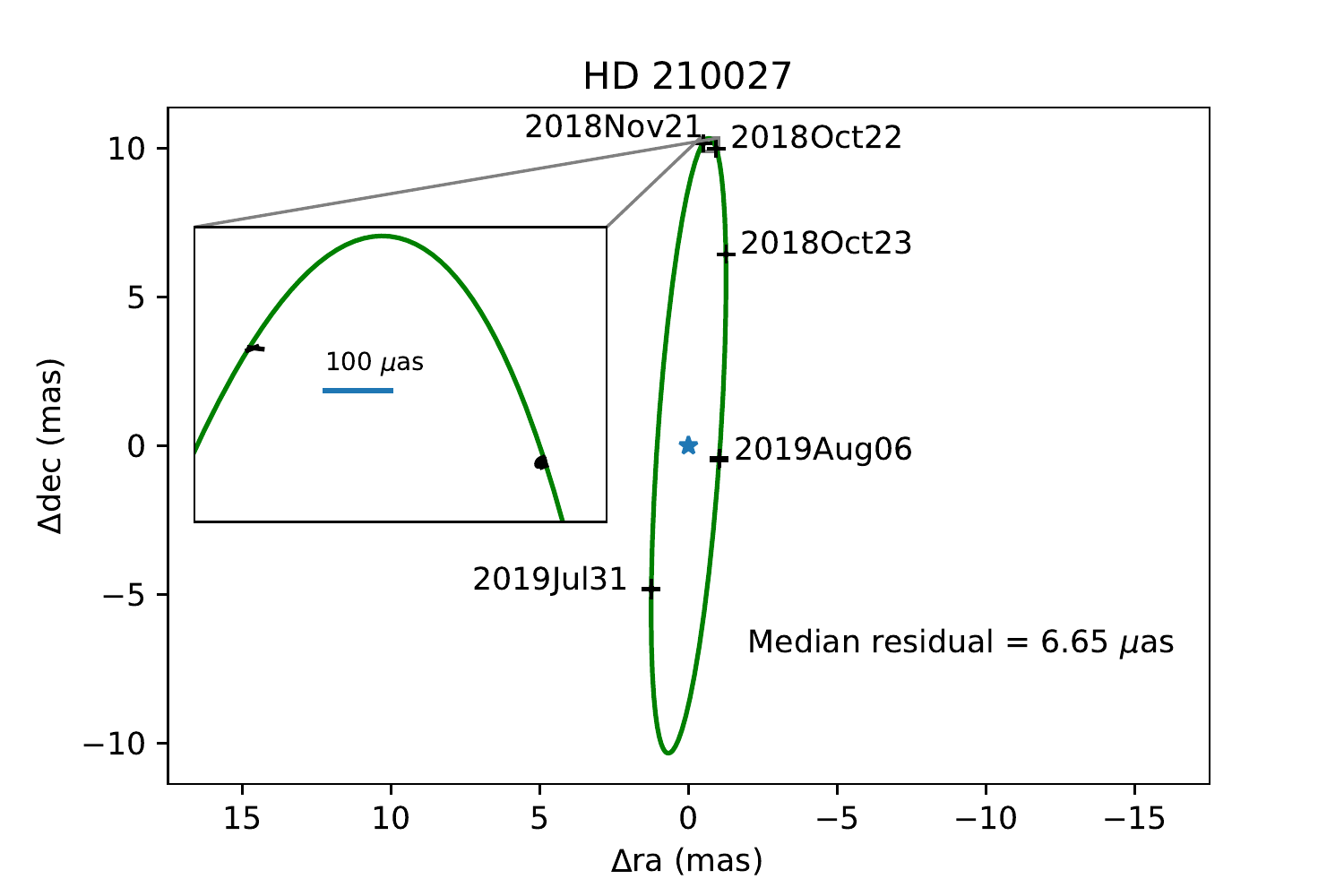}
\caption{Best-fit of the apparent orbit of $\iota$\,Peg, considering all the Campbell elements. The black marks with dates are the MIRC-X astrometric measurements. The green line is the best orbit with the median residual is 6.65$\mu$as. The reference frame is centered on the primary star. North is up and East to the left. The zoomed plot shows the residual of orbital fit, visually.}
\label{Fig25_IotaPeg_orbit}
\end{figure*}

Table\,\ref{table1} summarizes the model fitting parameters. The \texttt{squeeze} resulted parameters are used as initial parameters for the model fitting. Bandwidth smearing is taken into account in the model fitting. These parameters are confirmed by the \texttt{CANDID} binary detection algorithm within $3.4~\sigma$.  Figure\,\ref{Fig25_IotaPeg_orbit} shows the binary orbit fitted using the least-squares approach, and the best-fit parameters are presented in Table\,\ref{table2}. The best-fit parameters very well match with spectroscopic observations \citet{Konacki2010}. We estimate a flux ratio of primary to secondary $F_{\rm pri/sec}$ of $4.6\pm0.1$ with a smaller error compared with the flux ratio of $5.0\pm0.5$ reported by \citet{Konacki2010}.

\begin{table*}
\caption{$\iota$\,Peg best-fit astrometry determined with model (hereafter $MF$) and \texttt{CANDID} fitting algorithms (hereafter $CF$). The diameters of the primary and secondary in the binary system are 1.05 and 0.6\,mas, respectively. The last three columns are the difference in two methods computed in  $\sigma$ units, i.e., $\frac{|MF - CF|}{\sqrt{ {\sigma_{MF}}^2 + {\sigma_{CF}}^2}}  $. }             
\label{table1}      
\centering                          
\begin{tabular}{c c c c c c c c c c}        
\hline\hline                 
 &     \multicolumn{3}{c}{\texttt{squeeze} + modeling}  & \multicolumn{3}{c}{\texttt{CANDID}} \\  
 UT date   &   $F_{\rm pri/sec}$ & sep, $\rho$ (mas) & pos ang, $PA$ (deg) &  $F_{\rm pri/sec}$ & $\rho$ (mas) & $PA$ (deg) & $\Delta F / \sigma$ & $\Delta \rho / \sigma$ & $\Delta PA/\sigma$ \\ 
\hline                        
2018-10-22 & $4.55 \pm 0.04$ & $10.0352 \pm 0.0176 $ & $354.7005 \pm0.019$ &4.68$\pm$0.08 & $10.0206 \pm 0.0051$ & $354.6822 \pm 0.0137$ & 1.45 & 0.79  & 0.78\\

2018-10-23 & $4.60 \pm 0.03$ & $6.5705  \pm 0.0103 $ & $348.914 \pm0.007$  &4.62$\pm$0.09 & $6.5555 \pm 0.0019$ & $348.8865 \pm 0.0094$ & 0.21 & 1.43 & 2.35 \\

2018-11-21 & $4.60 \pm 0.01$ & $10.1793 \pm 0.0269 $ & $357.2094 \pm0.029$ &4.50$\pm$0.08 & $10.1523 \pm 0.0017$ & $357.1934 \pm 0.0132$ & 1.24 & 1.00 & 0.50\\

2019-07-31 & $4.77 \pm 0.02$ & $4.979  \pm 0.0070 $ & $165.5565 \pm0.007$ &4.59$\pm$0.05 & $4.9549 \pm 0.0023$ & $165.3923 \pm 0.0234$ & 3.34 & 3.27 & 2.63\\

2019-08-06 & $4.54 \pm 0.03$ & $1.1297  \pm 0.0075 $ & $244.4933 \pm0.005$ &4.49$\pm$0.05 & $1.1229 \pm 0.0061$ & $244.4312 \pm 0.01813$ & 0.86 & 0.70 & 3.30\\
\hline                                   
\end{tabular}
\end{table*}

\begin{table*}
\caption{Best-fit orbital solutions of $\iota$\,Peg binary star orbit. }             
\label{table2}      
\centering                          
\begin{tabular}{c c c }        
\hline\hline                 
Parameter  & MIRC-X &  \citet{Konacki2010}\\
\hline                        
Semi-major axis, $a$ (mas) & $10.3609 \pm 0.0038$* &  $10.329 \pm0.016$\\
Period, $P$ (days) & $10.21277 \pm 0.00021$ & $10.2130253\pm 0.0000016 $\\
Eccentricity, $e$  & $0.0052 \pm 0.0011$  &      $0.001764 \pm 0.000063$\\
Time of periastron, $T$ & $52997.4686 \pm 0.19$ &  $52997.378\pm0.052$\\
Longitude of the periastron, $\omega$ (deg) & $272.45 \pm 1.95 $ & $272.8\pm1.8$ \\
Longitude of the ascending node, $\Omega$ (deg) & $176.194 \pm 0.015$ \textdagger & $176.262\pm0.075$\\
Inclination, $i$ (deg) &  $95.865 \pm 0.010 $  & $95.865\pm0.011$\\
diameter primary		&	1.05 &		1.06	\\				
diameter secondary		&	0.6	 &	0.6	 \\
\hline  
\multicolumn{3}{c}{* 0.5\% uncertainty in the semi-major due to uncertain absolute wavelength calibration.}\\
\multicolumn{3}{c}{\textdagger  Our orbit $\Omega$ is not in J2000, apparent North-East precession.       }\\

\end{tabular}
\end{table*}

\section{\label{sec:6} Summary and conclusions}

MIRC-X is the world's highest-resolution imaging facility in the near-infrared, primarily aimed at studying circumstellar or circumbinary environments around young stellar objects, star spot imaging, astrometric planet detection, post-AGB circumbinary imaging, multiplicity surveys, and orbital monitoring of binary systems. MIRC-X is the upgrade of MIRC, redesigned and rebuilt substantially for sensitivity,  precision, broader wavelength coverage (J and H-bands), polarization control, and observational efficiency. The development of MIRC-X capitalized from the advances in the recent detector technology, SAPHIRA, leveraging many technical lessons from MIRC and expertise from instruments such as PIONIER and GRAVITY.  MIRC-X is the first instrument to use a C-RED ONE camera as a science detector.

MIRC-X demonstrated up to two magnitudes of sensitivity improvements in comparison to MIRC, the exact value depends on the atmospheric seeing and coherence time, and the spectrograph in use. This improved sensitivity enables us to observe faint YSOs, detect the astrometric signature of exoplanet companions in wide-separation binaries, and record high-spectral resolution observations. So far, MIRC-X has demonstrated 1\% calibrated visibility and $1^\circ$ closure phase precision on brighter stars and under good conditions. On astrometric precision, we measure $\leq10~\mu$as residuals on 100 mas binary star orbits after applying our etalon wavelength calibration (Gardner et al., in prep). MIRC-X has been making successful science observations since June 2017 and is the most observer-requested instrument at the CHARA Array. Its imaging capability is illustrated with the $\iota$\,Peg binary star system.  As a general user instrument, MIRC-X can be used for observations by the international community through CHARA open-access time offered through the NSF OIR Lab\footnote{\href{https://www.noao.edu/gateway/chara/}{https://www.noao.edu/gateway/chara/}}.

Even if MIRC-X is delivering unprecedented sensitivity for 6-telescope combination, the current performance still falls short compared to the fundamental limit of a 1\,m-telescope interferometer ($\rm{mH}\approx13$ assuming SNR=3 for coherence time 10\,ms and total transmission of 3\%). Undoubtedly the most limiting aspect is the wavefront quality reaching the single-mode fibers of MIRC-X. The new CHARA Adaptive Optics is currently under commissioning \citep{tenBrummelaar2018} and is expected to be used in routine science operations starting in 2020. Nearly diffraction-limited image quality should increase the MIRC-X sensitivity by at least one magnitude fainter, but there are still several operational hurdles to be tackled to achieve this goal.

On the MIRC-X side, several developments are currently ongoing:
\begin{itemize}
\item A few on-sky observations of J+H and polarimetric observation have been acquired and are being analyzed. But the data reduction pipeline for these modes is currently experimental, and further efforts will be required in order to make this mode available to the community. The J-band version of the pipeline is ready and tested for the R=50 mode but still  requires testing with other spectral modes. For the split-polarization mode, we are still working to understand the instrumental polarization induced by all the reflections in the CHARA Array and MIRC-X beam paths. 
\item By the end of 2020, the MYSTIC instrument will be installed next to MIRC-X. We anticipate operating both instruments simultaneously by default, delivering routine H+K observations.
\item In collaboration with the MIRC-X team, the Observatoire de la C\^{o}te d'Azur (France) currently develops the software module for a true phase-tracking mode for MIRC-X. The goal is to use MIRC-X as phase-tracker for MYSTIC and for the future six-telescopes visible beam combination instrument SPICA, expected to be commissioned at CHARA in 2021 (Denis Mourard, private communication).
\item Observatoire de la C\^{o}te d'Azur is also developing an alternate beam combiner for MIRC-X. This is a pairwise, integrated optics chip which codes each pair into four interference states (ABCD) as implemented, for instance, in PIONIER and GRAVITY. Since MIRC-X is currently limited by dark and background noise, at least under the most favorable atmospheric coherence time, such a design could improve the sensitivity by reducing the number of pixels (120 spatial pixels instead of 200 for the all-in-one design; and minimum spectral resolution of 20 instead of 50 for the all-in-one design). A prototype has been tested in March 2020, and a science-grade chip is expected for the end of 2020.
\item Last, the beam combiner optics of MIRC-X are currently at room temperature. Building a cryostat to keep the optics cold would reduce the background and increase the instrument sensitivity further.
\end{itemize}

Finally, MIRC-X is designed to conduct simultaneous observations with the forthcoming instruments MYSTIC~\citep{Monnier2018} and SPICA~\citep{Mourard2017}, providing observations spanning the R (SPICA), J and H (MIRC-X), and K band (MYSTIC).  The benefits are: (i) an increased (u,v)-coverage of the spatial frequencies enable snapshot imaging in the optical/near-infrared; (ii) a broader wavelength coverage to prove astrophysical conditions, such as the temperature structure of circum-stellar material.

\acknowledgments
 We thank the anonymous referee for remarks that improved the manuscript. MIRC-X has been built with support from the European Research Council (ERC) under the European Commission's Horizon 2020 program (Grant Agreement Number 639889). We acknowledge support from USA National Science Foundation (NSF-ATI 1506540) and NASA-XRP NNX16AD43G. N.A. acknowledges support from the Steward Observatory Fellowship in Instrumentation and Technology Development. B.R.S.\ acknowledges support by FINESST: NASA grant \#80NSSC19K1530 and by Michigan Space Grant Consortium: NASA grant \#NNX15AJ20H. A.L.\ acknowledges funding from the UK Science and Technology Facilities Council (STFC) through grant \#630008203. Furthermore, we acknowledge travel funds from STFC PATT grant \#ST/S005293/1.
 This work is based upon observations obtained with the Georgia State University Center for High Angular Resolution Astronomy Array at Mount Wilson Observatory.  The CHARA Array is supported by the National Science Foundation under Grant No. AST-1636624 and AST-1715788.  Institutional support has been provided from the GSU College of Arts and Sciences and the GSU Office of the Vice President for Research and Economic Development. This research has made use of the Jean-Marie Mariotti Center \texttt{Aspro} and \texttt{SearchCal} services. 


\appendix

\section{\label{sec:GDT}Group delay tracking algorithm}
\begin{figure*}
\centering
\includegraphics[width=\textwidth]{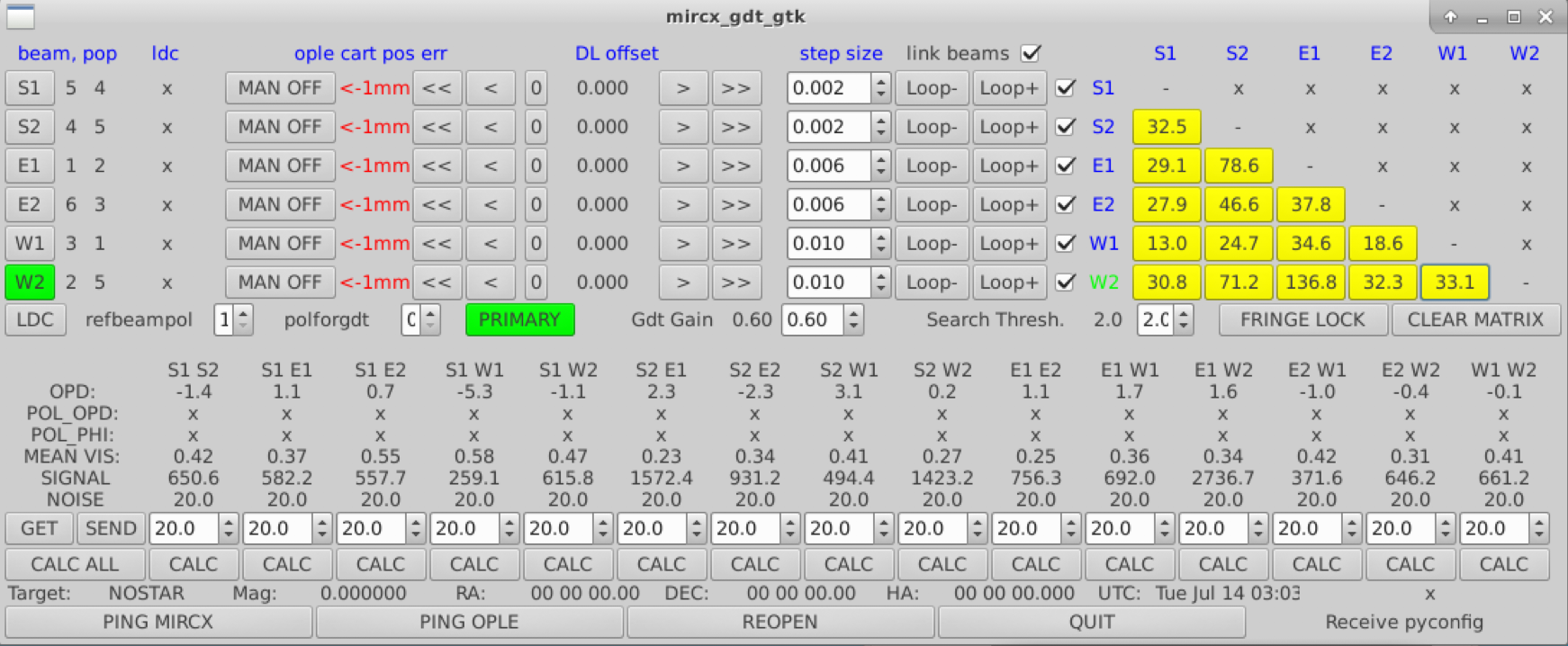}
\caption{\texttt{mircx\_gdt\_gtk}, the MIRC-X group delay tracking GUI. The \texttt{CALC ALL} button measures the thresholds of fringe detection by moving the carts away from fringes. \texttt{LOOP+} and \texttt{LOOP-} buttons enable the fringe search with step sizes entered adjacent to these buttons. Once the fringes are found on a particular pair of baselines, the buttons in the matrix are colored, and these fringes can be locked by pressing the \texttt{FRINGE LOCK} button. The \texttt{PRIMARY} button allows switching between the CHARA delay lines and the internal delay lines when sending the OPD offsets for correction. The wavelength dispersion correction is activated with \texttt{LDC} button. The GUI shows the status of beams, pops, and cart position error. It also shows the group delay measurement parameters OPD, SNR in the measurement, polarization OPD, wavelength-averaged visibility, etc. 
}
\label{Fig26_gdt}
\end{figure*}

The group delay calculation involves the following steps in the real-time server:

\begin{itemize}
\item Grab the raw detector images and carry out background subtraction

\item Add  several frames coherently

\item Remove electronic interference noise using edge columns in the frame

\item Carry out flat fielding (optional for bright targets)

\item Stretch the fringe window in the spatial direction to correct the different spatial scale of the different spectral channel

\item Compute 1-dimensional (1D) Fast Fourier transform (FFT) in the spatial direction on the fringes and get $FT^{\rm1d}$

\item Co-add the $FT^{\rm1d}$ of each polarization window if the Wollaston is in use

\item Compute 1D FFT in the spectral direction on $FT^{\rm1d}$, and get $FT^{\rm2d}$

\item Compute the power spectrum and integrate it incoherently over N coherent integration \\
$PSD =\sum_N FT^{\rm2d} FT^{{\rm2d},*}$

\item Estimate the SNR and group delay by measuring the amplitude and displacement of the peak in the $PSD$ for each baseline. Figure\,\ref{Fig26_gdt} is the group delay tracking, \texttt{mircx\_gdt\_gtk} GUI.

\item Perform bootstrapping over all triangles to obtain the final SNR and group delay estimate for each beam 

\end{itemize}

\section{\label{sec:bbias}Computing the Bispectrum Bias}

At low light levels, we need to account for a bias correction in the bispectrum \citep{Basden2004,Garcia2016}. We follow a similar procedure to that outline in Appendix C2 in \citet{Basden2004}. The bispectrum bias $\beta_{ijk}$ of triangle telescopes $ijk$,  is modeled by
\begin{equation}
\beta_{ijk}(\lambda)=C_0(\lambda)+C_1(\lambda)\,N_{\rm pho}(\lambda) + C_2(\lambda)\, \left[ D^2_{ij}(\lambda)+D^2_{jk}(\lambda)+D^2_{ki}(\lambda) \right]
\label{bbias_eqn}
\end{equation}
where $C_0$, $C_1$, and $C_2$ are the coefficients to be computed by the pipeline, $N_{\rm pho}$ is the total photometry, and 
\begin{equation*}
D^2_{ij}(\lambda) =  \big| \langle  I_{ij}( \lambda, t) \times I_{ij}(\lambda, t-1)^*  \,-\,\beta \rangle \big|
\end{equation*}
are the bias-corrected fringe power spectrum of the baselines forming the triangle, following the notation of \citet{Basden2004}. This term corresponds to the upper part of equation \ref{Eq.2}.

The BACKGROUND and FOREGROUND files are used to constrain $C_0$ and $C_1$ (offset and slope), since the $D^2$ term is necessarily zero in these dataset. The DATA files (with fringes) are used to constrain $C_2$. To estimate this final coefficient, we compute the bispectrum at closing spatial frequencies where there should be no true bispectrum signal from the source (at least one of the three frequencies is not occupied by a fringe peak).

One expects the $C_0$, $C_1$ and $C_2$ coefficients to vary depending on the number of frames which are coherently integrated $N_{\rm coh}$.  This is an additional complication because this parameter is tuned during the data reduction. Fortunately, we verified that we can get rid of this dependence with the following simple relationships:
\begin{equation}
\begin{aligned}
    C_{0,N_{\rm coh}} = C_0 / N_{\rm coh}^3 , \\
    C_{1,N_{\rm coh}} = C_1 / N_{\rm coh}^2 , \\
    C_{2,N_{\rm coh}} = C_2 / N_{\rm coh} .  
\end{aligned}
\end{equation}

\begin{figure}
\centering
\includegraphics[width=0.95\textwidth]{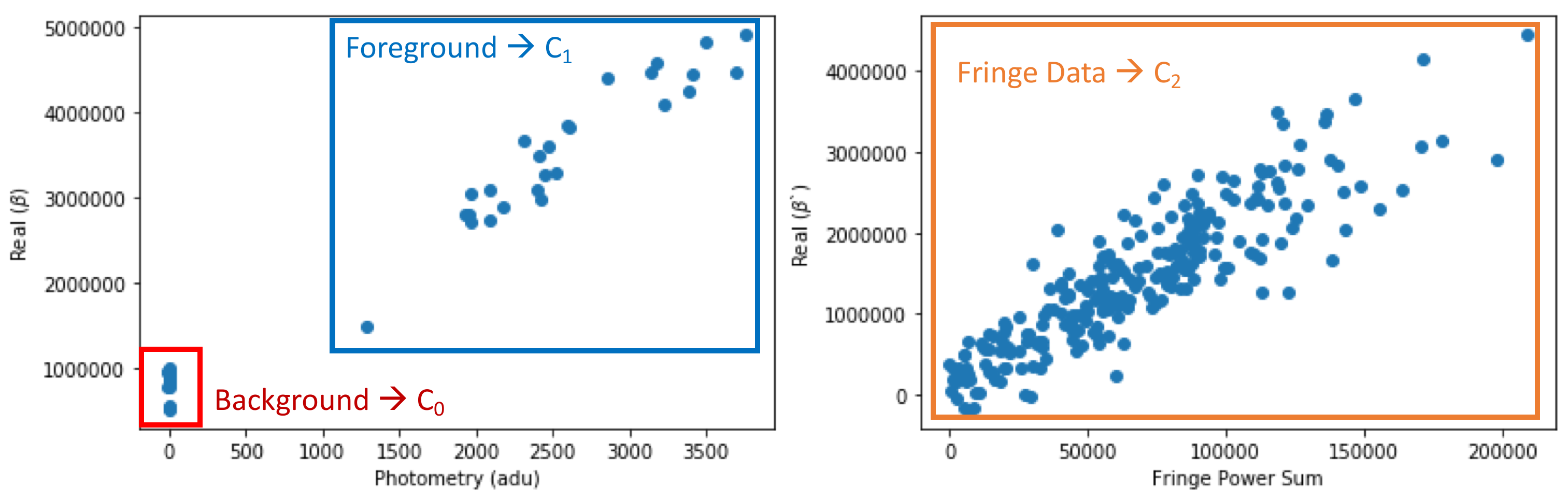}
\caption{Left panel: real part of the bispectrum for all BACKGROUND and FOREGROUND files of the night to compute the bispectrum bias coefficients $C_0$ and $C_1$ from equation\,\ref{bbias_eqn}. Right panel:  measurement of $\beta` = \beta - C_0 - C_1 N$  in the DATA files of the same night, to show the relationship for the $C_2$ term. We compute these coefficients for each spectral channel,  and perform a least-squares fit of the entire dataset to measure the coefficients. }
\label{Fig29_bbias_fig}
\end{figure}

The $C_0$, $C_1$ and $C_2$ coefficients are calibrated for each spectral channel and each detector setup (e.g., gain, frames per reset). We average over frames, ramps, and triangles to clean up our bispectrum signal in the pipeline. Figure\,\ref{Fig29_bbias_fig} shows an example of BACKGROUND, FOREGROUND, and on-sky fringe DATA used to compute the bias coefficients from equation \ref{bbias_eqn}. We perform a least-squares fit of this data to compute the coefficients.

\section{\label{sec:IotaPegObs}Observations, (u, v)-coverage and fitting residuals of $\iota$\,Peg}

\begin{figure*}
\centering
\includegraphics[width=0.45\textwidth]{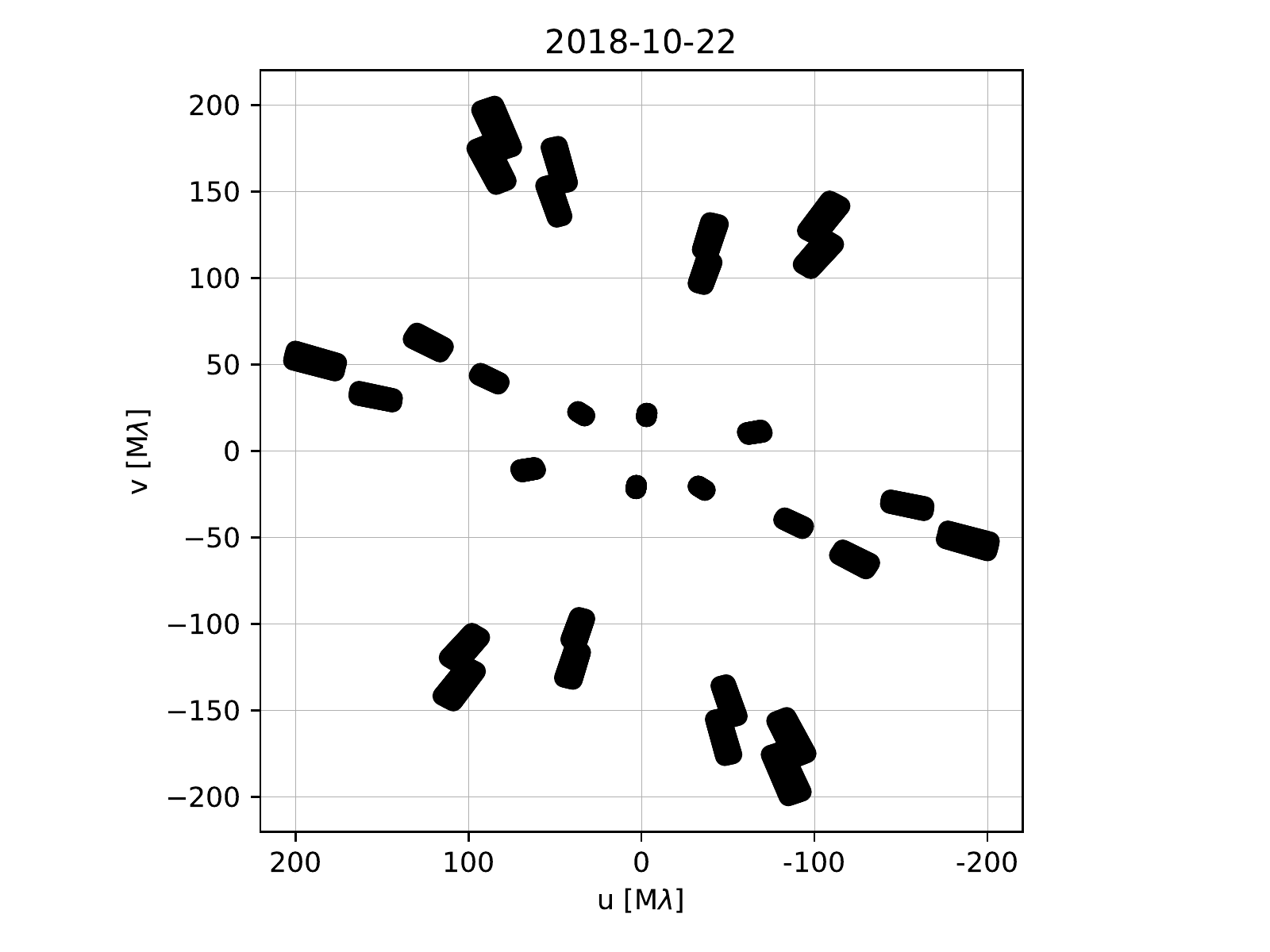}
\includegraphics[width=0.45\textwidth]{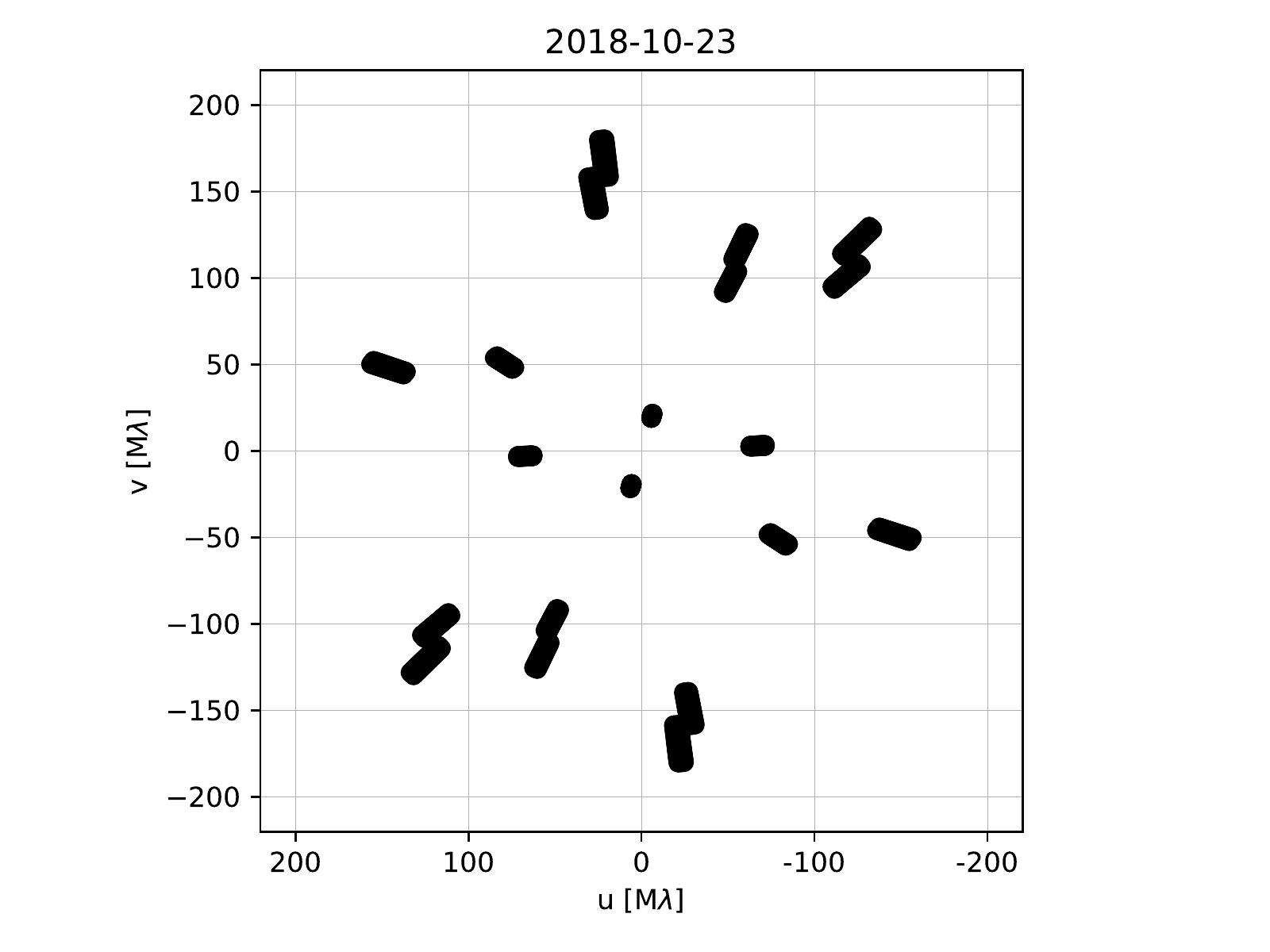}
\includegraphics[width=0.45\textwidth]{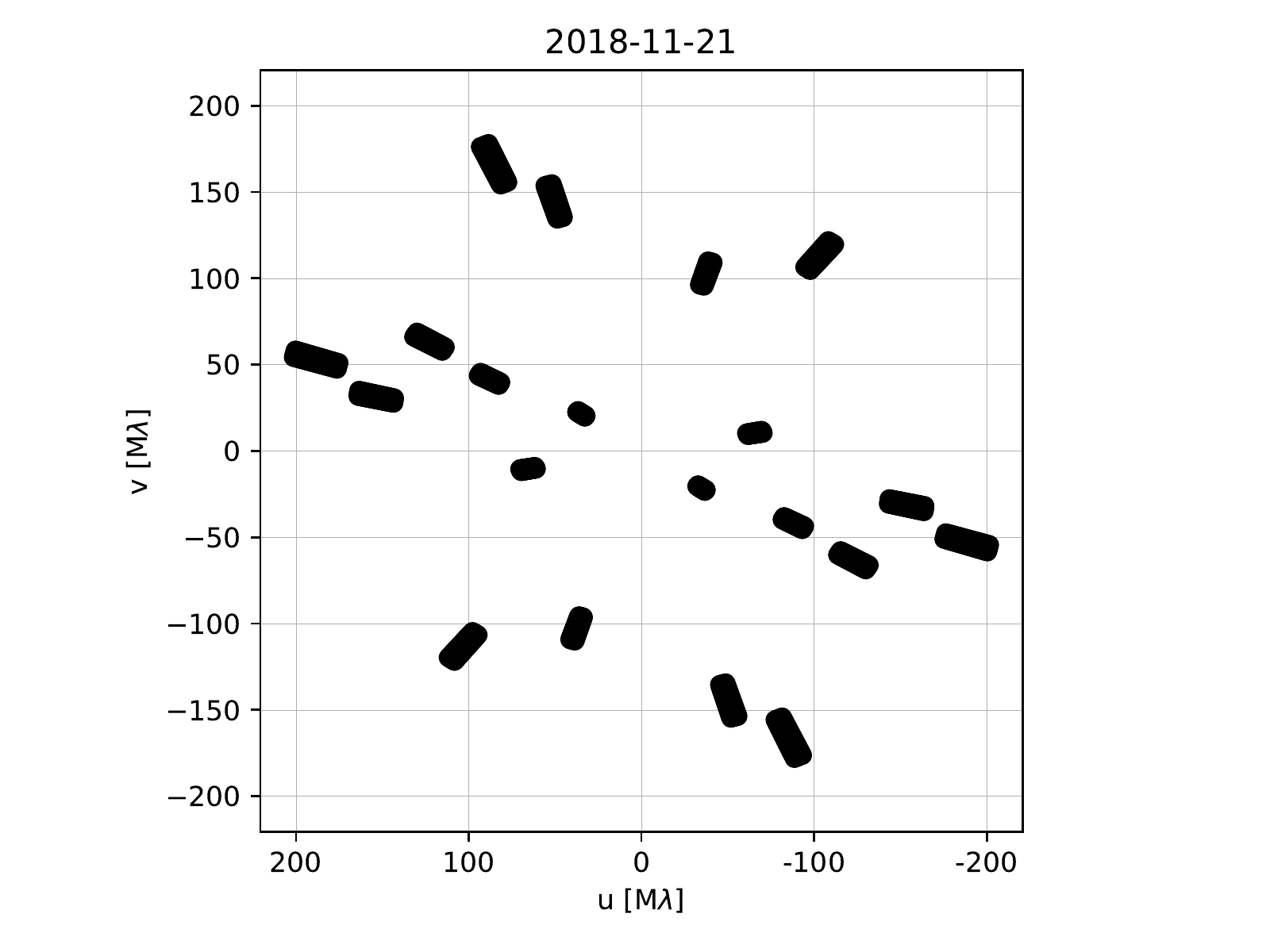}
\includegraphics[width=0.45\textwidth]{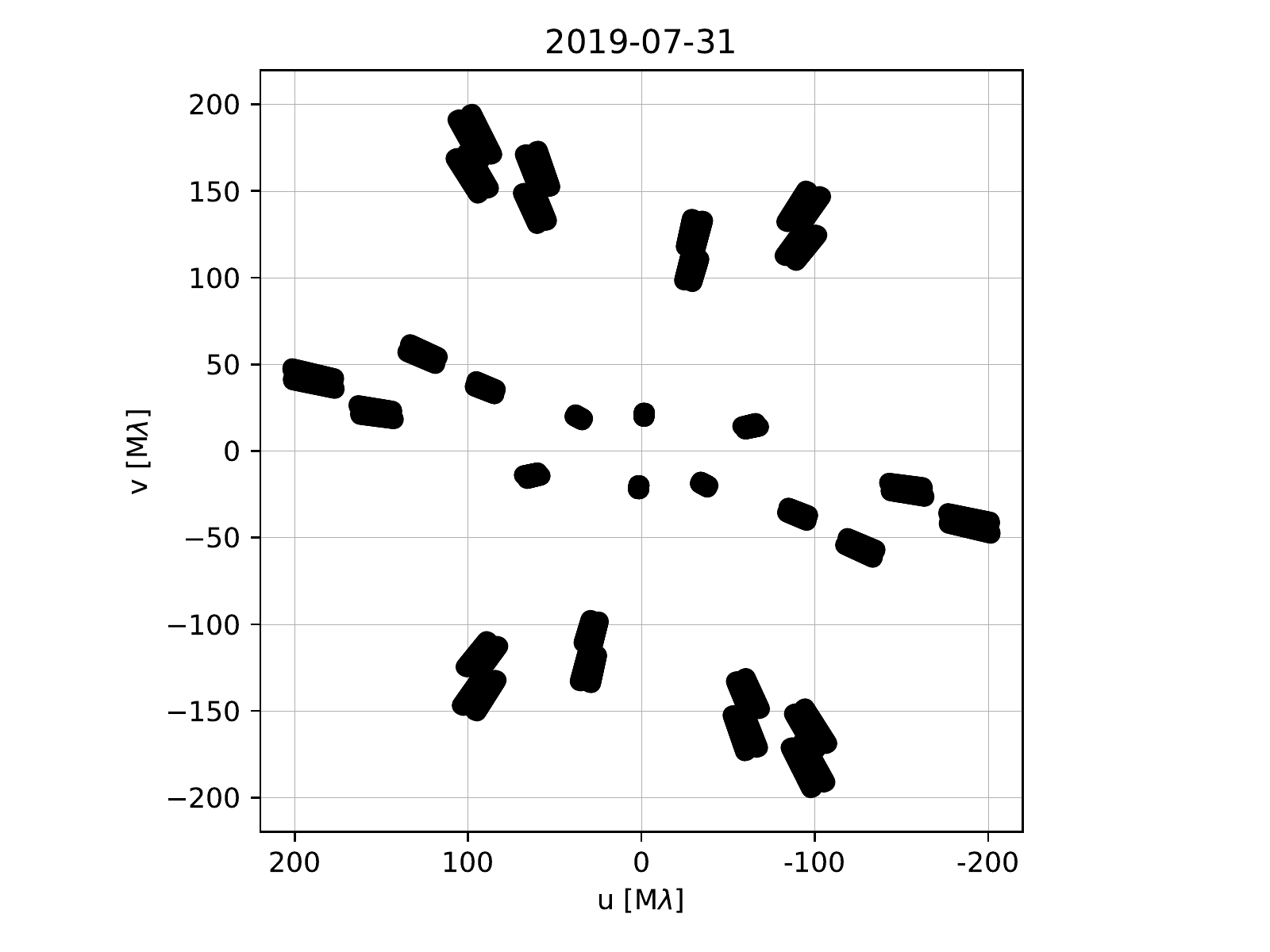}
\includegraphics[width=0.45\textwidth]{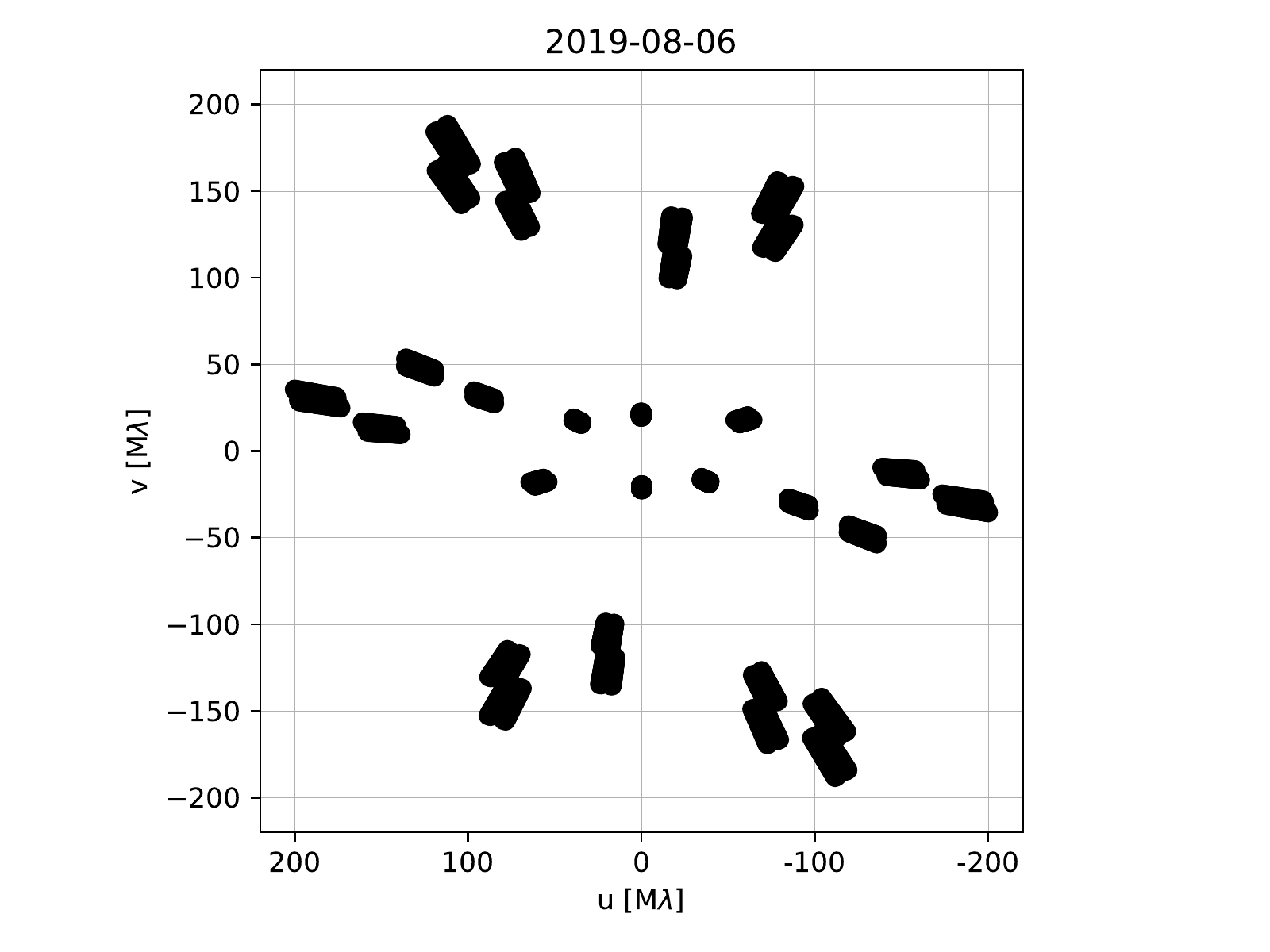}
\caption{(u, v)-coverage of observations of object $\iota$\,Peg are shown. Each panel corresponds to an epoch.  The radial tracks represent spectral dispersion over 34 spectral channels observed with grism R~=~190.}
\label{Fig27_UVcoverage}
\end{figure*}

\begin{figure*}
\centering
\includegraphics[width=0.45\textwidth]{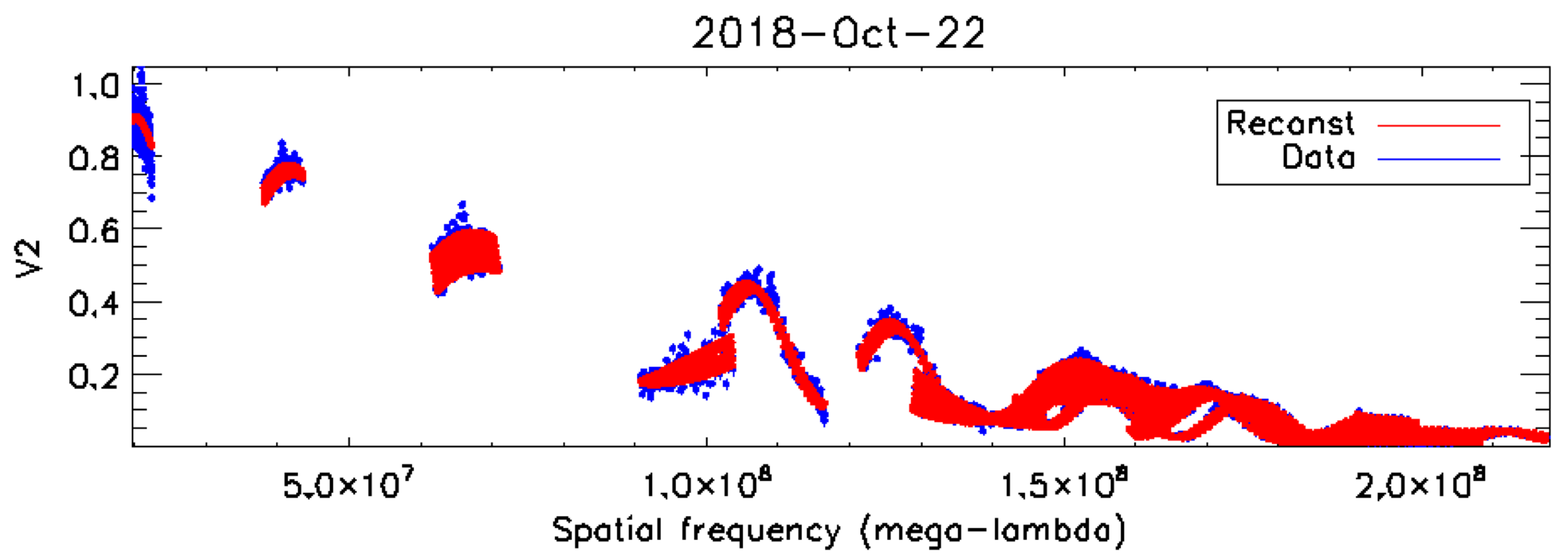}
\includegraphics[width=0.45\textwidth]{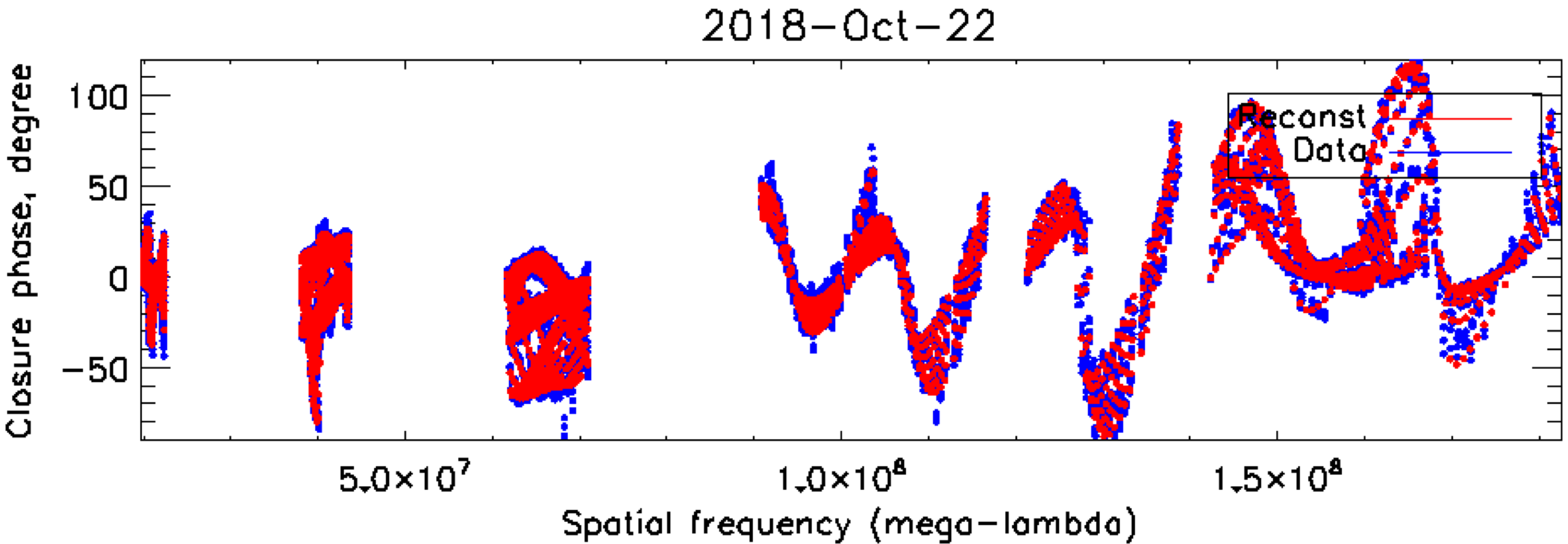}
\includegraphics[width=0.45\textwidth]{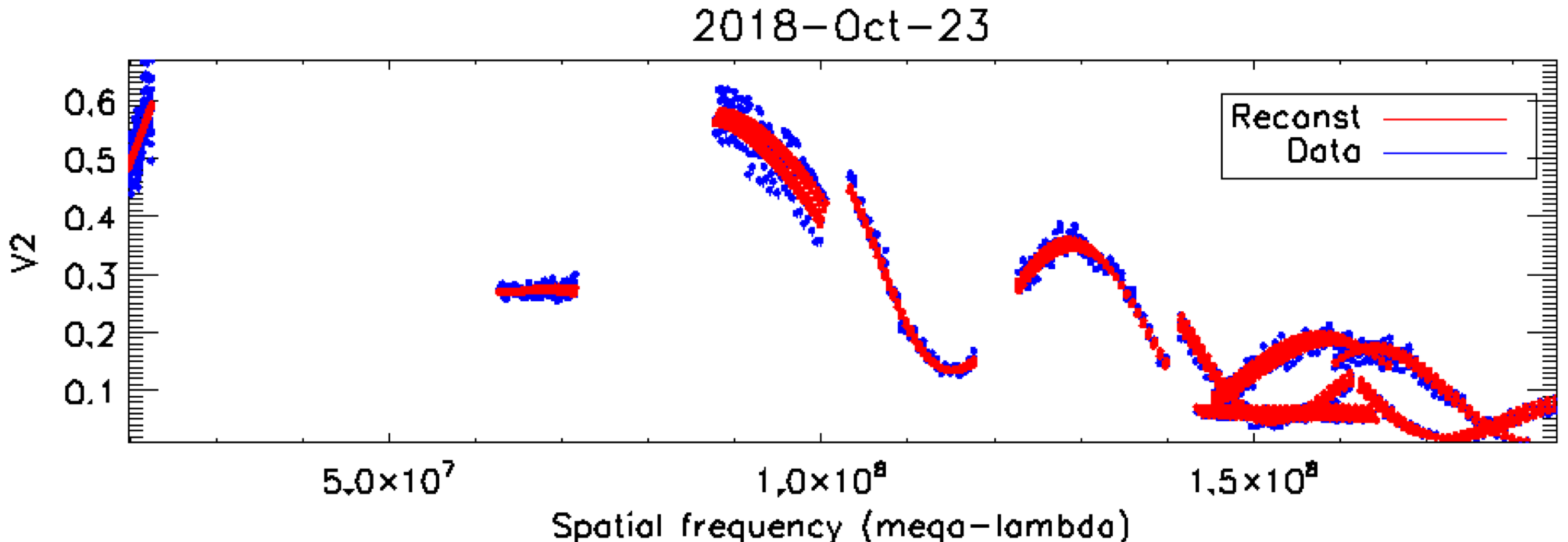}
\includegraphics[width=0.45\textwidth]{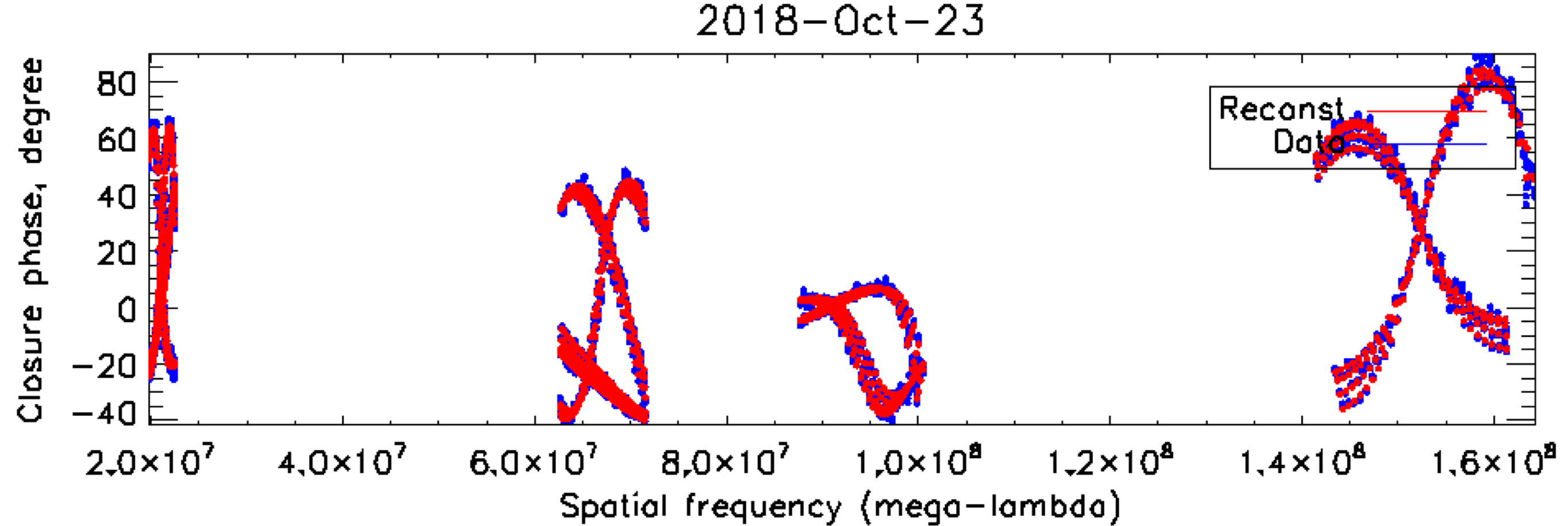}
\includegraphics[width=0.45\textwidth]{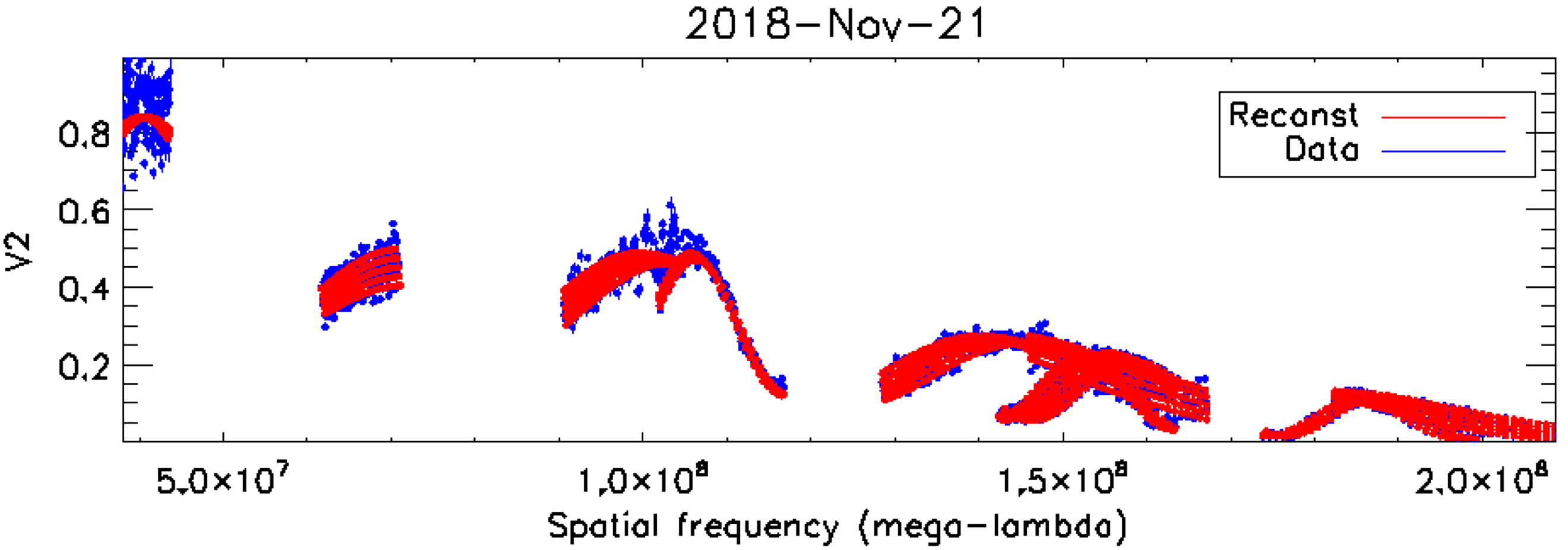}
\includegraphics[width=0.45\textwidth]{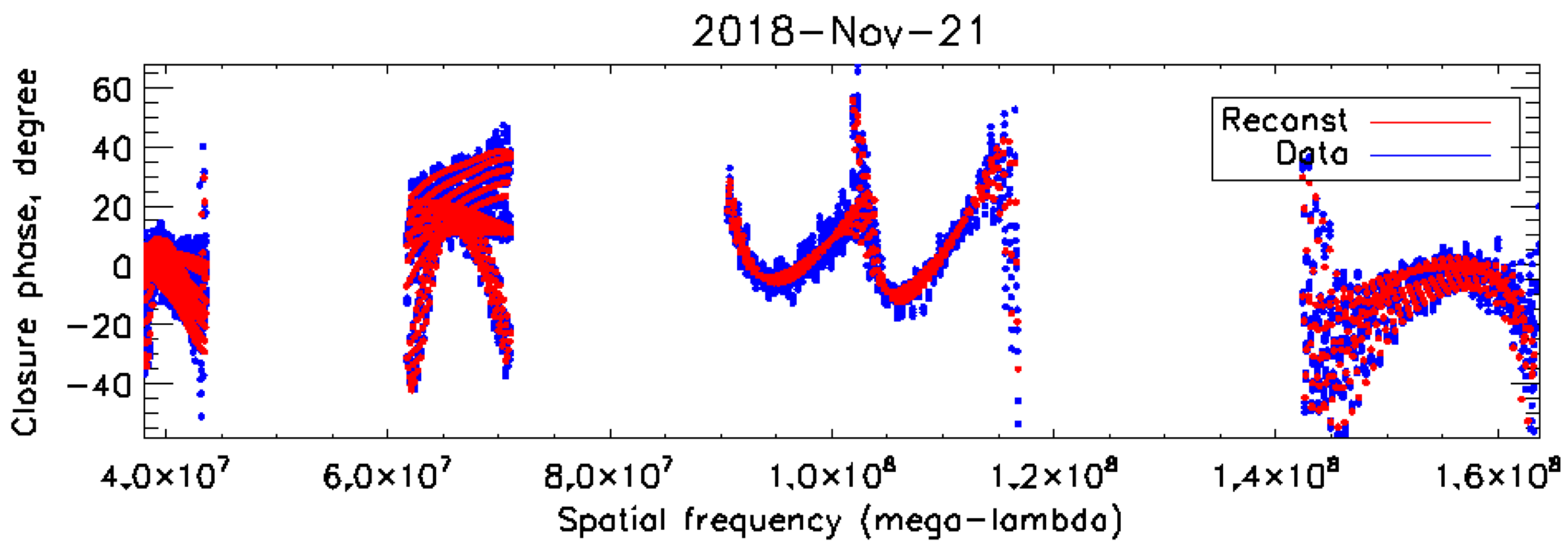}
\includegraphics[width=0.45\textwidth]{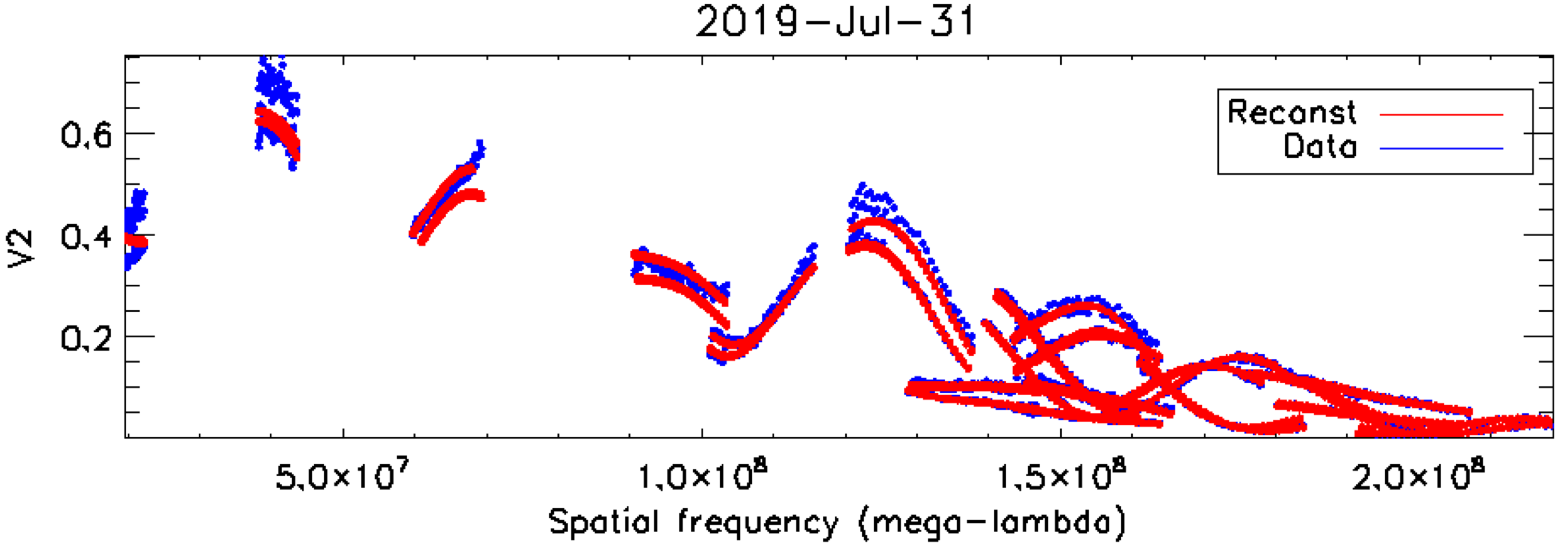}
\includegraphics[width=0.45\textwidth]{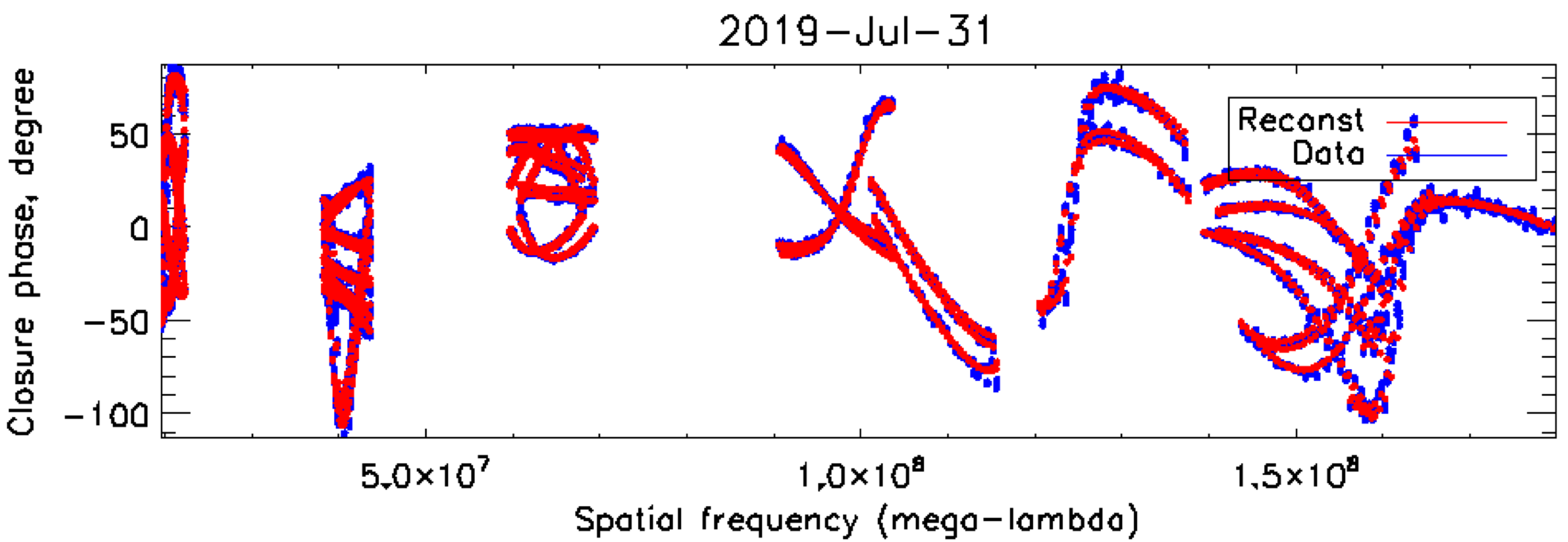}
\includegraphics[width=0.45\textwidth]{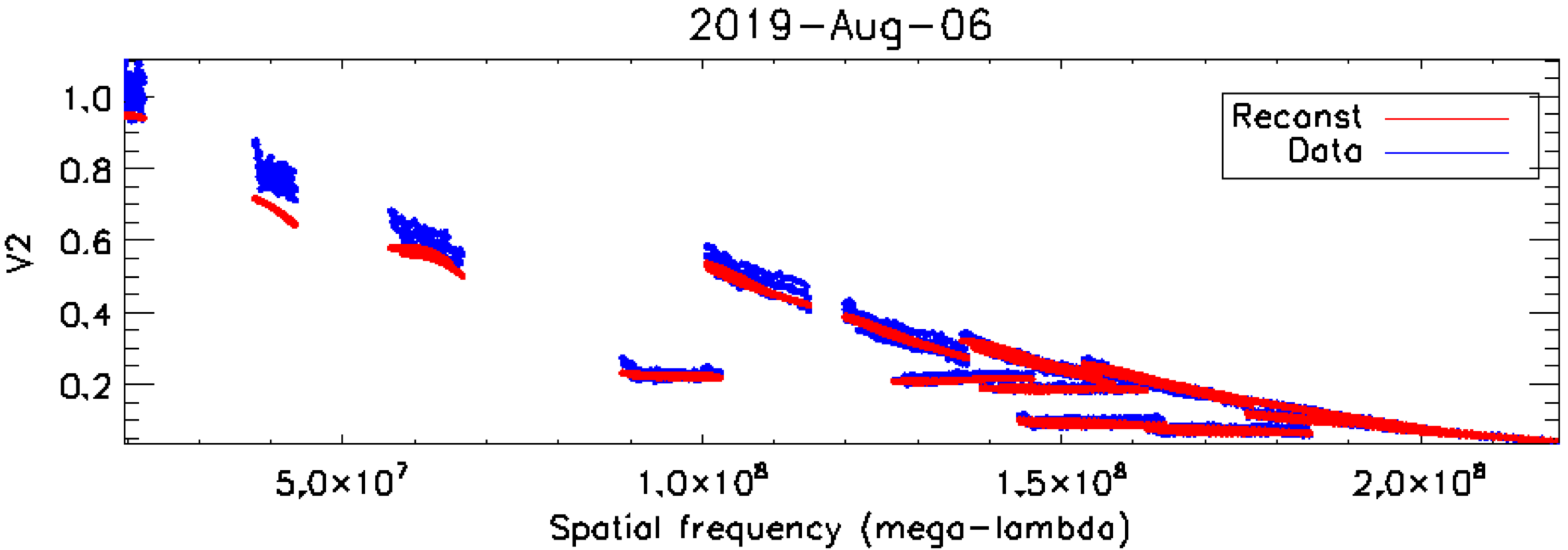}
\includegraphics[width=0.45\textwidth]{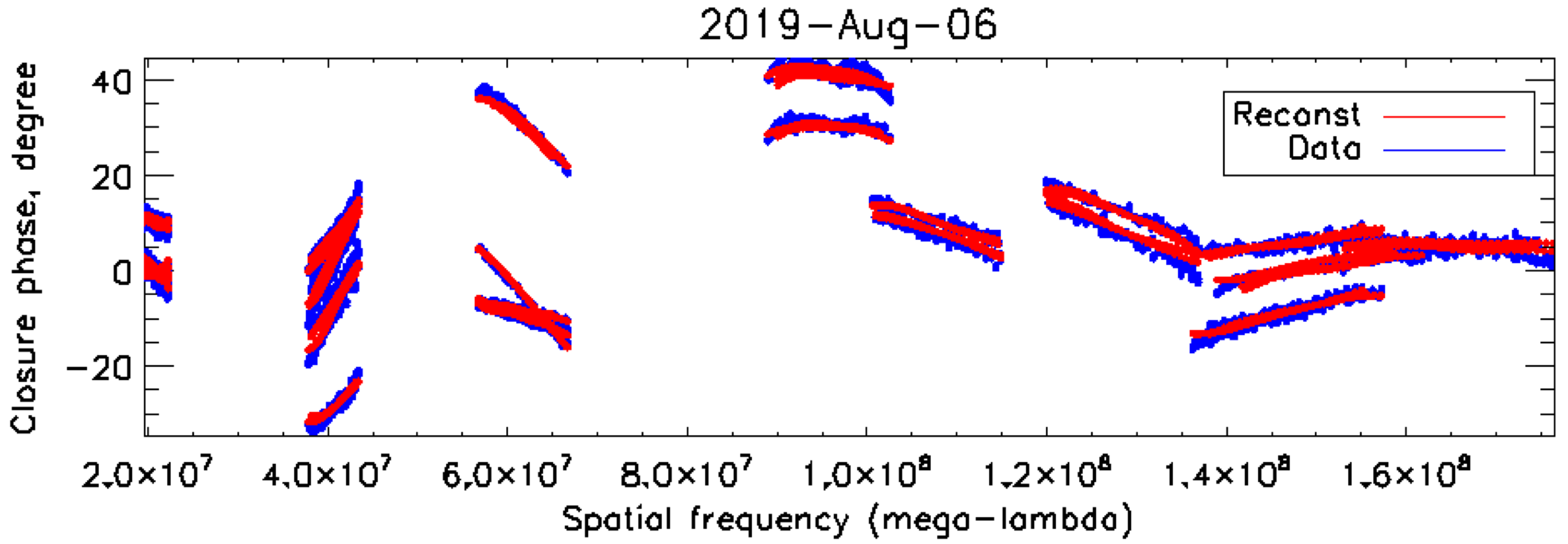}
\caption{The squared visibility $V^2$ and closure phase $C$ of $\iota$\,Peg for different epochs. The blue and red curves are the data and fitting of image reconstruction. $V^2$ is a measure of spatial resolution of the source by a given baseline at a
given wavelength.  The closure phases are related to the asymmetry of the binary system: the closure phase is null for a centro-symmetric target. }
\label{Fig28_Iota_peg_v2_c3}
\end{figure*}

$\iota$\,Peg was observed on UT dates 2018-10-22, 2018-10-23, 2018-11-21, 2019-07-31 and 2019-08-06 with its calibrator $\gamma$\,Peg  \citep[diameter in the H-band = $0.3719123\pm 0.02$, JSDC;][]{Bourges2017}. Figure\,\ref{Fig27_UVcoverage} presents the (u, v)-coverage. Figure\,\ref{Fig28_Iota_peg_v2_c3} shows the $V^2$ and $C$. All the observations were made with the same instrument configuration and detector avalanche gain 40. 
  
\bibliography{main}{}
\bibliographystyle{aasjournal}
\end{document}